\documentclass[useAMS,usenatbib]{mnras}
\pdfoutput=1

\usepackage{amssymb} 
\usepackage{amsmath} 
\usepackage{graphicx}
\usepackage{longtable,lscape}

\title[Proper motions of jets in Carina]{Proper motions of collimated jets from intermediate-mass protostars in the Carina Nebula} 

\author[Reiter et al.]{
Megan Reiter$^{1}$\thanks{email: mreiter@umich.edu (MR)}, 
Megan M. Kiminki$^{2}$, Nathan Smith$^{2}$, and John Bally$^{3}$ \\
$^{1}$University of Michigan, Ann Arbor, MI 48109, USA \\
$^{2}$Steward Observatory, University of Arizona, Tucson, AZ 85721, USA \\
$^{3}$Center for Astrophysics and Space Astronomy, University of Colorado, 389 UCB, Boulder, CO 80309, USA}

\begin{document}

\date{Accepted XXX. Received YYY; in original form ZZZ}
\pagerange{\pageref{firstpage}--\pageref{lastpage}} \pubyear{2017}
\maketitle
\label{firstpage}


\begin{abstract}
We present proper motion measurements of 37 jets and HH objects in the Carina Nebula measured in two epochs of H$\alpha$ images obtained $\sim 10$~yrs apart with \emph{HST}/ACS. 
Transverse velocities in all but one jet are faster than $\gtrsim 25$~km~s$^{-1}$, confirming that the jet-like H$\alpha$ features identified by \citet{smi10} trace outflowing gas. 
Proper motions constrain the location of the jet-driving source and provide kinematic confirmation of the intermediate-mass protostars that we identify for 20/37 jets. 
Jet velocities do not correlate with the estimated protostar mass and embedded driving sources do not have slower jets. 
Instead, transverse velocities (median $\sim 75$~km~s$^{-1}$) are similar to those in jets from low-mass stars. 
Assuming a constant velocity since launch, we compute jet dynamical ages (median $\sim 10^4$~yr). 
If continuous emission from inner jets traces the duration of the most recent accretion bursts, then these episodes are sustained longer (median $\sim 700$~yr) than the typical decay time of an FU~Orionis outburst. 
These jets can carry appreciable momentum that may be injected into the surrounding environment. 
The resulting outflow force, $dP/dt$, lies between that measured in low- and high-mass sources, despite the very different observational tracers used. 
Smooth scaling of the outflow force argues for a common physical process underlying outflows from protostars of all masses. 
This latest kinematic result adds to a growing body of evidence that intermediate-mass star formation proceeds like a scaled-up version of the formation of low-mass stars.
\end{abstract}

\begin{keywords}
stars: formation --- jets --- outflows 
\end{keywords}


\section{Introduction}\label{s:intro}

Bipolar outflows appear to be a ubiquitous feature of star formation.
Outflows are observed from both low- and high-mass sources, where the active accretion of the embedded protostars must power the outflows. 
The underlying physical mechanisms responsible for jet launch and collimation are not well-understood, especially for higher-mass sources where the geometry of the circumstellar accretion disk may be appreciably different than in low-mass stars \citep[e.g.,][]{vin02}. 
Differences in the outflow morphology seen between low- and high-mass sources have been cited to suggest that the dominant physical processes may not be the same across all ZAMS masses \citep[e.g.,][]{she03}. 
Highly collimated jets have been observed from many low-mass ($<2$~M$_{\odot}$) sources \citep[e.g.,][]{ray90,rei98,bal01}. 
In contrast, there are few reports of small opening-angle outflows from high-mass ($>8$~M$_{\odot}$) sources \citep[e.g.][]{lacy07,guz11,guz16}.

In this paper, we make the distinction between jets and outflows according to their morphology and kinematics \citep[e.g.,][]{fra14}. 
Fast, narrow \textit{jets} have small opening angles ($\sim 5^{\circ}$) and higher velocities than the typical CO outflow (on the order of $\sim 100$~km~s$^{-1}$). 
Jets appear to be launched near the protostar and quickly collimated into narrow streams above the poles of the star.
Many collimated jets have been observed emanating from low-mass protostars where they propagate outside the natal cloud and can be observed in near-IR and optical images (e.g., HH~34, HH~46/47, HH~111).

In contrast to jets, \textit{outflows} are slower ($\sim 10$s of km~s$^{-1}$), often with wider opening angles. 
Outflows may be launched from a variety of radii in the disk, and/or entrained from the circumstellar environment by the underlying jet, or a combination of the two.
Molecular outflows are seen from sources of all masses, often in CO \citep[e.g.,][]{gue96,gue99,zap05,hul16} or SiO \citep[e.g.][]{gue98,hir06,cod07,lop11}, and occasionally in less abundant molecules \citep[e.g.][]{lee07,tap08,pod15}. 
Outflows are observed to have a variety of opening angles, from narrow, jet-like flows \citep[e.g., HH~212,][]{gue99} to extremely wide features more characteristic of the widening of the circumstellar envelope. 
Indeed, the opening angle of the outflow may be an evolutionary indicator \citep[e.g.,][]{arc06}.  
Higher mass sources tend to be deeply embedded in their early evolution, so their mass loss tends to be observed as molecular outflows \citep[e.g.,][]{fue01,bel08,bro16}.

\begin{figure*}
\centering
\includegraphics[trim=0mm 20mm 0mm 10mm,angle=0,scale=0.675]{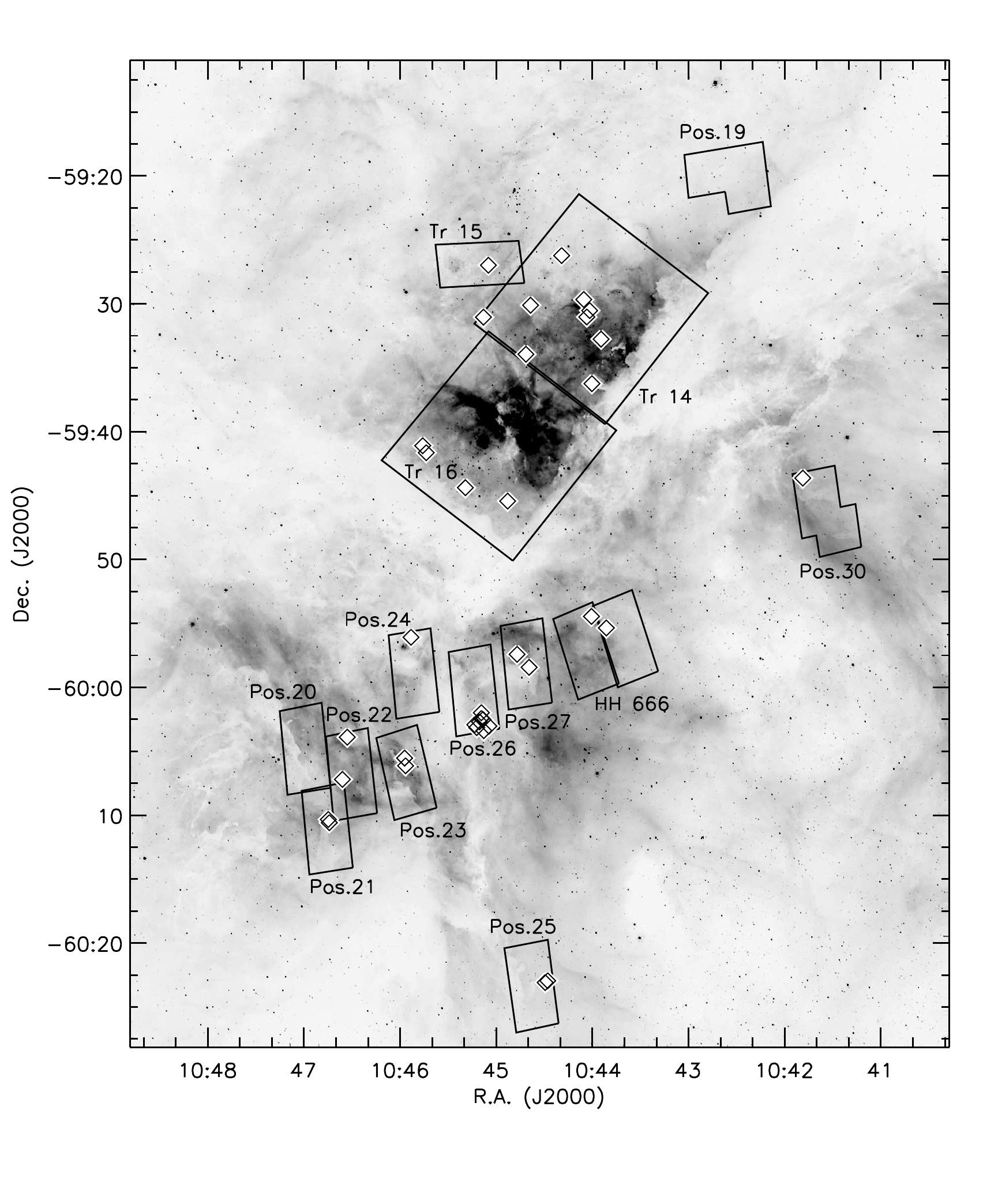} 
\caption{Ground-based H$\alpha$ image of the Carina Nebula with boxes showing the position of our \emph{HST} pointings. 
Position names correspond to \citet{smi10}. 
The location of the jets are marked with white diamonds (outlined in black).
}\label{fig:hst_pointings} 
\end{figure*}

Similar outflow behavior regardless of the mass of the driving protostar argues strongly for a common physical mechanism governing their launch and collimation. 
This is of particular interest for higher-mass driving sources where the accretion mechanism is not understood in detail.
However, the complexity of typical high-mass star-forming regions makes it challenging to observe individual sources with the same techniques used for isolated low-mass protostars. 
High-mass stars are rare, and therefore found in more distant regions. 
Large column densities obscure the accreting protostars during their brief pre-main-sequence evolution, requiring long-wavelength observations where several sources may be unresolved in the beam \citep[see, e.g.,][]{mau15}.

Fortunately, intermediate-mass ($\sim 2-8$~M$-{\odot}$) stars provide a bridge between low- and high-mass sources in both the physics of formation and observational accessibility. 
Compared to high-mass stars, they are more numerous in nearby regions, with examples in relatively unobscured environments that can be studied with the same techniques used in low-mass stars \citep[e.g.,][]{mcg04,ell13}. 
At the same time, observations of molecular outflows from embedded intermediate-mass sources \citep[e.g.,][]{bel08,van12,van16} provide a more direct comparison with the typical observation of an outflow from a high-mass star \citep[e.g.,][]{mau15}.

The Carina Nebula hosts the largest known population of jets driven by intermediate-mass protostars in a single region.
\citet{smi10} discovered 40 jets and candidate jets in their H$\alpha$ survey with \emph{HST}/ACS.
More than 65 O-type stars form a giant H~{\sc ii} region within Carina \citep{smi06a}.
Their combined UV radiation lights up the nearby jets, revealing material in the jet body that lies between shock fronts and would remain invisible in more quiescent regions.
\citet{smi10} infer that the jet-driving protostars are likely intermediate-mass sources, based on the high mass-loss rates estimated from the H$\alpha$ emission measure. 
Bright [Fe~{\sc ii}] emission observed in every jet targeted for near-IR follow-up indicates high densities in the bodies of these jets \citep[required to prevent ionization to Fe$^{++}$, see][]{rei13,rei16}. 
Jet densities inferred from the [Fe~{\sc ii}] emission are an order of magnitude higher than estimated from H$\alpha$, leading to a corresponding order of magnitude increase in the mass-loss rate. 
Spectra and proper motions measured for four of the most powerful jets reveal outflow velocities similar to those measured from low-mass stars \citep{rei14}.
Proper motions also provide kinematic confirmation of candidate jet-driving sources that have been identified near the axis of the jet \citep{ohl12,rei13,rei16}.

In this paper, we present proper motions of 37 jets and HH objects in the Carina Nebula.
Unlike \citet{rei14} which only considered four jets with multi-epoch imaging, we report transverse velocities for all of the HH jets in Carina discovered by \citet{smi10}.
To do so, we use two epochs of images obtained with the same narrowband filter and camera separated by a longer time baseline \citep[$\sim 8-10$~yr, compared to $\sim 4$~yr in][]{rei14} in order to minimize systematic uncertainties. 
First results from this program were presented in \citet{rei15b,rei15a,rei17}.


\section{Observations}\label{s:obs}

\begin{table*}
\caption[ACS observations of HH jets in the Carina Nebula]{ACS observations of HH jets in the Carina Nebula}
\vspace{5pt}
\centering
\vspace{3pt}
\begin{tabular}{llllllll}
\hline\hline
Target & RA & DEC & Pos & Date & Date & $\Delta$t \\ 
 & J2000 & J2000 & num & $1^{st}$ Obs. & $2^{nd}$ Obs. & [yr] \\ 
\hline 
HH~666$^{* \dagger}$ & 10:43:51.3 & --59:55:21 & HH~666 & 2005 Mar 30 & 2014 Feb 21 & 8.9 \\
HH~900$^{\star}$  & 10:45:19.3 & --59:44:23 & Tr~16 & 2005 Jul 18 & 2014 Aug 04 & 9.1 \\
HH~901$^*$ & 10:44:03.5  & --59:31:02 & Tr~14 & 2005 Jul 17 & 2015 Jan 16 & 9.5 \\
HH~902$^*$ & 10:44:01.7  & --59:30:32 & Tr~14 & 2005 Jul 17 & 2015 Jan 16 & 9.5 \\
HH~903  & 10:45:56.6 & --60:06:08 & Pos~23 & 2005 Jun 26 & 2014 Feb 22 & 8.9 \\
HH~1004 & 10:46:44.8 & --60:10:20 & Pos~21 & 2006 Mar 13 & 2015 Mar 11 & 9.0 \\
HH~1005 & 10:46:44.2 & --60:10:35 & Pos~21 & 2006 Mar 13 & 2015 Mar 11 & 9.0 \\
HH~1006 & 10:46:33.0 & --60:03:54 & Pos~22 & 2006 Mar 16 & 2015 Mar 11 & 9.0 \\
HH~1007 & 10:44:29.5 & --60:23:05 & Pos~25 & 2005 Sept 10 & 2015 Mar 11 & 9.5 \\
HH~1008 & 10:44:47.0 & --59:57:25 & Pos~27 & 2006 Mar 18 & 2015 Mar 12 & 9.0 \\
HH~1009 & 10:44:39.5 & --59:28:26 & Pos~27 & 2006 Mar 18 & 2015 Mar 12 & 9.0 \\
HH~1010 & 10:41:48.7 & --59:43:38 & Pos~30 & 2005 Sept 11 & 2015 Mar 12 & 9.5 \\
HH~1011 & 10:45:04.9 & --59:26:59 & Tr~15 & 2006 Nov 26 & 2014 Nov 23 & 8.0 \\
HH~1012 & 10:44:38.6 & --59:30:07 & Tr~14 & 2005 Jul 17 & 2015 Jun 28 & 10.0 \\
HH~1013 & 10:44:19.2 & --59:26:14 & Tr~14 & 2005 Jul 17 & 2015 Jun 28 & 10.0 \\
HH~1014 & 10:45:45.9 & --59:41:06 & Tr~16 & 2005 Jul 18 & 2014 Aug 04 & 9.1 \\
HH~1015 & 10:44:27.9 & --60:22:57 & Pos~25 & 2005 Sept 10 & 2015 Mar 11 & 9.5 \\
HH~1016 & 10:45:53.2 & --59:56:05 & Pos~24 & 2006 Mar 15 & 2015 Mar 11 & 9.0 \\
HH~1017 & 10:44:41.5 & --59:33:57 & Tr~14 & 2005 Jul 17 & 2015 Jan 16 & 9.5 \\
HH~1018 & 10:44:52.9 & --59:45:26 & Tr~16 & 2005 Jul 18 & 2014 Aug 04 & 9.1 \\
HH~1019 & 10:44:00.3 & --59:36:15 & Tr~14 & 2005 Jul 17 & 2015 Jun 28 & 10.0 \\
HH~1066$^*$ & 10:44:05.4 & --59:29:40 & Tr~14 & 2005 Jul 17 & 2015 Jan 16 & 9.5 \\
HH~1156 & 10:45:45.9 & --59:41:06 & Tr~16 & 2005 Jul 18 & 2014 Aug 04 & 9.1 \\
HH~1159 & 10:45:08.3 & --60:02:31 & Pos~26 & 2006 Mar 2 & 2015 Mar 12 & 9.0 \\
HH~1160 & 10:45:09.3 & --60:01:59 & Pos~26 & 2006 Mar 2 & 2015 Mar 12 & 9.0 \\
HH~1161 & 10:45:09.3 & --60:02:26 & Pos~26 & 2006 Mar 2 & 2015 Mar 12 & 9.0 \\
HH~1162 & 10:45:13.4 & --60:02:55 & Pos~26 & 2006 Mar 2 & 2015 Mar 12 & 9.0 \\
HH~1163 & 10:45:12.2 & --60:03:09 & Pos~26 & 2006 Mar 2 & 2015 Mar 12 & 9.0 \\
HH~1164 & 10:45:10.5 & --60:02:42 & Pos~26 & 2006 Mar 2 & 2015 Mar 12 & 9.0 \\
HH~1166  & 10:44:45.3 & --59:55:50 & Pos~27 & 2006 Mar 18 & 2015 Mar 12 & 9.0 \\
HH~1167  & 10:45:04.6 & --60:03:02 & Pos~26 & 2006 Mar 2 & 2015 Mar 12 & 9.0 \\
HH~1168  & 10:45:08.0 & --59:31:03 & Tr~14 & 2005 Jul 17 & 2015 Jun 28 & 10.0 \\
HH~1169 & 10:45:56.6 & --60:06:08 & Pos~23 & 2005 Jun 26 & 2014 Feb 22 & 8.9 \\
HH~1170 & 10:46:36.0 & --60:07:12 & Pos~22 & 2006 Mar 16 & 2015 Mar 11 & 9.0 \\
HH~1171 & 10:45:07.8 & --60:03:23 & Pos~26 & 2006 Mar 2 & 2015 Mar 12 & 9.0 \\
HH~1172 & 10:44:00.6 & --59:54:27 & HH~666 & 2005 Mar 30 & 2014 Feb 21 & 8.9 \\
HH~1173 & 10:43:54.6 & --59:32:46 & Tr~14 & 2005 Jul 17 & 2015 Jan 16 & 9.5 \\
\hline
\multicolumn{7}{l}{see also $^*$ \citet{rei14}, $^{\dagger}$ \citet{rei15b}, $^{\star}$ \citet{rei15a} } \\ 
\end{tabular} 
\label{t:obs}
\end{table*}


We present new \emph{HST}/ACS H$\alpha$ images of the Carina Nebula obtained between 21 February 2014 and 28 June 2015 under programmes GO-13390 and GO-13791. 
These second epoch observations were designed to duplicate the observational setup of the original \emph{HST}/ACS H$\alpha$ survey of Carina presented in \citet{smi10} as closely as possible. 
The new data cover all of the fields imaged in Carina and include 35 jets and candidate jets (see Figure~\ref{fig:hst_pointings}). 
NGC~3324 was not included in the second epoch observations, so we are unable to measure proper motions of the two jets and two candidate jets in that region. 
However, we include two jets in Carina that were discovered in subsequent analysis of the original survey data (HH~1164, \citealt{rei16} and HH~1019, \citealt{rei17}) for a sample of 37 jets. 

The first epoch of ACS observations was taken in 2005--2006 \citep[programmes GO-10241 and GO-10475; see][]{smi10}.
Observations were organized in groups by orbit, with each orbit composed of three pairs of {\tt CR-SPLIT} exposures.
In most cases, the three pairs were arranged to form a 205\arcsec $\times$ 400\arcsec tile, with the exposure offsets designed to fill in the inter-chip gaps.
We repeated these observations in 2014--2015 (programmes GO-13390 and GO-13791, PI: N.\ Smith), aiming to replicate the pointings and position angles of the original observations as closely as possible in order to limit position-dependent systemic effects when measuring proper motions.
Due to changes in the \emph{HST} Guide Star Catalog over the decade between observations, we were forced to rotate a total of six tiles by $\sim$180\degr.
This rotation introduced an additional component of uncertainty into measurements of jets located in Tr14, POS~25, and POS~30 (see Table~\ref{t:obs}). 

In both epochs, images were obtained with an exposure time of $1000$~s using the F658N filter (which transmits both H$\alpha$ and [N~{\sc ii}] $\lambda 6583$). 
Comparing the new data with first epoch observations allows us to measure proper motions over a time baseline of $\sim 8-10$~yr; specific dates of observation and time intervals are listed for each jet in Table~\ref{t:obs}.

We align and stack the images of each orbit's tile as described in \citet{rei15b,rei15a}, adapting the method of \citet{anderson2008a,anderson2008b}, \citet{andersonvandermarel2010}, and \citet{sohn2012}.
The twelve images of each tile (six per epoch) are aligned to a common master reference frame via the positions of stars measured with point spread function (PSF) photometry.
All reference frames have 50-mas pixels and are aligned so the $y$ axis points north.
Since the reference frame for each tile is based on the average position of the stars in that tile, these reference frames are not tied to an external proper-motion zero-point.
Instead, they are in the frame of the Carina Nebula, allowing us to measure the local motions of jet features without having to correct for Galactic rotation and other large-scale motions.

PSF photometry was performed using the program {\tt img2xym\_WFC.09x10} \citep{andersonking2006}, which employs a library of chip-position-dependent effective PSFs.
Stellar positions are then corrected for geometric distortion as per \citet{anderson2006}.
The positions of well-measured, uncrowded stars are used to find the linear transformations from each exposure into the master reference frame.
We repeat the fitting process a total of three times, refining the reference-frame positions based on the results of each iteration.
Finally, the six images from each epoch are resampled and stacked into a single image.
The stacked images from two epochs of one observed tile are thus aligned to the same reference frame and are directly comparable.
The formal alignment error across one tile is $<$2 mas ($\sim$2 km s$^{-1}$ over 9--10 years at the distance of the Carina Nebula).

\begin{figure*}
\centering
\includegraphics[trim=0mm 0mm 0mm 0mm,angle=0,scale=0.5]{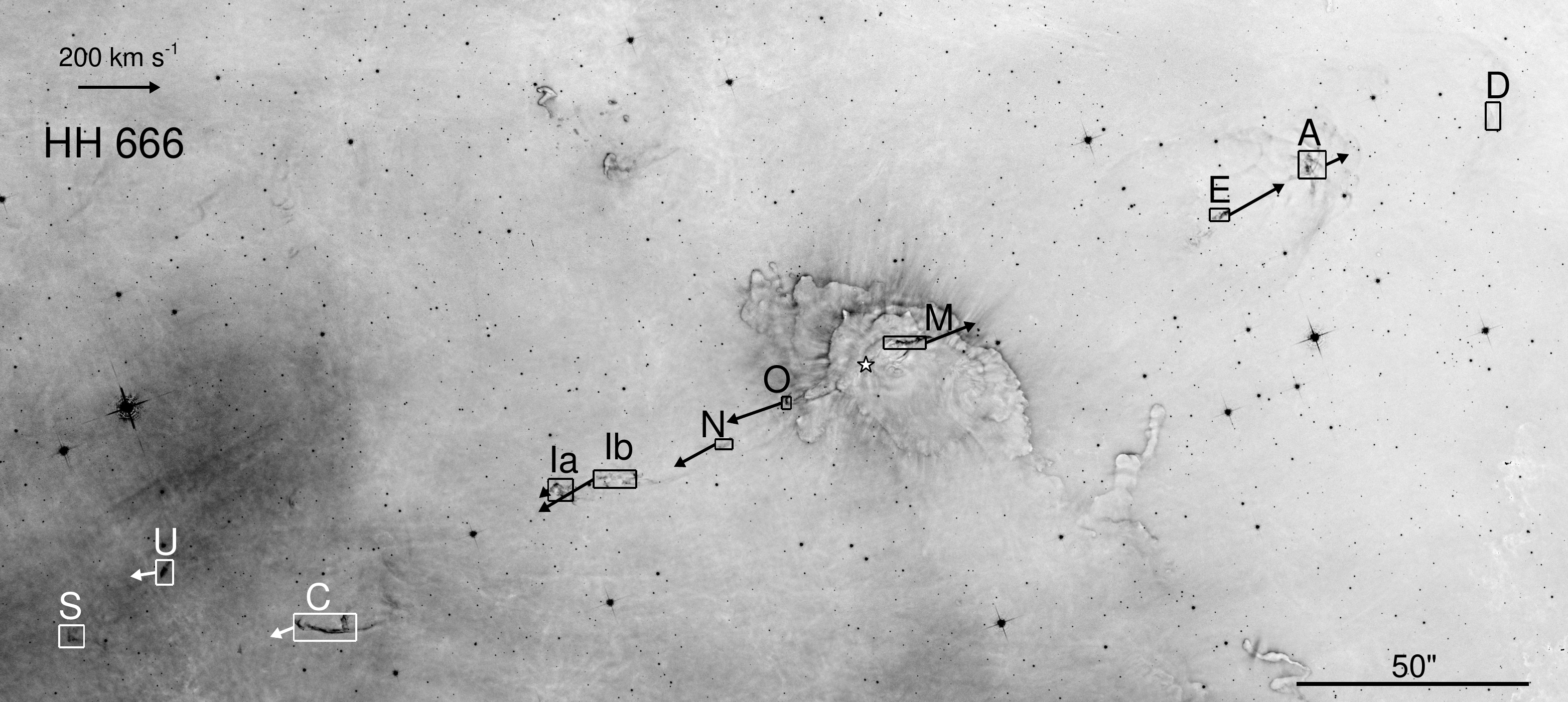} 
\caption{H$\alpha$ image of HH~666 showing the full extent of the jet. In this paper, we present proper motion measurements of the large knots identified by \citet{smi04,smi10}. Boxes indicate the features used to measure proper motions and the arrows mark the direction and magnitude of the transverse velocity. Feature velocities are listed in Table~\ref{t:jet_pm}. The white star indicates the location of the IR-bright driving source. 
}\label{fig:hh666_boxes} 
\end{figure*}
Several features of interest were spread across multiple tiles, necessitating an extra step.
The eastern part of HH~666 was observed in a different orbit from the rest of the object, and the HH~901/902 complex was split across four orbits.
In these cases, we create stacked images of each tile as described above, then use the stars in common where the two tiles overlap to derive the linear transformation from one reference frame to another.
Once the positions of all the stars in the relevant stacked images have been transformed to a single master reference frame, we find the transformations from the original exposures to the final stitched reference frame and re-stack.
The alignment accuracy between stitched orbits is degraded slightly because of the relatively small overlap between tiles.
We estimate it to be $\lesssim$5 km s$^{-1}$ over the 9--10 year baseline at this distance.

In many cases, the optimal alignment of the two epochs as determined from the position of the stars differs from that determined by aligning the dust pillars. 
Since we are interested in the velocities of jets from intermediate-mass stars, we measure knot motions in the frame of the driving source, which we assume remains stationary. 
We therefore further refine the alignment by applying small, linear shifts to the second epoch to minimize residuals in a difference image. 
If the driving source is optically visible, we align the star. 
For embedded (unseen) driving sources, we align the images to minimize subtraction residuals from bright emission along the irradiated edges of the natal pillars. 
Thus all proper motions reported in Table~\ref{t:jet_pm} are relative to the (assumed stationary) jet origin. 
The fact that these jets are externally irradiated requires that they lie close to the high-mass stars in Carina, so we assume that all jets are at the distance of $\eta$~Carinae \citep[$2300\pm50$~pc,][]{smi06b}. 

Aligning nebular features is less precise than aligning point sources, particularly where the contrast between the irradiated pillar edges and the bright nebular emission of the H~{\sc ii} region is poor. 
Allowing for a typical alignment uncertainty of $\sim 0.2$~pix adds an additional $\sim 11-14$~km~s$^{-1}$ uncertainty in the measured transverse velocity. 
We therefore only consider features faster than 15~km~s$^{-1}$ in the following analysis.

\begin{figure}
\centering
\includegraphics[trim=35mm 35mm 35mm 35mm,angle=0,scale=0.325]{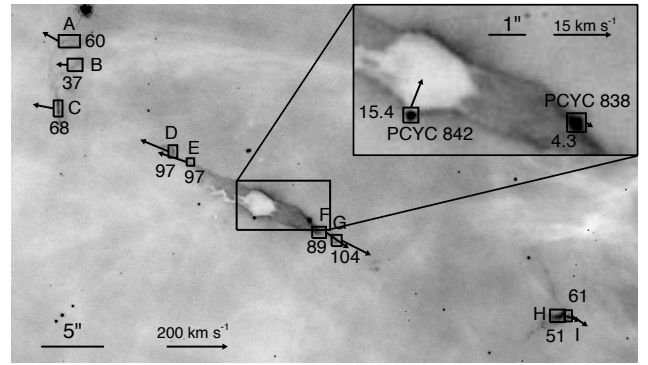} 
\caption{Same as Figure~\ref{fig:hh666_boxes} for HH~900. These H$\alpha$ proper motions were originally presented in \citet{rei15a}. }\label{fig:hh900_boxes} 
\end{figure}

We measure proper motions of jet knots using the modified cross-correlation technique developed by \citet{cur96,har01,mor01}. 
For each jet, we identify knots that do not change significantly in brightness and morphology between epochs. 
After subtracting a median-filtered image (with a 25~pix kernel) to suppress background emission, we extract the jet knot in a small box optimized for that feature (box sizes used for each knot are shown in Figures~\ref{fig:hh666_boxes}-\ref{fig:hhc15_boxes}). 
We generate an array with the total of the square of the difference between the two images for small shifts (0.1~pixel) of this box relative to the second-epoch image. 
The minimum of this array corresponds to the offset that best matches the two images.
We repeat this procedure for a variety of box sizes to estimate the uncertainty in the measured pixel offset. 
The typical uncertainty is a few km~s$^{-1}$, although this varies with the brightness of the feature and the complexity of the background emission. 
Proper motions and their associated uncertainties are listed in Table~\ref{t:jet_pm}.   
Previous applications of our implementation of this technique to measure proper motions are presented in \citet{rei14,rei15b,rei15a,kim16}.

\section{Results}\label{s:results}

We present proper motion measurements of 37 jets driven by intermediate-mass young stars in the Carina Nebula.
With a $\sim 10$~yr baseline between images, we are sensitive to knot velocities faster than $\sim 15$~km~s$^{-1}$, assuming that everything is at the distance of $\eta$~Carinae \citep[measured to be $2300\pm50$~pc by][]{smi06b}. 
All but one of the jets identified by \citet{smi10} have at least one knot with a transverse velocity $\gtrsim 50$~km~s$^{-1}$. 
Each of the 15 candidate jets identified in \citet{smi10} also have transverse velocities $\gtrsim 50$~km~s$^{-1}$ allowing us to confirm them as bonafide jets. 
Kinematics validate the seven candidate jets that we confirmed previously based on the morphology of their near-IR [Fe~{\sc ii}] emission \citep[see][]{rei13,rei16}; this is the first confirmation of the remaining eight candidate jets. 
Proper motion measurements for all 37 jets are listed in Table~\ref{t:jet_pm}, including first results from this survey that have already been published \citep{rei15a,rei15b,rei17}.

Proper motions reveal the velocity structure in the jet, and in some cases also provide kinematic confirmation of the jet-driving source. 
Many of these sources were identified as candidate protostars in the Pan-Carina YSO Catalog \citep[PCYC,][]{pov11}, allowing us to compare to the stellar parameters obtained from model fits to the IR Spectral Energy Distribution (SED). 
In the following, we briefly summarize proper motion results for each jet. 

%
\subsection{New ACS proper motions}\label{ss:new_pm}

\textit{HH~666 -- extended features:} 
\citet{rei15b} presented detailed proper motions of the innermost part of HH~666. We include the proper motions of the small knots that comprise HH~666~M and O in Table~\ref{t:jet_pm}. 
In this paper, we add the proper motions of the other knots identified by \citet{smi04,smi10} and first measured by \citet{rei14} (see Figure~\ref{fig:hh666_boxes}).

We measure transverse velocities in features A,E,N,I,C that are slower than those reported in \citet{rei14}, but within the uncertainties. 
Some discrepancy in the velocities measured in the two studies may be expected given that \citet{rei14} match images taken from two different instruments with two different distortions which will increase the systematic uncertainty in the proper motions. 

New \emph{HST}/ACS images include features that fell outside the \emph{HST}/WFC3-UVIS image used by \citet{rei14}. 
This allows us to measure the proper motions of features D,U,S for the first time. 
\citet{smi10} proposed that C,U,S might trace a fragmented bow shock and indeed proper motion vectors of both C and U point toward S. 
Both D and S are the most distant knots of their respective sides of HH~666 and trace the slowest velocities measured in that limb of the jet.   

\begin{figure*}
\centering
\includegraphics[trim=0mm 0mm 0mm 0mm,angle=0,scale=0.675]{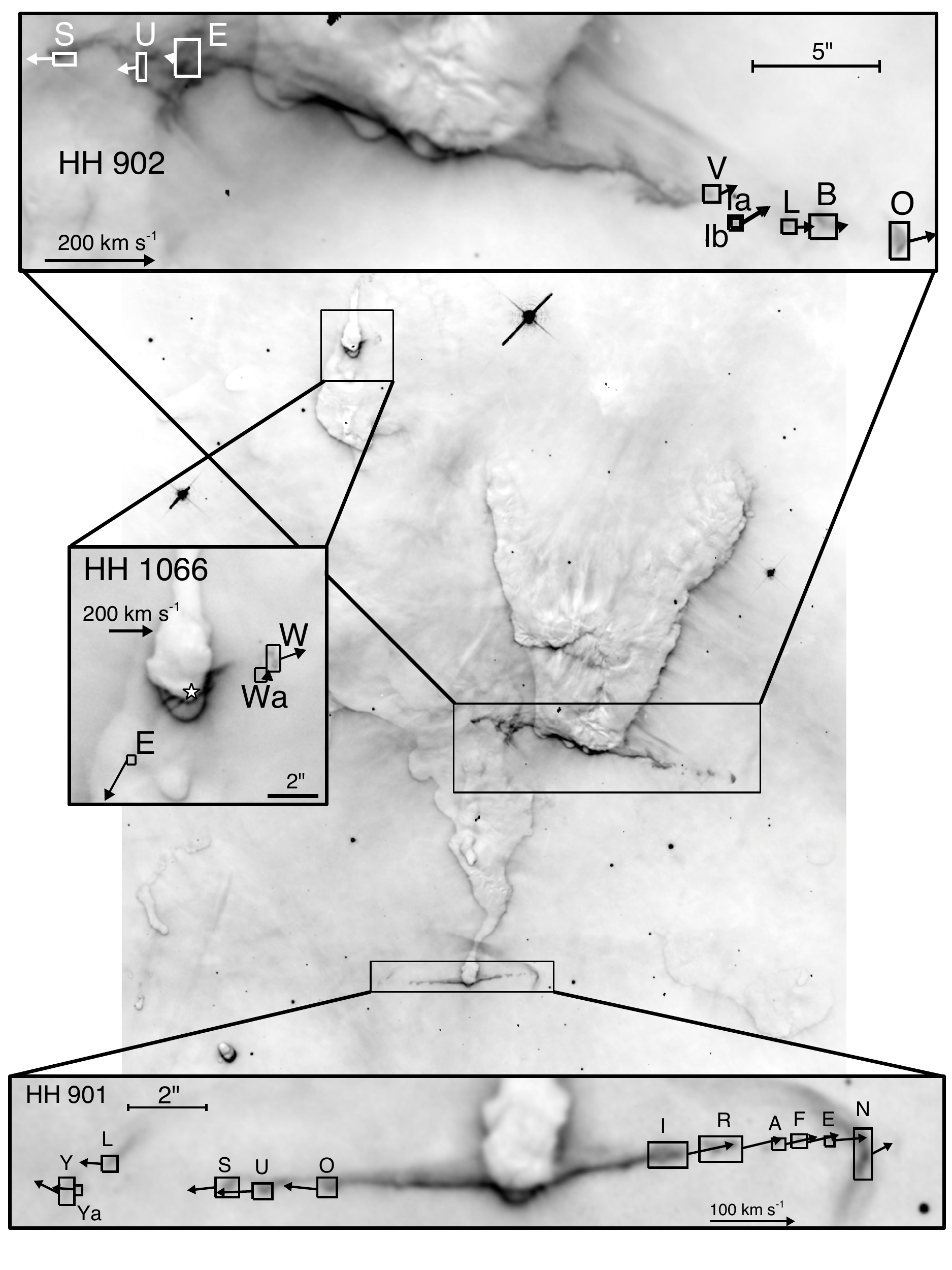} 
\caption{Same as Figure~\ref{fig:hh666_boxes} showing the cloud complex that contains HH~901, 902, and 1066. We measure proper motions in the same H$\alpha$ knots as \citet{rei14}. }\label{fig:hh901_complex_boxes} 
\end{figure*}
\textit{HH~900:} 
\citet{rei15a} presented the proper motion of HH~900 using \emph{HST}/ACS images from this program (see also Figure~\ref{fig:hh900_boxes}). 
We include transverse and radial velocities reported in that paper in Table~\ref{t:jet_pm} and the analysis in Section~\ref{s:discussion}.

\textit{HH~901:} 
HH~901 is located in the Tr14 image mosaic, and therefore is subject to an additional uncertainty in the proper motions due to imperfect distortion correction to match the first epoch image to the rotated second epoch. 
Despite this complication, the proper motions we measure in HH~901 agree with those obtained by \citet{rei14} within the uncertainties (see Figure~\ref{fig:hh901_complex_boxes} and Table~\ref{t:jet_pm}).

\textit{HH~902:} 
HH~902 lies to the northwest of HH~901, in the same cloud complex that is irradiated by Tr14 (see Figure~\ref{fig:hh901_complex_boxes}).
As in HH~901, transverse velocities are within the uncertainty of the velocities presented in \citet{rei14}. 
Knot velocities do not monotonically decrease with increasing distance from the driving source, as seen in some other jets (e.g., HH~666).

\textit{HH~903:} 
\begin{figure*}
\centering
\includegraphics[trim=0mm 0mm 0mm 0mm,angle=0,scale=0.775]{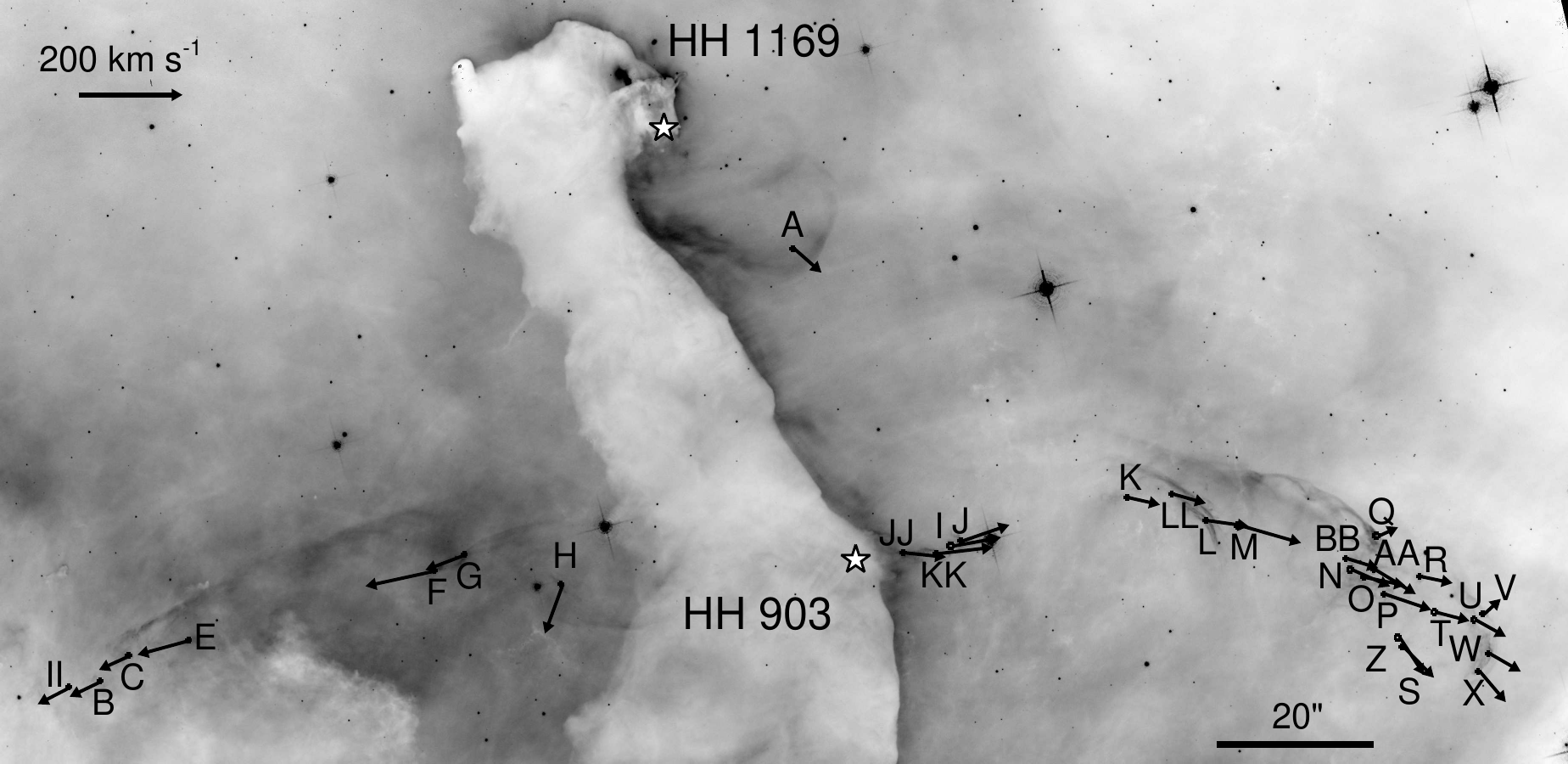} 
\caption{Same as Figure~\ref{fig:hh666_boxes} for HH~903 (the larger jet that emerges perpendicular to the pillar) and HH~1169 \citep[near the pillar head, HH~c-10 in][]{smi10}. }\label{fig:hh903_boxes} 
\end{figure*}
\begin{figure*}
\centering
\includegraphics[trim=0mm 0mm 0mm 0mm,angle=0,scale=0.675]{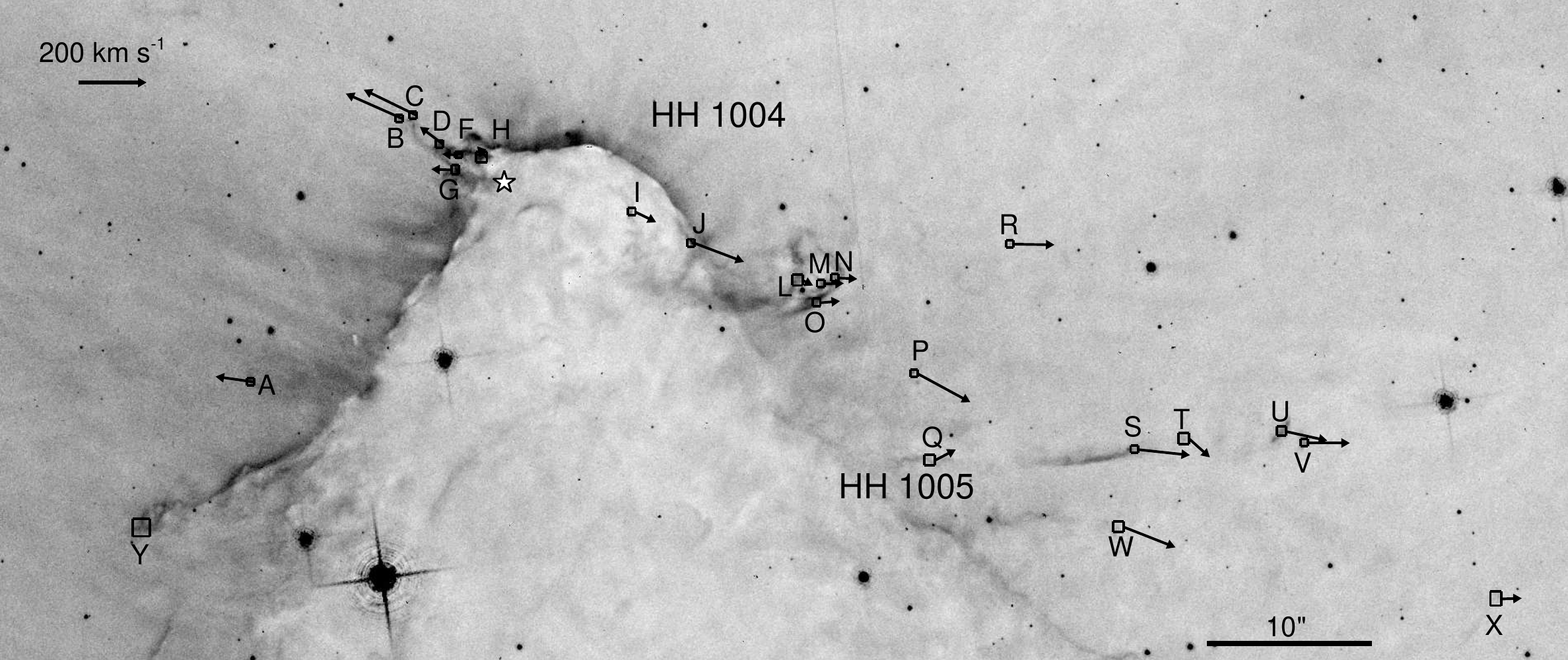} 
\caption{Same as Figure~\ref{fig:hh666_boxes} for HH~1004 (top of the pillar) and HH~1005 (middle of the pillar). HH~1004 shows clear bipolar, outwardly directed gas motions. None of the H$\alpha$ features associated with the eastern side of HH~1005 show transverse motion in the plane of the sky. }\label{fig:hh1004_boxes} 
\end{figure*}
Proper motions of the jet knots in HH~903 trace the bipolar outflow moving in opposite directions away from the driving source that is deeply embedded in the western edge of the parent dust pillar. 
\citet{rei16} report the detection of a [Fe~{\sc ii}] jet that threads the more diffuse H$\alpha$ emission in HH~903. 
In the inner jet, the two emission lines appear to trace the same knots. 
Features I,J,JJ,KK trace the inner jet (see Figure~\ref{fig:hh903_boxes}) where [Fe~{\sc ii}] emission is also bright and continuous. 
On the opposite side of the dust pillar, features F,G,H trace the H$\alpha$ counterparts to features that \citet{rei16} identified in [Fe~{\sc ii}]. 
Together, the knots seen in [Fe~{\sc ii}] trace an S-shaped flow (the WFC3-IR images cover the portion of the jet from feature F in the east to M in the west). 
Proper motion vectors reflect this ``S'' shape with the knots in the inner jet (I,J,JJ,KK) pointing slightly northwest while the knots on the opposite side (F,G,H) point to the southeast. 

More distant features in both limbs of the jet point southwest and southeast, giving the impression that the entire outflow bends away from the head of the pillar. 
Wispy, tenuous H$\alpha$ features are offset $\sim 5\arcsec$ above the [Fe~{\sc ii}] jet axis. 
\citet{rei16} note the two-component jet-outflow morphology, similar to other jets (e.g., HH~666, HH~900) where H$\alpha$ traces the wider angle, slower outflow. 
In light of this, it is possible that H$\alpha$ emission only traces the upper half of the outflow lobe closer to the source of ionizing radiation.

\textit{HH~1004:} 
H$\alpha$ knots in HH~1004 trace the movement of the two jet limbs away from the candidate YSO identified by \citet{rei16}. 
We measure faster proper motions in jet knots in the eastern limb (features B,C) and slower motions where the jet crosses the pillar ionization front (E,F,G,H; see Figure~\ref{fig:hh1004_boxes}). 
In the opposite limb, features K,L,M,N,O, trace the western bow shock. 
Two knots, P,R, lie beyond this bow shock and also move toward the west, away from the HH~1004 driving source, but with motions that seem to expand away from the jet axis.  
Knot U lies $\sim 30\arcsec$ beyond the western bow shock of HH~1004, near the edge of Figure~\ref{fig:hh1004_boxes}. 
\citet{smi10} identify this feature as part of HH~1005, which emerges from the same pillar as HH~1004, albeit $\sim 15\arcsec$ further south (we describe this jet in the next section).
Knot U lies along the HH~1004 jet axis and moves away from the putative driving source.
This raises the possibility that knot U is part of HH~1004, although proper motions that deviate from a single smooth stream are seen in many of the jets in Carina. 
 
\textit{HH~1005:} 
HH~1005 lies just below HH~1004, bisecting the same dust pillar $\sim 15\arcsec$ further south. 
Few of the H$\alpha$ features associated with the jet show measurable motion. 
The H$\alpha$ limb that extends off the eastern edge of the pillar that \citet{smi10} identify as HH~1005 has a transverse velocity $\lesssim 15$~km~s$^{-1}$ (feature HH~1005~Y, see Figure~\ref{fig:hh1004_boxes} and Table~\ref{t:jet_pm}). 
\citet{rei16} identified an [Fe~{\sc ii}]-bright portion of the jet offset $\sim 1\arcsec$ below this H$\alpha$ feature. 
Such a morphology is reminiscent of two-component outflows like HH~666, where [Fe~{\sc ii}] also traces markedly higher velocities than H$\alpha$. 

Above the slow-moving feature Y, just east of the pillar edge lies knot A.
It has a jet-like velocity of $\sim 115$~km~s$^{-1}$ and moves away from the pillar.
Knot A is the solitary moving feature to the east of the pillar, so it is unclear if it is part of a coherent jet.
Furthermore, no driving source can be unambiguously identified for HH~1005 (\citealt{ohl12} suggest a source, but no jet emission can be seen emerging from the protostar, so we regard this as a candidate, see \citealt{rei16}).
We therefore cannot establish whether knot A and other features in HH~1005 are consistent with a common origin. 

A few knots in the western limb of HH~1005 move away from the pillar (features S,T,U,V,Q,W,X). 
Like features P and R in HH~1004 (see above), features Q and W point away from the HH~1005 jet axis delineated by smooth H$\alpha$ and [Fe~{\sc ii}] emission. 
Features S,T,U,W all point to the southwest and may plausibly be associated with HH~1004. 
Near the edge of the image, feature V appears to trace the apex of the jet.

\textit{HH~1006:} 
\begin{figure}
\centering
\includegraphics[trim=0mm 0mm 0mm 0mm,angle=0,scale=0.5]{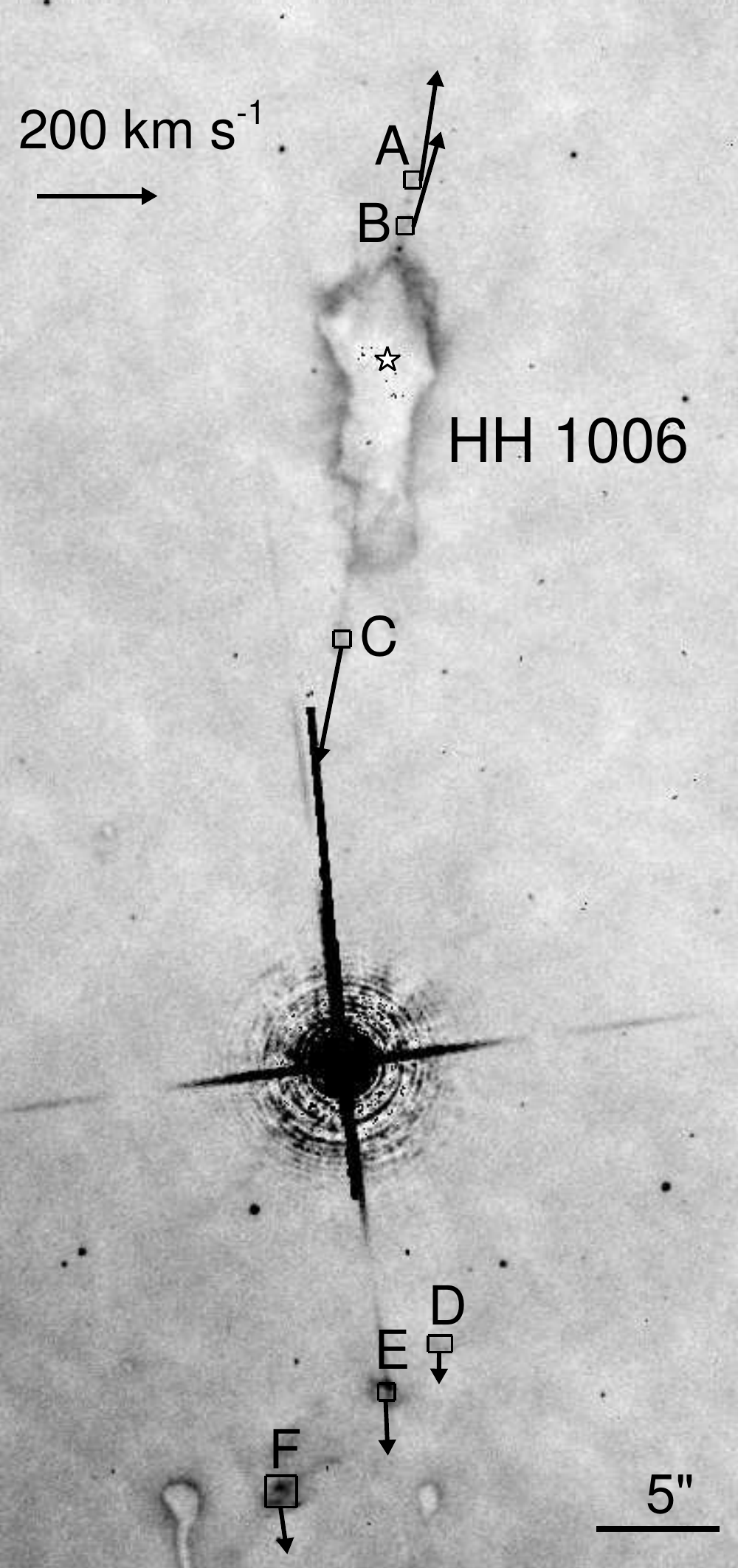} 
\caption{Same as Figure~\ref{fig:hh666_boxes} for HH~1006. Fast jet velocities and a high mass-loss rate make HH~1006 one of the stronger jets in Carina. }\label{fig:hh1006_boxes} 
\end{figure}
The highly collimated bipolar outflow, HH~1006, emerges from a small globule in the south pillars. 
Fast knots at the leading edge of both limbs of the inner jet point toward more distant bow shocks. 
The northern bow shock falls outside of the H$\alpha$ image area, but is seen in [Fe~{\sc ii}] \citep{rei16}. 
Features D,E,F trace what appears to be the wing and tip of the southern bow shock (see Figure~\ref{fig:hh1006_boxes}). 
With an average transverse velocity of $\sim 150$ km s$^{-1}$, HH~1006 is one of the fastest jets in the sample. 
Knots in the southern bow shock are slower (with an average velocity of $\sim 60$ km s$^{-1}$), and all point slightly away from the jet axis defined by the inner jet. 
Like other jets in this sample (e.g., HH~666, HH~903), the three knots of the bow shock appear to trace one part of the shock wing. 
Archival adaptive optics images at K$_s$ presented by \citet{sah12} show an hourglass-shaped nebulosity at the position of the driving source that suggests that the jet lies near the plane of the sky.

\textit{HH~1007 and HH~1015:} 
\begin{figure}
\centering
\includegraphics[trim=0mm 0mm 0mm 0mm,angle=0,scale=0.425]{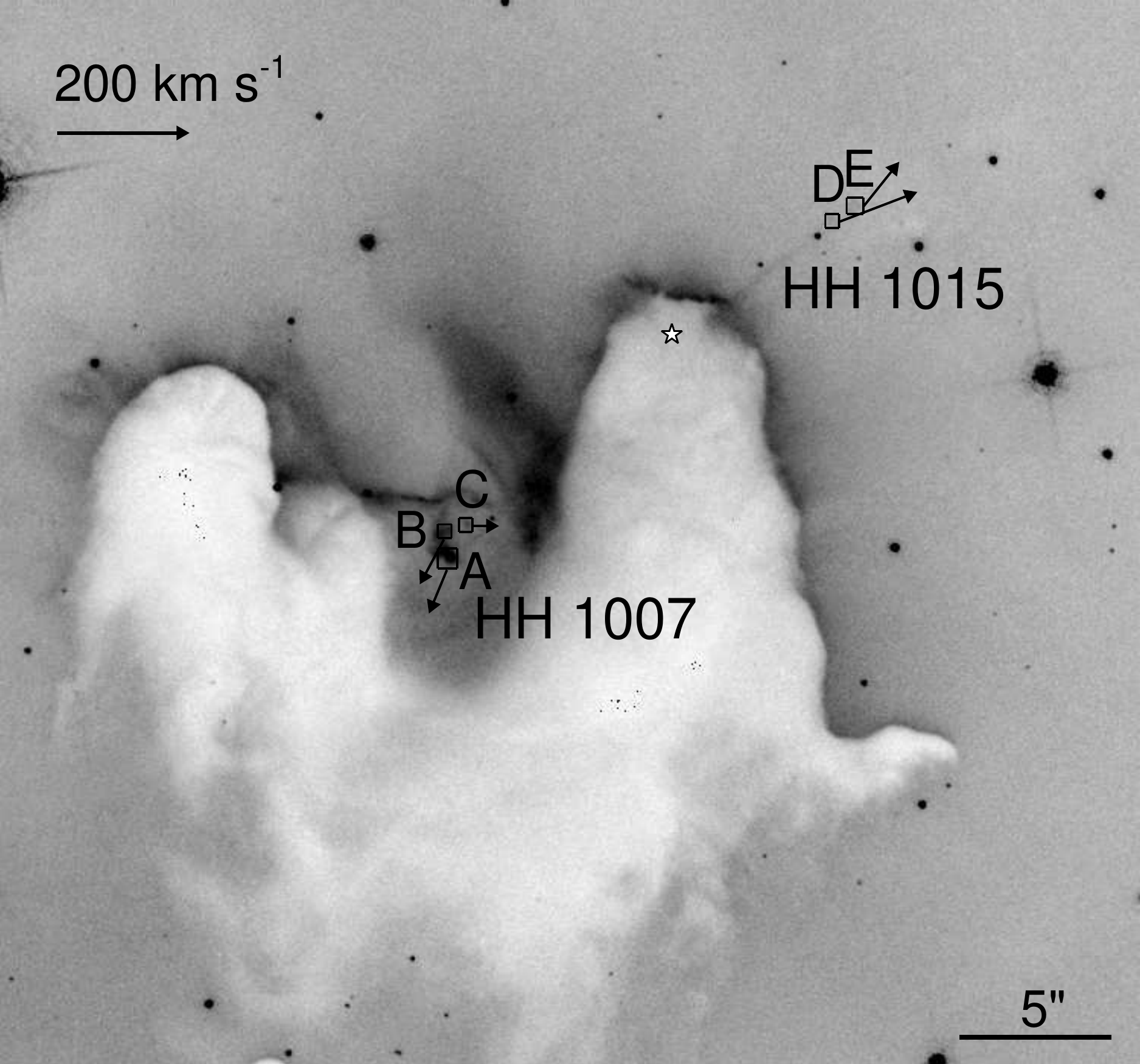} 
\caption{Same as Figure~\ref{fig:hh666_boxes} for HH~1007 and HH~1015. 
Proper motions suggest a common origin in the western pillar, and appear to trace the two bipolar limbs of a single jet. }\label{fig:hh1007_boxes} 
\end{figure}
\begin{figure}
\centering
\includegraphics[trim=0mm 0mm 0mm 0mm,angle=0,scale=0.425]{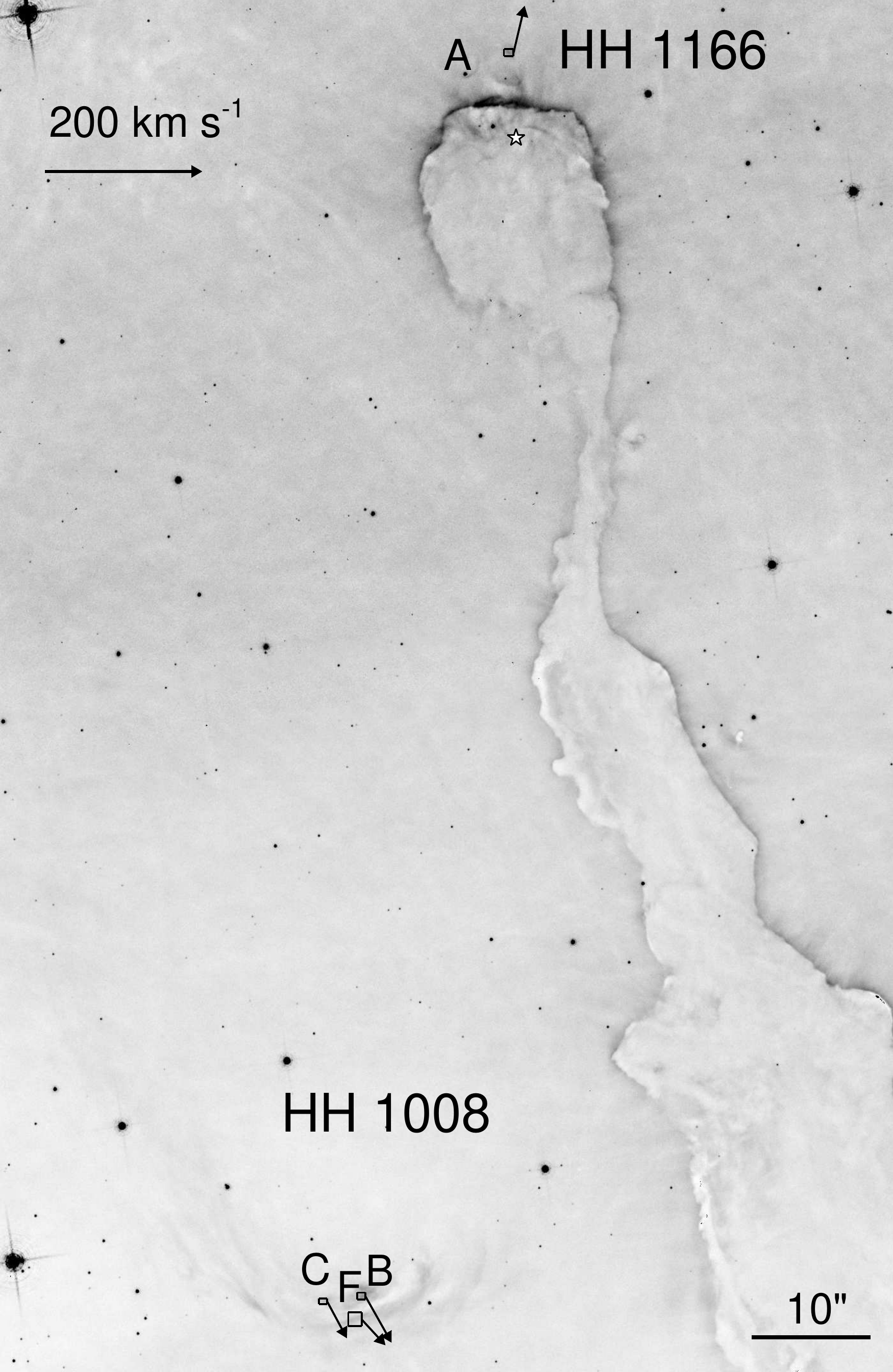} 
\caption{Same as Figure~\ref{fig:hh666_boxes} for HH~1008 and HH~1166. Proper motions of knot A show it moving away from the pillar head (and the protostar embedded within it), demonstrating that HH~1166 is, in fact, a jet. 
Knots in HH~1008 propagate parallel to the pillar major axis. }\label{fig:hh1008_boxes} 
\end{figure}
\begin{figure}
\centering
\includegraphics[trim=0mm 0mm 0mm 0mm,angle=0,scale=0.625]{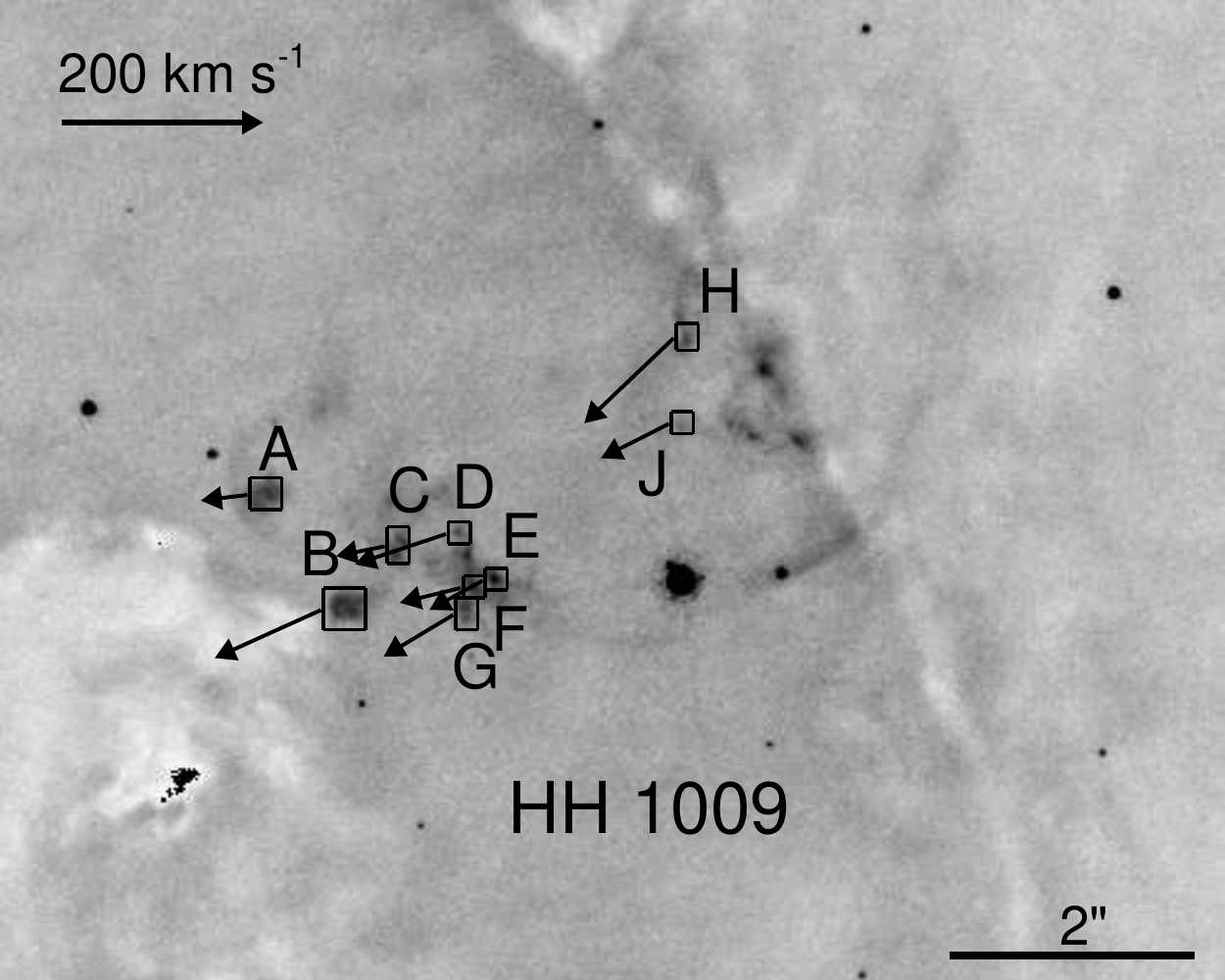} 
\caption{Same as Figure~\ref{fig:hh666_boxes} for HH~1009. Several knots comprise HH~1009, all moving away from the nearby dust pillar. The feature lies $\sim 90\arcsec$ southwest of HH~1008, but their kinematics demonstrate that the two features are unrelated. }\label{fig:hh1009_boxes} 
\end{figure}
\begin{figure}
\centering
\includegraphics[trim=0mm 0mm 0mm 0mm,angle=0,scale=0.475]{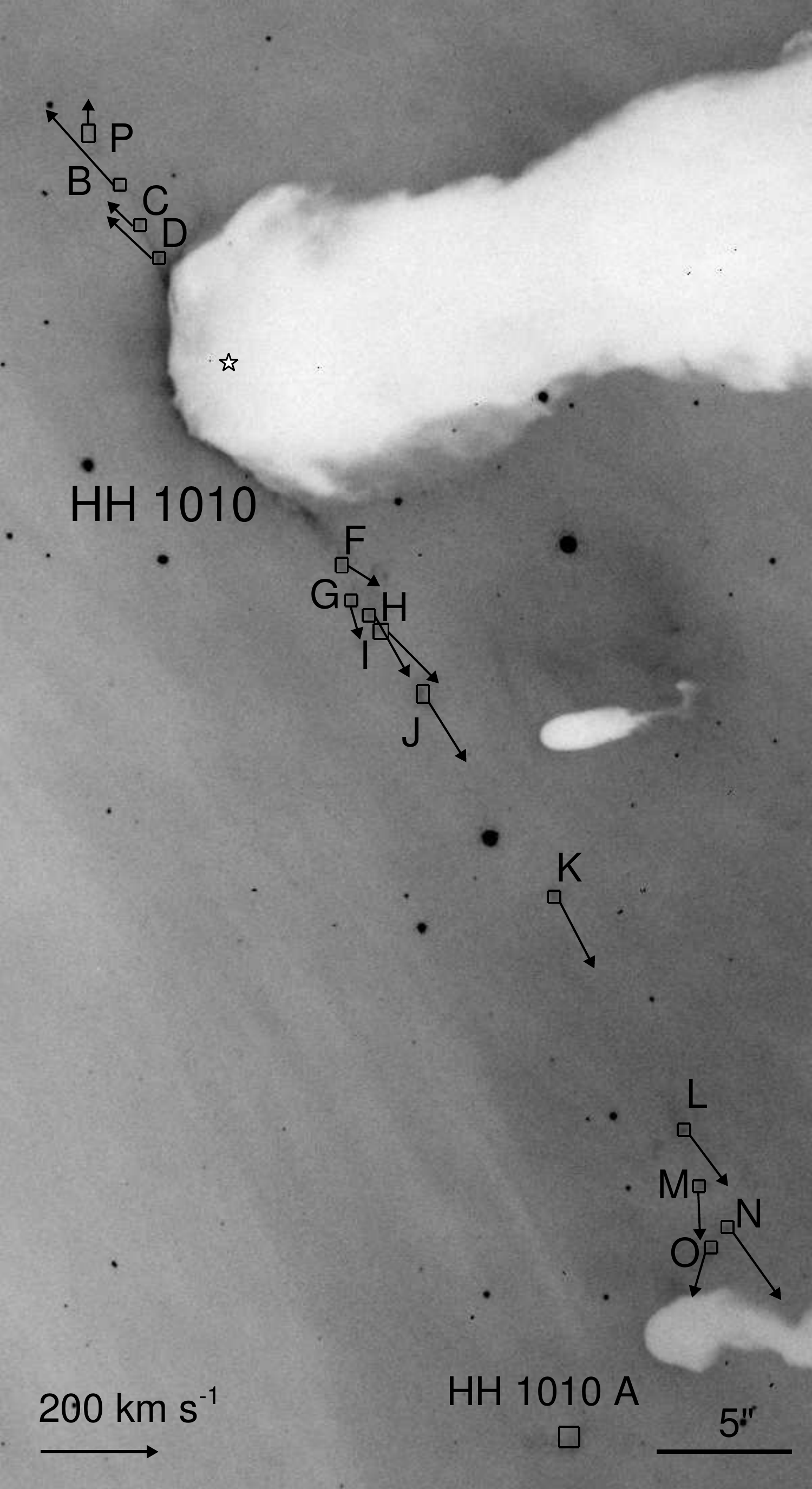} 
\caption{Same as Figure~\ref{fig:hh666_boxes} for HH~1010. The bipolar jet HH~1010 emerges from a pillar head and propagates away, along an axis nearly perpendicular to the pillar major axis. Proper motions confirm the [Fe~{\sc ii}] shock feature identified by \citet{rei16}, but rule out HH~1010~A as being associated with the jet. }\label{fig:hh1010_boxes} 
\end{figure}
\citet{rei16} noted the absence of [Fe~{\sc ii}] emission along the H$\alpha$ feature that \citet{smi10} identify as HH~1007. 
However, HH~1007 and HH~1015 lie along the same jet axis, leading \cite{rei16} to suggest that these two features may actually represent the two limbs of the same jet. 
Indeed, H$\alpha$ proper motions from HH~1007 and HH~1015 suggest a common origin inside the western pillar (see Figure~\ref{fig:hh1007_boxes}). 
The linear H$\alpha$ emission that extends off the eastern pillar (in the saddle between the two pillars) does not appear to move during the 9.5~yrs between epochs.

We can, however, measure the proper motions of a few H$\alpha$ knots that lie just beneath the smooth, apparently stationary H$\alpha$ limb. 
HH~1007~C points toward the pillar and HH~1015 driving source. 
Features A and B more clearly lie along the same jet axis as HH~1015 and move away from the putative driving source identified by \citet{rei16}.

HH~1015 emerges from the apex of the western pillar.
We measure the proper motion of two knots that move along the jet axis away from the driving source embedded in the tip of the pillar, in the opposite direction of knots A and B in HH~1007.

\textit{HH~1008:} 
HH~1008 is a bow shock located $\sim 25 \arcsec$ east of a narrow dust pillar (G287.73-0.92). 
\citet{smi10} speculated that HH~1008 may be part of the same outflow as candidate jet HH~c-2 (now HH~1166, see below) that emerges from the tip of the nearby dust pillar (see Figure~\ref{fig:hh1008_boxes}). 
H$\alpha$ proper motions point parallel to the major axis of the pillar and may trace the HH~1166 counterjet if the outflow is bent by the large-scale flow of plasma in the H~{\sc ii} region, \citep[see, e.g.,][]{bal06}.

\textit{HH~1009:} 
Like HH~1008, HH~1009 is a series of knots tracing a bow shock (see Figure~\ref{fig:hh1009_boxes}). 
HH~1009 lies $\sim 1.5$\arcmin\ southwest of HH~1008, near the edge of the same dust pillar (G287.73-0.92). 
Unlike HH~1008, proper motions point southeast, away from the pillar. 
No clear jet body indicates the origin of HH~1009, but knot kinematics point back to PCYC~599, a class 0/I ($2.2 \pm 1.6$ M$_{\odot}$, log(L$_{bol}$) $= 1.9 \pm 1.3$ L$_{\odot}$ protostar embedded in the pillar $\sim 25$\arcsec\ northwest of the knots.

\textit{HH~1010:} 
Proper motions show knots in the bipolar jet HH~1010 moving away from the driving source embedded in the pillar head (see Figure~\ref{fig:hh1010_boxes}). 
Additional knots in the southwest limb, features L,M,N,O move in the same direction as the inner jet, confirming that the [Fe~{\sc ii}] knots identified by \citet{rei16} are part of the jet.
\citet{smi10} identified a feature, HH~1010~A, that appears to cap the southwest limb of the jet.
However, HH~1010~A does not lie along the same axis as other jet knots, nor does it appear to move, making it unlikely that this feature is part of HH~1010. 

HH~1010 emerges from one of the pillars observed with the MUSE Integral Field Unity (IFU) spectrograph by \citet{mcl16}. 
Radial velocities extracted from those observations confirm that the northern limb is blueshifted ($v_{rad} \approx -2$~km~s$^{-1}$) and the southern limb is redshifted ($v_{rad} \approx 6$~km~s$^{-1}$), respectively.

\textit{HH~1011:} 
\begin{figure}
\centering
\includegraphics[trim=0mm 0mm 0mm 0mm,angle=0,scale=0.475]{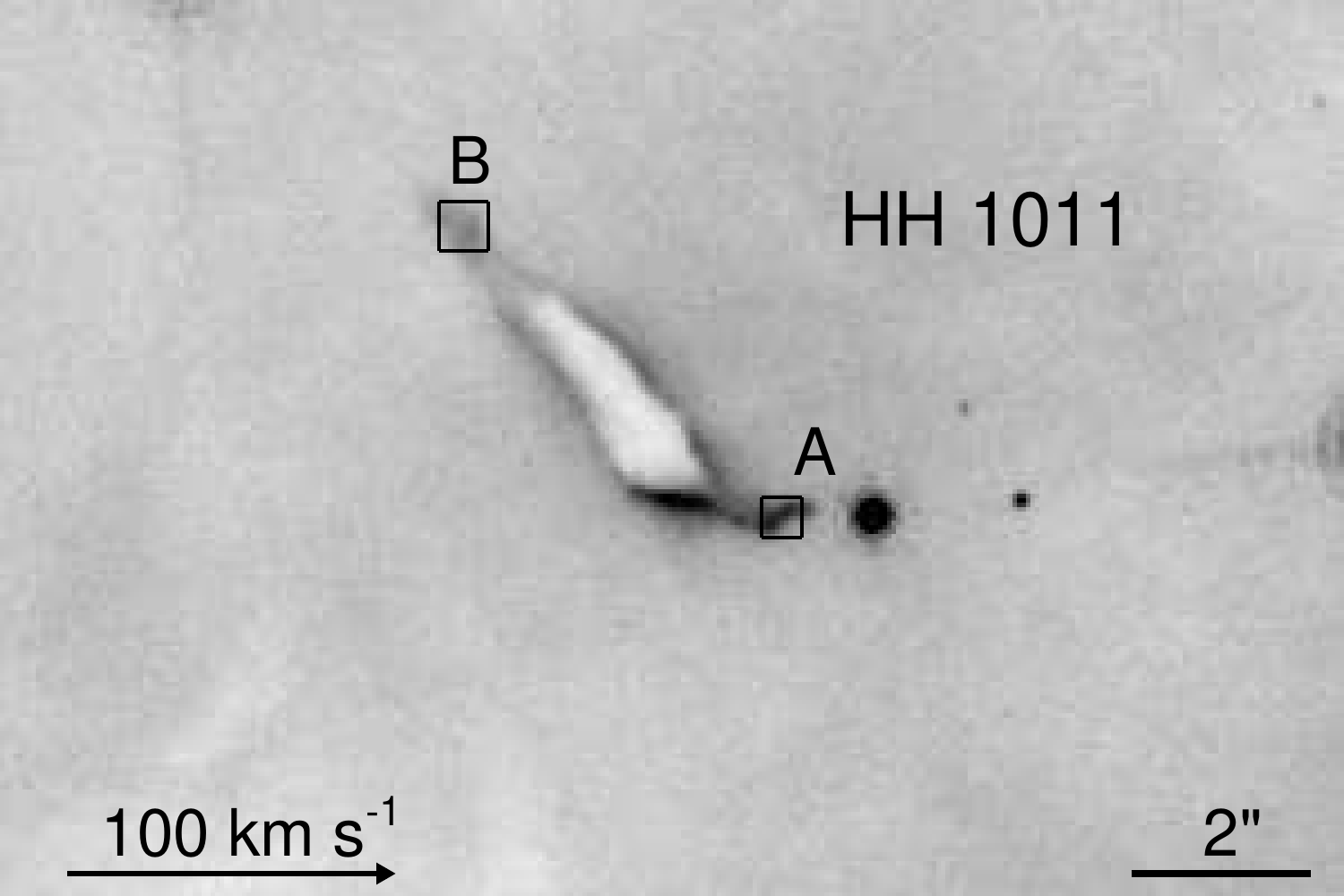} 
\caption{Same as Figure~\ref{fig:hh666_boxes} for HH~1011. \citet{smi10} identify HH~1011 as a small microjet that extends off of a small globule in Tr~15. 
However, any transverse motion is below our sensitivity. }\label{fig:hh1011_boxes} 
\end{figure}
\citet{smi10} identify bright H$\alpha$ emission extending off a small globule in Tr15 as the one-sided jet HH~1011. 
The proper motion of the H$\alpha$ knot at the end of the putative jet is $\lesssim 15$, slower than most jet-like features in this sample (see Figure~\ref{fig:hh1011_boxes}).

\textit{HH~1012:} 
\begin{figure}
\centering
\includegraphics[trim=0mm 0mm 0mm 0mm,angle=0,scale=0.575]{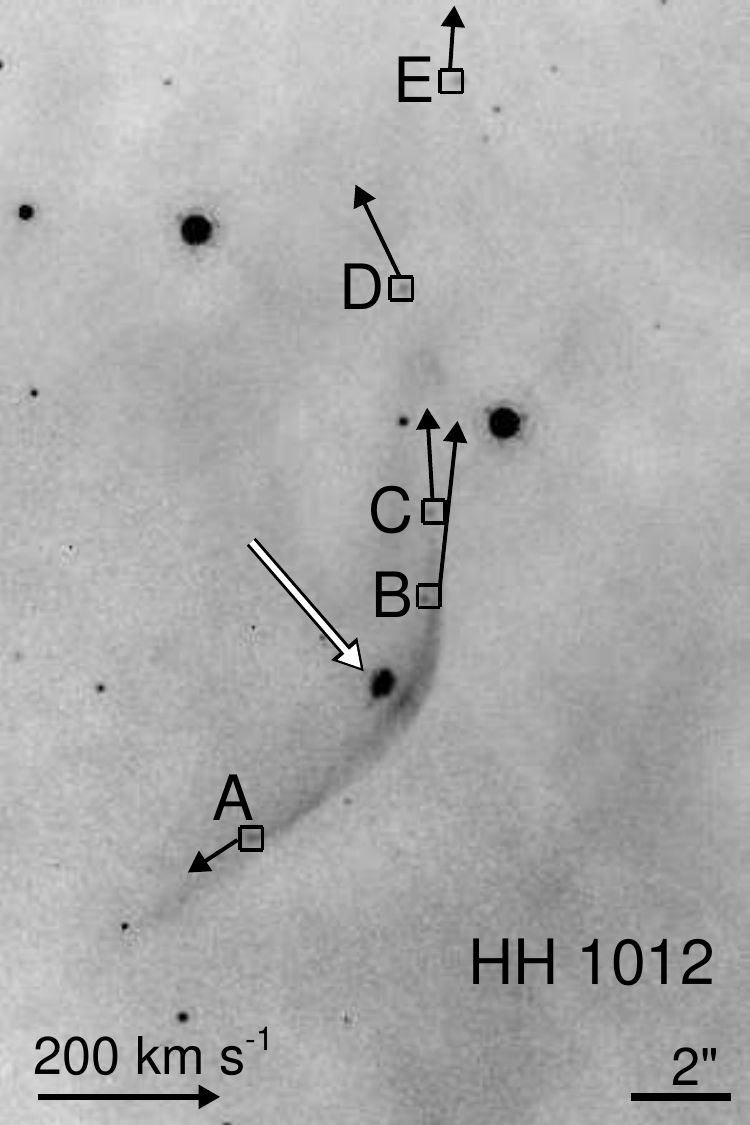}
\caption{Same as Figure~\ref{fig:hh666_boxes} for HH~1012. The LL~Ori-like object HH~1012 flows away from the optically-visibly protostar (denoted with a white arrow). Jet knots move away from the protostar in a curved path, following the curvature of the diffuse arc of emission to the southwest. }\label{fig:hh1012_boxes} 
\end{figure}
HH~1012 appears to be a classic LL-Ori-type object, similar to many such curved flows seen in Orion \citep[e.g.,][]{bal01,bal06}. 
Both the jet knots and the shrouding H$\alpha$ bow shock bend away from Tr14. 
Jet knots follow this curved path as they propagate away from the unobscured star just inside the apex of the H$\alpha$ bow (see Figure~\ref{fig:hh1012_boxes}).  
Knot velocities from features B,C,D are fast ($\sim 100$~km~s$^{-1}$), similar to other jet-like objects in this sample (e.g., HH~1006).

\textit{HH~1013:} 
\begin{figure}
\centering
\includegraphics[trim=0mm 0mm 0mm 0mm,angle=0,scale=0.325]{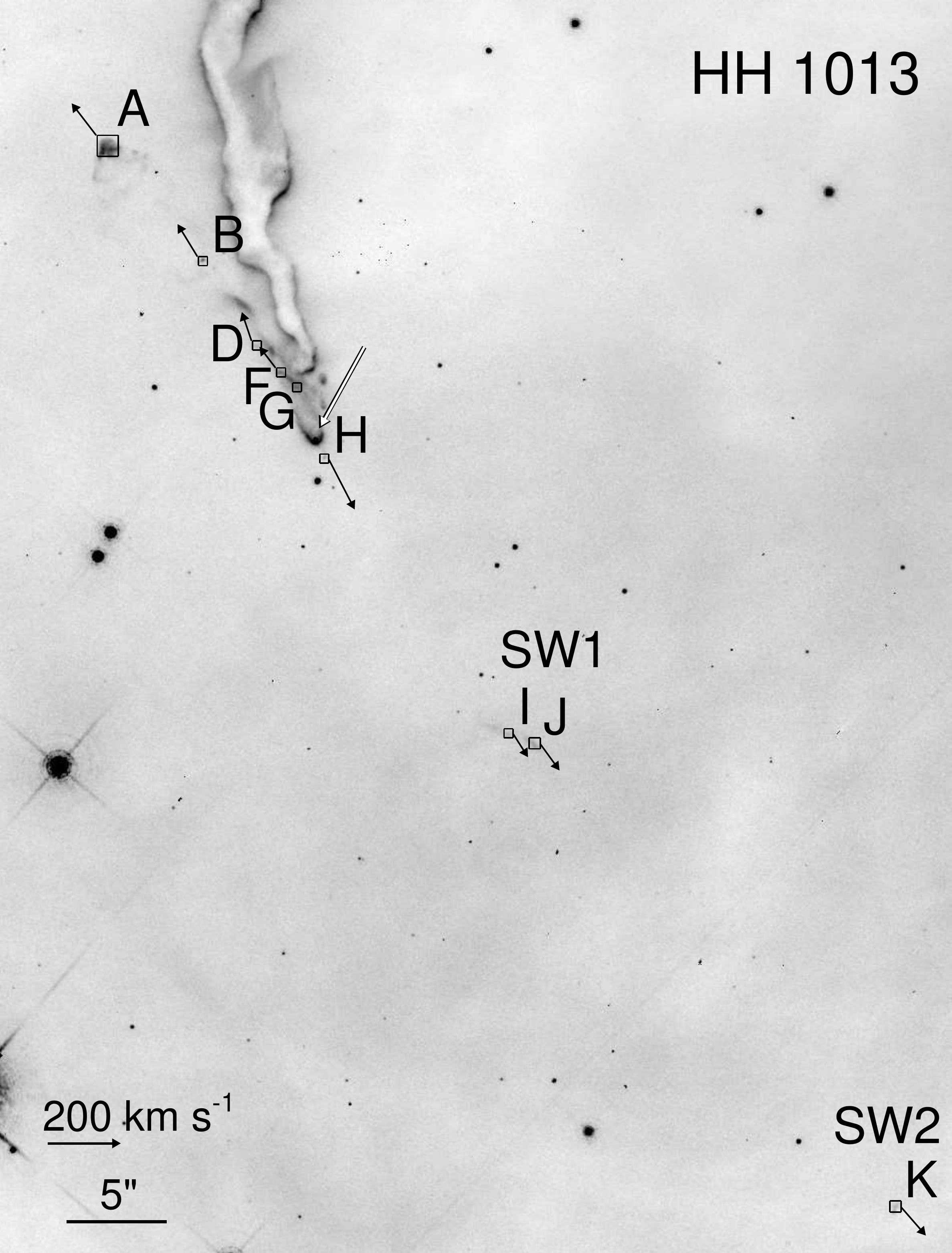} 
\caption{Same as Figure~\ref{fig:hh666_boxes} for HH~1013. HH~1013 emerges from the tip of a narrow pillar. The driving source (denoted with a white arrow) is enshrouded in H$\alpha$ emission. Unlike the many jets that are aligned nearly perpendicular to the major axis of the pillar, the axis of HH~1013 is tilted $\sim 20^{\circ}$ compared to the major axis of the pillar. }\label{fig:hh1013_boxes} 
\end{figure}
\begin{figure*}
\centering
\includegraphics[trim=0mm 0mm 0mm 0mm,angle=0,scale=0.5]{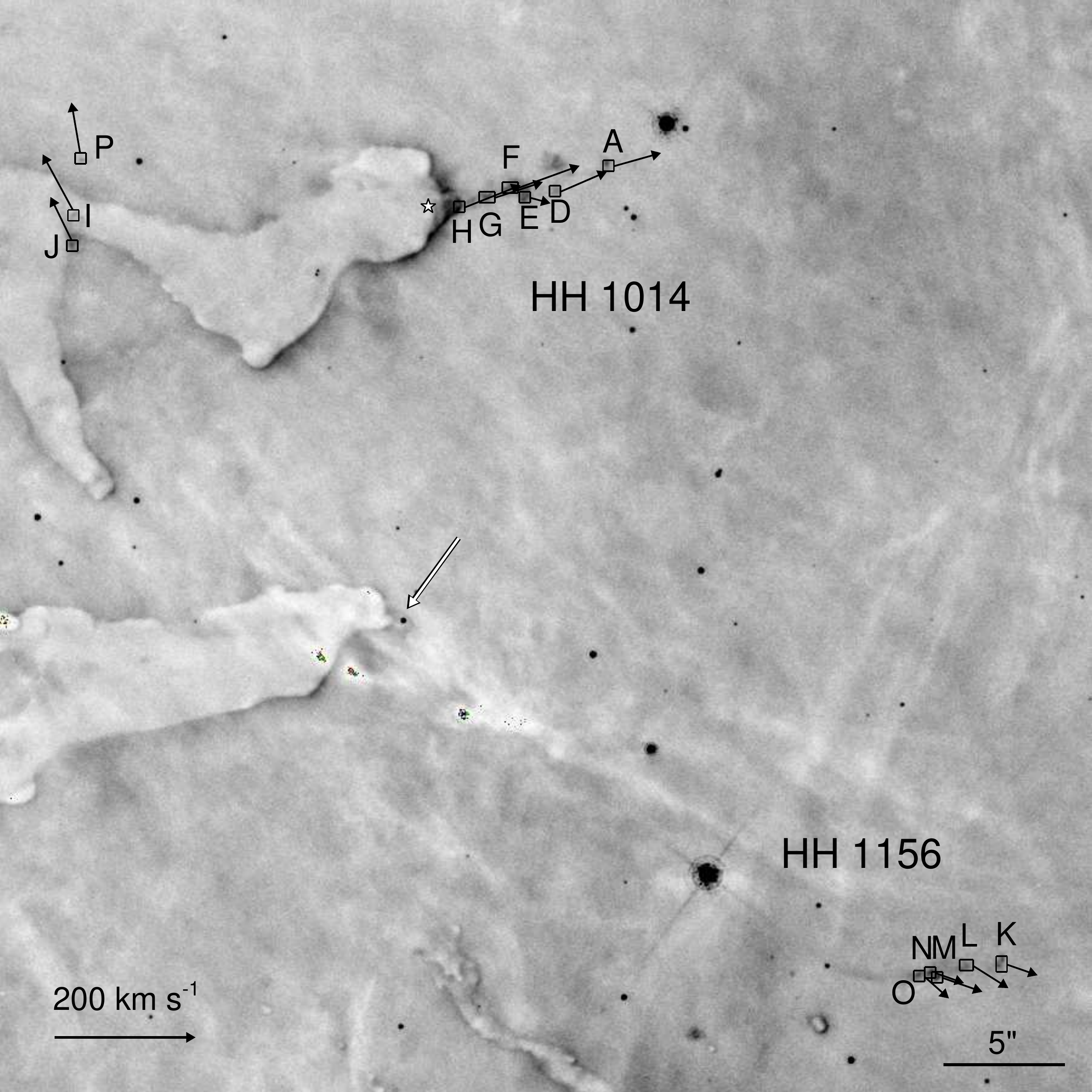} 
\caption{Same as Figure~\ref{fig:hh666_boxes} for HH~1014. Outward-directed velocities show the motion of HH~1014 away from the pillar tip. Further south in the same image are several knots tracing one of the bow shocks associated with HH~1156. The jet body remains invisible in H$\alpha$ and thus far have only been detected in near-IR [Fe~{\sc ii}]. }\label{fig:hh1014_boxes} 
\end{figure*}
Jet knots in HH~1013 move in opposite directions away from the driving source. 
To the northeast, multiple knots trace a straight jet axis that makes a $23^{\circ}$ angle with the major axis of the nearby dust pillar (see Figure~\ref{fig:hh1013_boxes}). 
HH~1013 is no longer embedded in a dust pillar, but a loop of H$\alpha$ emission surround the protostar, similar to the wide-angle H$\alpha$ that traces an outflow cavity around source still embedded in a dust pillar (e.g., HH~666 and HH~1164). 
H$\alpha$ emission at the apex of this feature has a transverse velocity $\lesssim 15$~km~s$^{-1}$, slower than the apices of the H$\alpha$ loops in HH~666 and HH~1164.

To the southwest, a microjet extends off the HH~1013 driving source itself. 
Features I and J, $\sim 15$\arcsec\ further southwest, trace the feature \citet{smi10} identify as SW1. 
Together with HH~1013~K (SW2, located $\sim 1$\arcmin\ southwest of SW1), these knots trace a jet axis that is aligned with the northeastern limb. 
\citet{smi10} identify two additional knots, HH~1013~SW3 and SW4, however, both of these fall outside the frame of the second-epoch image.

\textit{HH~1014:} 
H$\alpha$ proper motions in HH~1014 trace jet knots propagating away from the driving source located just inside the tip of the pillar (see Figure~\ref{fig:hh1014_boxes}).
The jet axis is nearly parallel to the major axis of the pillar, so the counterjet, if any, will be embedded in the dust pillar.
\citet{rei16} identified a few knots inside the pillar in near-IR [Fe~{\sc ii}] images that appear to lie along the HH~1014 jet axis. 
These [Fe~{\sc ii}] knots can also be identified in H$\alpha$ images (features I,J,P), allowing us to measure their proper motions. 
All three knots move \textit{perpendicular} to the HH~1014 jet axis.
In fact, proper motions fall along the jet axis of nearby HH~1156 which emerges from an unobscured protostar located $\sim 15\arcsec$ south of the HH~1014 pillar. 
H$\alpha$ proper motions of HH~1156 are described in the next section.

\textit{HH~1016:} 
\begin{figure}
\centering
\includegraphics[trim=0mm 0mm 0mm 0mm,angle=0,scale=0.5]{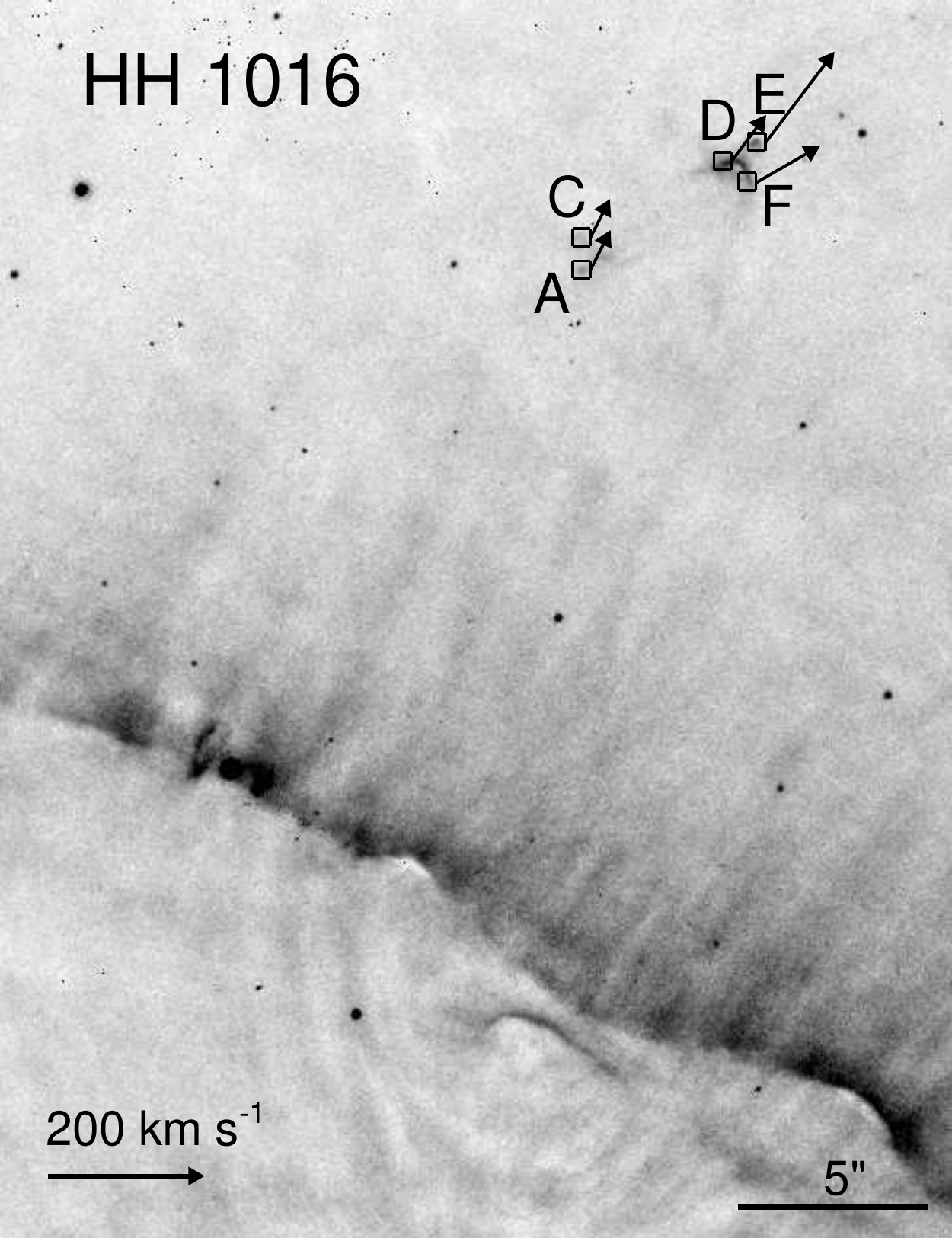} 
\caption{Same as Figure~\ref{fig:hh666_boxes} for HH~1016. HH~1016 emerges from the same pillar that contains the Treasure Chest cluster. }\label{fig:hh1016_boxes} 
\end{figure}
HH~1016 is a bow-shock-like feature that moves away from its natal pillar with a transverse velocity $> 100$~km~s$^{-1}$ (see Figure~\ref{fig:hh1016_boxes}). 
An additional knot located behind the shock, closer to the pillar (features A,C) propagate in the same direction as the shock feature and may trace the extended shock wing. 
Proper motions of the jet point back to the same dust pillar that houses the embedded Treasure Chest cluster, although no single protostar has been identified as the HH~1016 driving source.

\textit{HH~1017:} 
\begin{figure}
\centering
\includegraphics[trim=0mm 0mm 0mm 0mm,angle=0,scale=0.575]{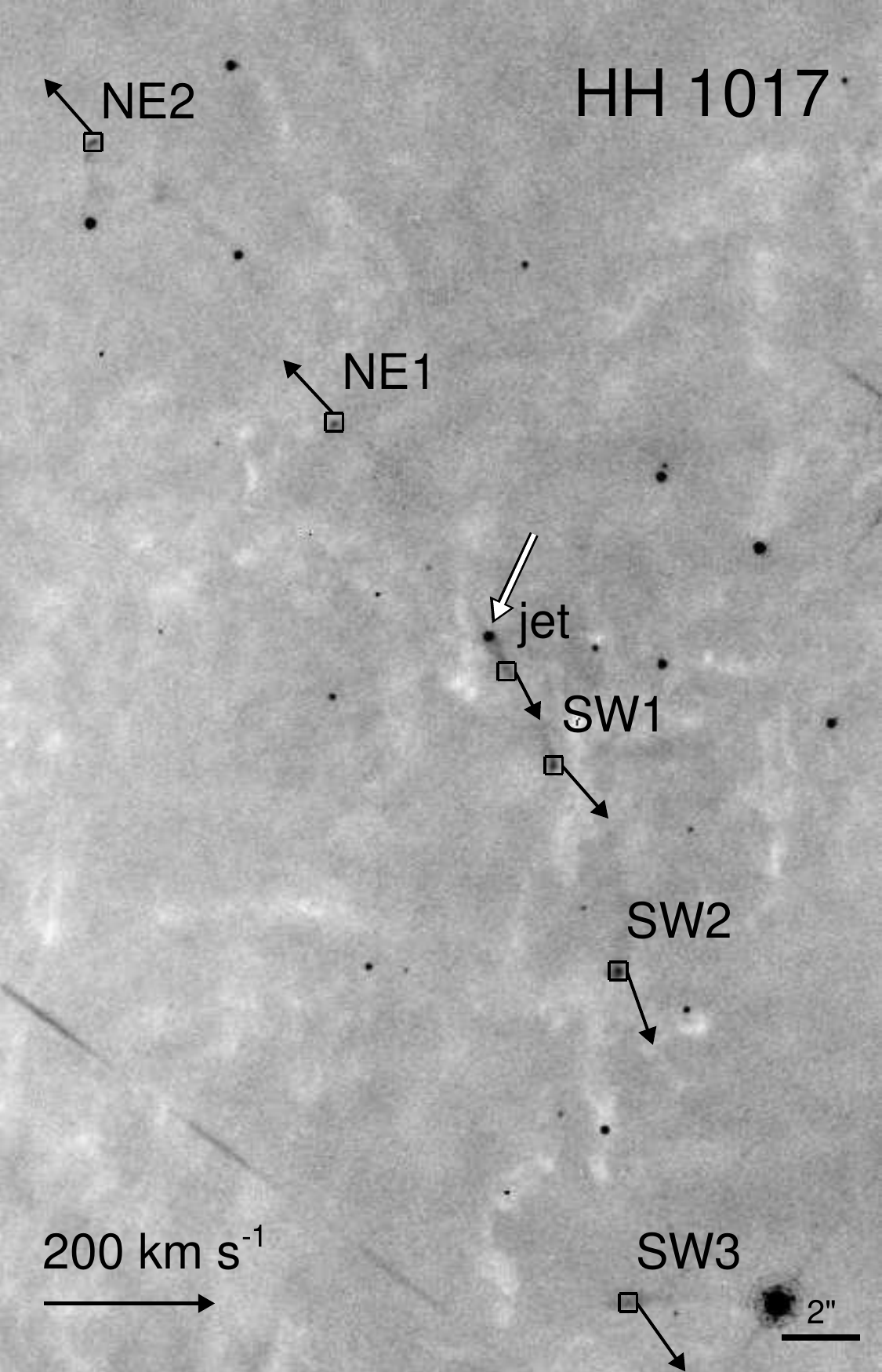}
\caption{Same as Figure~\ref{fig:hh666_boxes} for HH~1017. Like HH~1012, knots in HH~1017 move away from the driving protostar (denoted with a white arrow) along the arc of the curved jet. HH~1017 is located near Tr14, and the arc of the jet bends away from the cluster. }\label{fig:hh1017_boxes} 
\end{figure}
\begin{figure}
\centering
\includegraphics[trim=0mm 0mm 0mm 0mm,angle=0,scale=0.675]{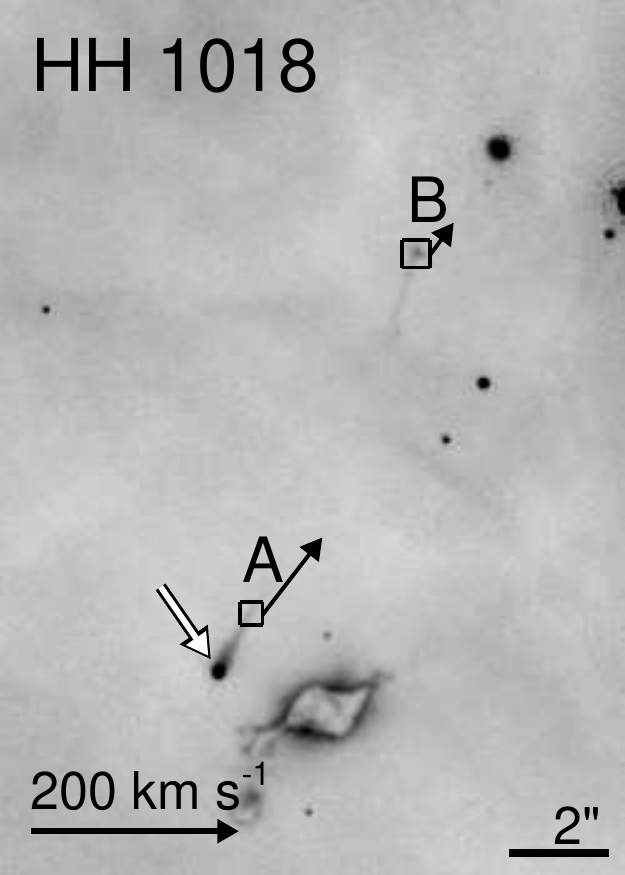} 
\caption{Same as Figure~\ref{fig:hh666_boxes} for HH~1018. }\label{fig:hh1018_boxes} 
\end{figure}
\begin{figure}
\centering
\includegraphics[trim=0mm 0mm 0mm 0mm,angle=0,scale=0.675]{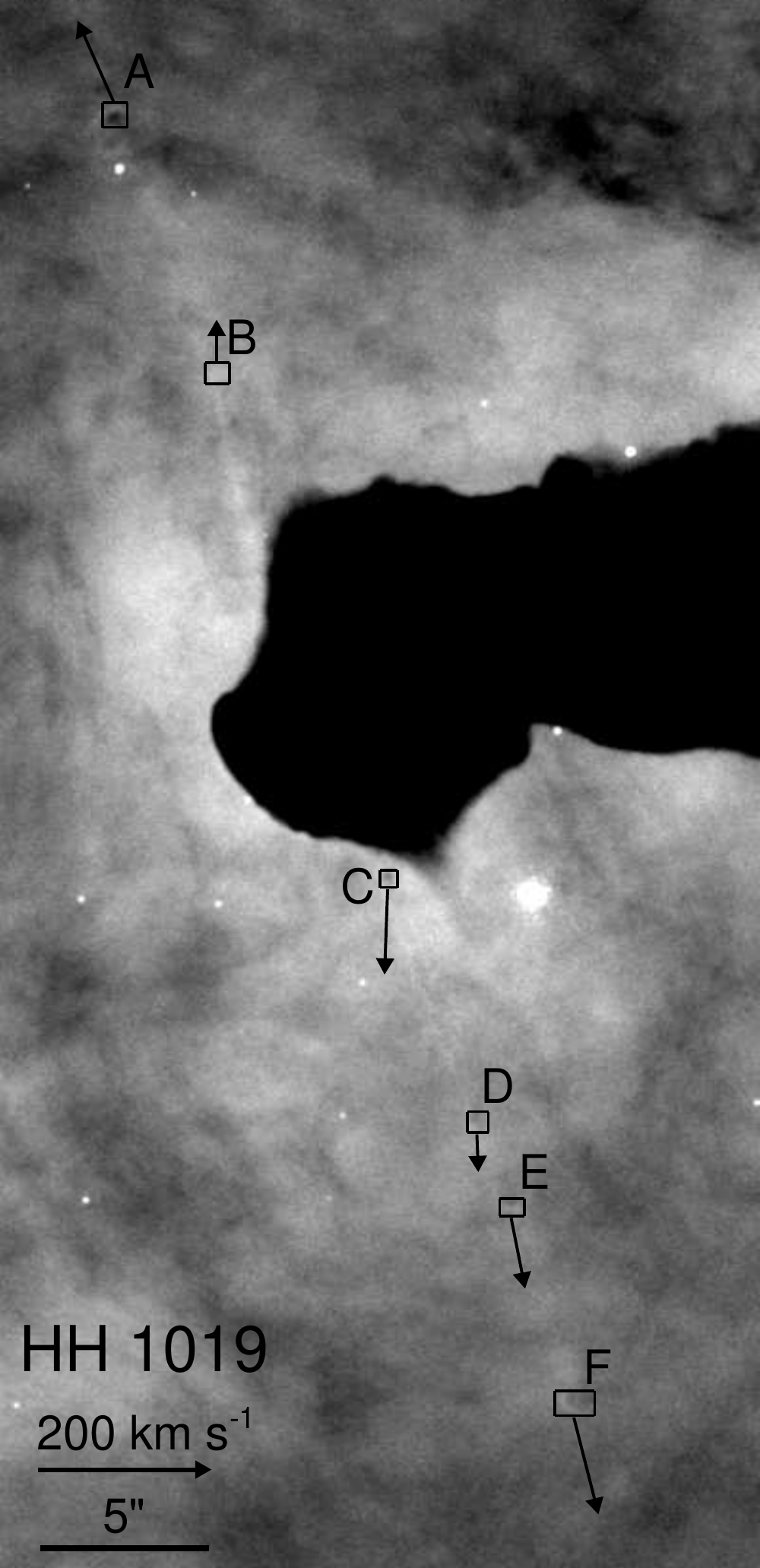} 
\caption{Same as Figure~\ref{fig:hh666_boxes} for HH~1019. Figure and proper motions of HH~1019 first reported in \citet{rei17}. }\label{fig:hh1019_boxes} 
\end{figure}
\begin{figure*}
\centering
\includegraphics[trim=20mm 0mm 20mm 0mm,angle=0,scale=0.75]{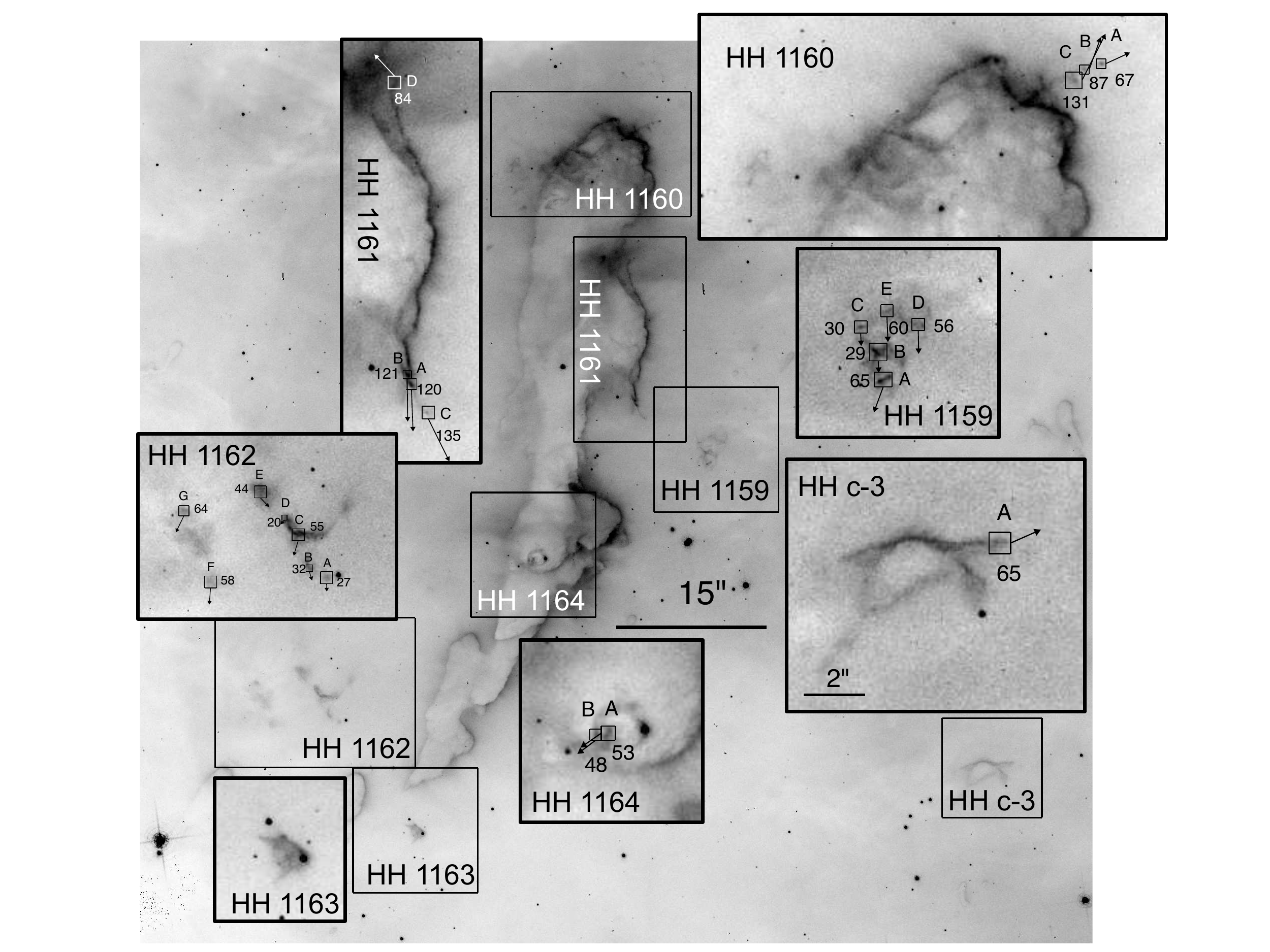} 
\caption{Same as Figure~\ref{fig:hh666_boxes} for HH~1159-1164, 1167. The seahorse-shaped globule (located in the South Pillars) houses at least 4 independent outflows. Boxes around each HH~jet / object show the region zoomed in to show the small boxes used to measure proper motions. White arrows / stars indicate the location of the driving sources. }\label{fig:hhc4_boxes} 
\end{figure*}
Unlike many jets in Carina that emerge from protostars embedded in dust pillars, both the jet and the driving source of HH~1017 are unobscured in the H~{\sc ii} region. 
HH~1017 is one of the few jets in Carina with a morphology similar to the HH jets seen in Orion \citep[e.g., HH~502,][]{bal01}. 
We measure the proper motions of several discrete jet knots that move away from the protostar seen at the center of Figure~\ref{fig:hh1017_boxes}. 
Like HH~1012, transverse velocities are fast and jet-like ($\gtrsim 80$~km~s$^{-1}$) and the motion of the knots follows the curved jet morphology seen in single-epoch images.

\textit{HH~1018:} 
Like HH~1017, HH~1018 is unobscured in the H~{\sc ii} region (see Figure~\ref{fig:hh1018_boxes}). 
Knots A and B move along the jet axis defined by the microjet. 
We measure a faster transverse velocity in knot A ($\sim 90$~km~s$^{-1}$) at the apex of the microjet and a slower velocity from the more distant knot B, as seen in some other HH jets \citep[e.g.,][]{dev97}.

\textit{HH~1019:} 
\citet{smi10} did not include HH~1019 among the jets and candidate jet identified in the original H$\alpha$ survey of Carina. 
However, several jet knots can be seen \textit{in silhouette} against the bright background of the H~{\sc ii} region. 
\citet{rei17} report the proper motions of the silhouette knots (see Figure~\ref{fig:hh1019_boxes}). 
Transverse velocities of $\sim 100$ km s$^{-1}$ on either side of the pillar trace fast, jet-like motions.

\textit{HH~1066:} 
HH~1066 \citep[HH~c-1 in][]{smi10} resides in the same cloud complex as HH~901 and HH~902 (see Figure~\ref{fig:hh901_complex_boxes}).
Transverse velocities presented here agree well with those in \citet{rei14}. 
With the second epoch \emph{HST}/ACS images, we measure the proper motion of an additional knot, HH~1066~Wa, in the western limb of the jet. 
Like HH~666~U, this small feature appears to be an off-axis portion of the bow shock (in this case, HH~1066~W).

\textit{HH~1156:} 
\citet{rei16} identify the collimated body of the HH~1156 jet in near-IR [Fe~{\sc ii}] images. 
While there is no H$\alpha$ emission from the inner jet, the southwestern bow shock \citep[HH~c-14 in][]{smi10} does have H$\alpha$ emission that allows us to measure its motion away from HH~1156 driving source (see Figure~\ref{fig:hh1014_boxes}). 
To the northeast, features I,J,P lie nearly equidistant ($\sim 20$\arcsec) from the driving source and appear to trace the complementary bow shock. 
If these features do trace the apices of the bipolar jet, then the total length of HH~1156 is $\gtrsim 40\arcsec$, or nearly $\sim 0.5$~pc.

\textit{HH~1159 -- HH~1164:} 
HH~1159 \citep[HH~c-4 in][]{smi10} is one of six HH objects associated with a large, seahorse-shaped globule in the South Pillars (see Figure~\ref{fig:hhc4_boxes}). 
A small cluster of H$\alpha$ knots compose HH~1159, creating a C-shaped arc that lies to the west of the globule. 
H$\alpha$ proper motion show the entire complex of knots moving to the south (see Figure~\ref{fig:hhc4_boxes}). 
No other features seen in the complex follow this direction of motion, suggesting that HH~1159 does not originate from a source inside the globule. 

%
HH~1160 \citep[HH~c-5 in][]{smi10} emerges from the apex of the seahorse-shaped globule. 
Tenuous H$\alpha$ emission hints at a coherent flow while strong [Fe~{\sc ii}] emission clearly demonstrates that HH~1160 is a collimated bipolar jet \citep[see Figure~\ref{fig:hhc4_boxes} and][]{rei16}. 
Three H$\alpha$ knots trace the northwestern limb of the jet, with transverse proper motions $\sim 65-130$~km~s$^{-1}$. 

Near-IR [Fe~{\sc ii}] reveals the HH~1160 counterjet inside the pillar where H$\alpha$ no longer traces the jet. 
The arc of H$\alpha$ emission identified as feature~D (see Figure~\ref{fig:hhc4_boxes}) is confused with nebulous emission in the region and is not obviously part of the jet. 
However, it is located at the southeastern tip of the counterjet seen in [Fe~{\sc ii}] and appears to trace the terminal shock. 
Proper motions show that feature~D moves in the opposite direction of features~A,B,C, away from the driving source identified by \citet{rei16}. 

%
\citet{smi10} identify two streams of H$\alpha$ emission extending off a protrusion from the globule (the belly of the seahorse) that \citet{rei16} confirm to be a bipolar jet with [Fe~{\sc ii}] images. 
H$\alpha$ knots in the north and south limbs of HH~1161 \citep[HH~c-6 in][]{smi10} move in opposite directions with transverse velocities of $\sim 100$ km s$^{-1}$ pointed away from the presumed location of the driving source (see Figure~\ref{fig:hhc4_boxes}). 

%
Like HH~1159, HH~1162 \citep[HH~c-7 in][]{smi10} lies outside the seahorse globule and is composed of a series of knots.
Together, these features trace the arc of a shock that moves to the south (see Figure~\ref{fig:hhc4_boxes}). 
HH~1162 resides on the opposite side of the seahorse from HH~1159, but both features move to the south, and are therefore likely driven by different protostars. 

%
HH~1163 emerges from an unobscured protostar located just south of the seahorse globule. 
Fan-like H$\alpha$ emission encompasses a collimated [Fe~{\sc ii}] jet \citep[][this feature was identified as HH~c-8 by \citealt{smi10}]{rei16}. 
The [Fe~{\sc ii}] jet traces the top of the fan, where H$\alpha$ emission is also brightest. 
An H$\alpha$ knot on the top of the fan coincides with the tip of the [Fe~{\sc ii}] jet and traces a transverse velocity of $\sim 60$~km~s$^{-1}$ away from the driving source (see Figure~\ref{fig:hhc4_boxes}). 

%
\citet{rei16} identified a new jet located in the middle of the seahorse-shaped globule where a bright, collimated [Fe~{\sc ii}] jet threads through a loop of H$\alpha$ emission.
In HH~1164, as in HH~666~O and HH~900, [Fe~{\sc ii}] and H$\alpha$ trace two different morphologies and presumably two different outflow components.
The [Fe~{\sc ii}] jet converges with the H$\alpha$ emission on the far side of the ``loop'' allowing us to measure the transverse velocity of the jet.
Two H$\alpha$ knots located opposite the driving source (see Figure~\ref{fig:hhc4_boxes}) with transverse velocities of $\sim 50$ km s$^{-1}$ trace the eastern limb of this apparently one-sided jet.

\textit{HH~1166 (previously HH~c-2):} 
\citet{smi10} identify candidate jet HH~c-2 emerging from the apex of a large, twisted dust pillar in the South Pillars.
Knot~A in Figure~\ref{fig:hh1008_boxes} propagates into the H~{\sc ii} region with proper motions tracing a direct line back to a protostar embedded in the head of the pillar (PCYC~666 from \citealt{pov11}, see Table~\ref{t:jet_props}). 
This confirms that HH~c-2 is, in fact, a jet. It has therefore been assigned an HH number of HH~1166. 

HH~1008 lies near the same pillar, $\sim 105\arcsec$ south of the pillar head. 
It is possible to draw a line connecting knot A from HH~1166 to the knots in HH~1008, and indeed \citet{smi10} speculated that HH~1166 is the HH~1008 counterjet. 
This may be the case, especially if the jet is bent by winds from Tr16 \citep[see discussion of jet bending in, e.g.,][]{bal06}. 

\textit{HH~1167 (previously HH~c-3):} 
\citet{smi10} identify three extensions off of a small pillar head in the South Pillars as candidate jet HH~c-3. 
This object lies near the seahorse-shaped globule that contains HH~1159-1164 (see the lower right of Figure~\ref{fig:hhc4_boxes}). 
\citet{rei16} note [Fe~{\sc ii}] emission in HH~c-3~A that may extend from the candidate protostar located at the pillar tip. 
While the nature of HH~c-3 is ambiguous in images alone, the outward motion of H$\alpha$ emission from knot A indicates its jet-like nature. 
This kinematic evidence demonstrates that HH~c-3 is, in fact, a jet and has been assigned an HH number of HH~1167.

\textit{HH~1168 (previously HH~c-9):} 
\begin{figure}
\centering
\includegraphics[trim=0mm 0mm 0mm 0mm,angle=0,scale=0.75]{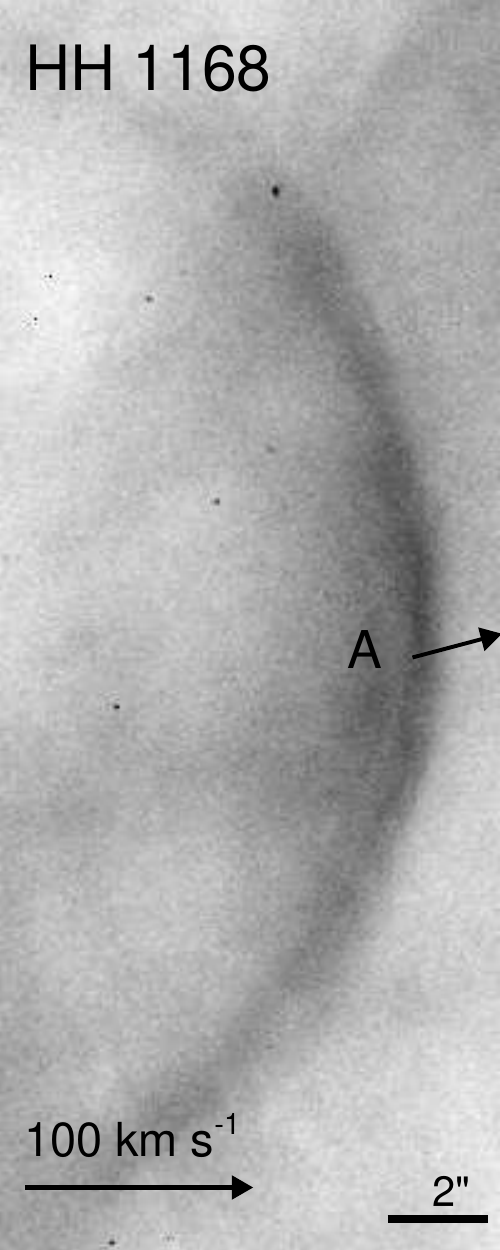} 
\caption{Same as Figure~\ref{fig:hh666_boxes} for HH~c-9. HH~c-9 lies near the edge of the frame, and as such, its driving source and point of origin remain unknown. }\label{fig:hhc9_boxes} 
\end{figure}
Candidate jet feature HH~c-9 is a smooth, curved bow shock located near the edge of an image frame in the large Tr14 image mosaic. 
Second-epoch H$\alpha$ images allow us to trace the transverse velocity of $\sim 40$~km~s$^{-1}$ in this bow shock as it propagates to the west (see Figure~\ref{fig:hhc9_boxes}).
This fast velocity argues for an origin from a jet and protostar located outside the imaged area.
Based on this, HH~c-9 in now assigned the HH number HH~1168.

\textit{HH~1169 (previously HH~c-10):} 
\citet{smi10} identified candidate jet HH~c-10 based on bow-shock-like H$\alpha$ emission that suggests a jet that emerges from the head of the same dust pillar that houses HH~903 (see Figure~\ref{fig:hh903_boxes}).
Proper motions show that the bow shock moves to the southwest of the pillar head, implying an outflow axis that is nearly perpendicular to the axis of HH~903.
No clear jet body traces HH~c-10 in H$\alpha$ images, although there is some tentative evidence in [Fe~{\sc ii}] \citep[see][]{rei16}. 
Nevertheless, the bow shock has a transverse velocity of $> 50$ km s$^{-1}$ along an axis that implies an origin in the dust pillar.
Given this evidence for jet-like motions, HH~c-10 has been assigned an HH number of HH~1169.

\textit{HH~1170 (previously HH~c-11):} 
\begin{figure}
\centering
\includegraphics[trim=0mm 0mm 0mm 0mm,angle=0,scale=0.75]{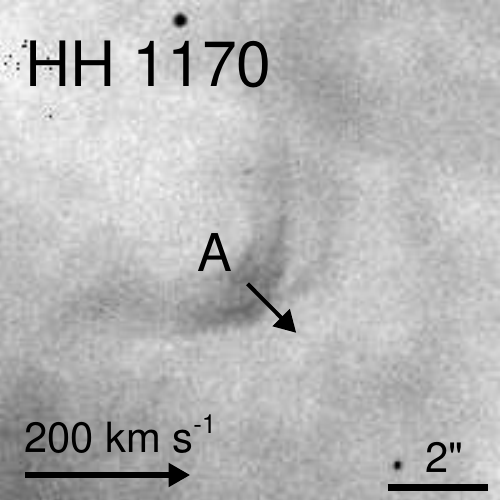} 
\caption{Same as Figure~\ref{fig:hh666_boxes} for HH~1170. Like HH~1168, HH~1170 looks like a jet bow shock, but does not have an obvious driving source or point of origin. }\label{fig:hhc11_boxes} 
\end{figure}
HH~1170 \citep[candidate jet HH~c-11 in][]{smi10} has a bow-shock-shaped morphology, but is not associated with a nearby collimated jet, much like HH~1168. 
Proper motions confirm that HH~1170 moves to the west (the direction suggested by its morphology, see Figure~\ref{fig:hhc11_boxes}) with a transverse velocity $\sim 50$~km~s$^{-1}$. 
Based on the high velocity of this bow shock, we confirm it as a bonafide HH object.

\textit{HH~1171 (previously HH~c-12):} 
\begin{figure}
\centering
\includegraphics[trim=0mm 0mm 0mm 0mm,angle=0,scale=0.75]{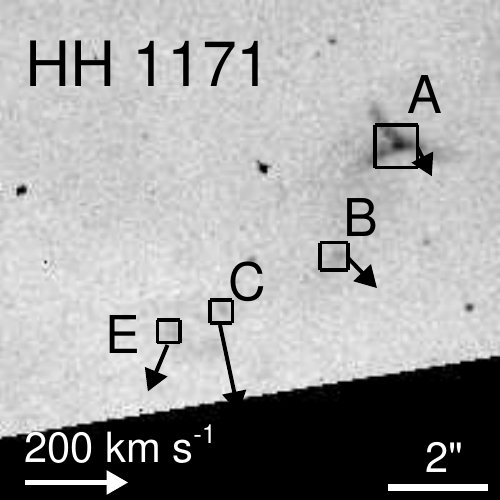} 
\caption{Same as Figure~\ref{fig:hh666_boxes} for HH~1171. HH~1171 lies near the HH~1159-1164 complex of outflows (see Figure~\ref{fig:hhc4_boxes}). Multiple knots trace a coherent feature, but their proper motions do not clearly identify the driving source. }\label{fig:hhc12_boxes} 
\end{figure}
Multiple knots make up the orphan shock structure that \citet{smi10} identify as candidate jet HH~c-12 (see Figure~\ref{fig:hhc12_boxes}). 
Like HH~1008 and HH~1009, there is no associated collimated jet and no obvious jet source. 
Nevertheless, the arc of knots is moving with transverse velocities $\gtrsim 50$~km~s$^{-1}$.
Given the likely origin from a young star, HH~c-12 has been assigned an HH number of HH~1171. 
Proper motions point back toward the seahorse-shaped globule that contains HH~1159-1164, but do not align with any of the jet axes in the region. 

\textit{HH~1172 (previously HH~c-13):} 
\begin{figure}
\centering
\includegraphics[trim=0mm 0mm 0mm 0mm,angle=0,scale=0.75]{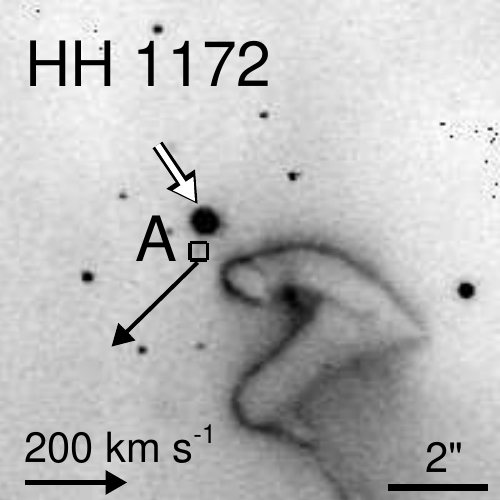} 
\caption{Same as Figure~\ref{fig:hh666_boxes} for the microjet HH~1172 that emerges from a star located to the northeast of HH~666.  }\label{fig:hhc13_boxes} 
\end{figure}
\citet{smi10} identify a candidate microjet HH~c-13 extending off a star located to the northeast of HH~666. 
Model-fits to the IR SED of the point source \citep[from the PCYC][]{pov11} support the inference that it is a young stellar object. 
Proper motions show a jet-like stream of H$\alpha$ emission moving away from the star with a transverse velocity $\gtrsim 100$~km~s$^{-1}$ (see Figure~\ref{fig:hhc13_boxes}). 
Given the clear protostar-jet morphology, and kinematic confirmation of the jet with proper motions, this object has been assigned an HH number of HH~1172.

\textit{HH~1173 (previously HH~c-15):} 
\begin{figure}
\centering
\includegraphics[trim=0mm 0mm 0mm 0mm,angle=0,scale=0.75]{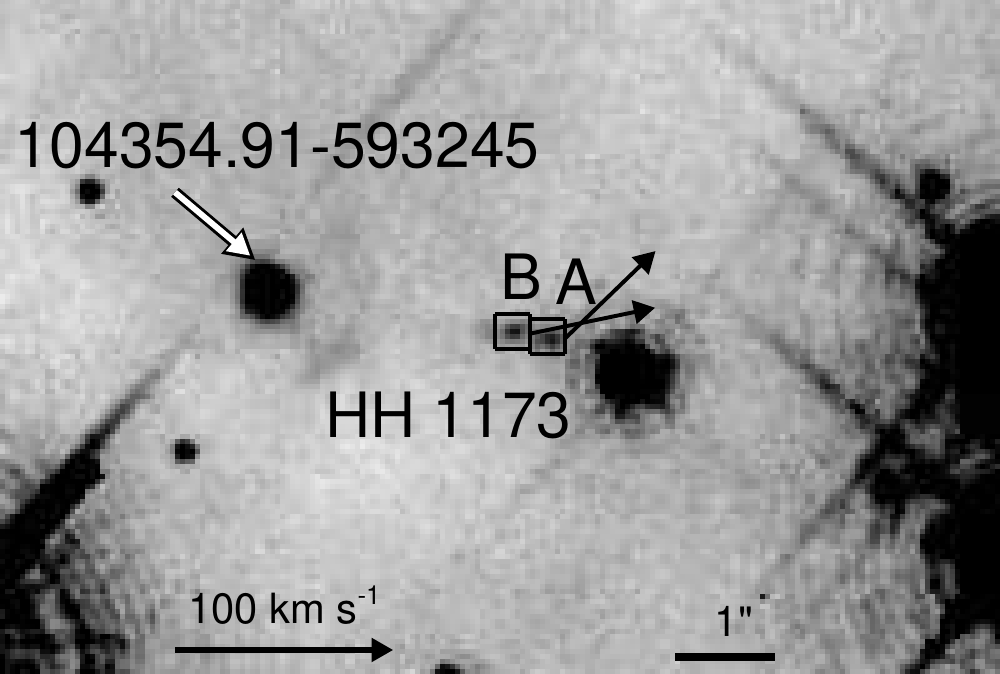} 
\caption{Same as Figure~\ref{fig:hh666_boxes} for HH~1173. Two knots trace the microjet HH~1173 in the core of Tr14. Proper motions suggest an origin from the nearby object 104554.91-593245 (marked with a white arrow) that \citet{smi10} identify as a possible proplyd. }\label{fig:hhc15_boxes} 
\end{figure}
\citet{smi10} identify two knots of H$\alpha$ emission amid the harsh UV environment of Tr14 as candidate jet HH~c-15. 
These knots lie immediately east of a candidate protostar identified in the PCYC \citep{pov11}.
However, proper motions indicate that the jet knots move \textit{toward} the candidate protostar, ruling it out as the driving source (see Figure~\ref{fig:hhc15_boxes}). 
Proper motions are more consistent with an origin from the star immediately east of the knots (104354.91-593245 in Figure~\ref{fig:hhc15_boxes}). 
An arc of H$\alpha$ emission traces a limb-brightened shell, wind shock, or proplyd envelope leading \citet{smi10} to identify 104354.91-593245 as a possible proplyd, although this star is not classified as a YSO in the PCYC. 
Proper motions favor this object as the jet-driving source and rule out the PCYC-identified protostar to the west. 
Given the fast transverse velocities and association with a young star, HH~c-15 has been assigned an HH number of HH~1173.

\begin{figure*}
\centering
$\begin{array}{cc}
\includegraphics[trim=15mm 10mm 15mm 10mm,angle=0,scale=0.35]{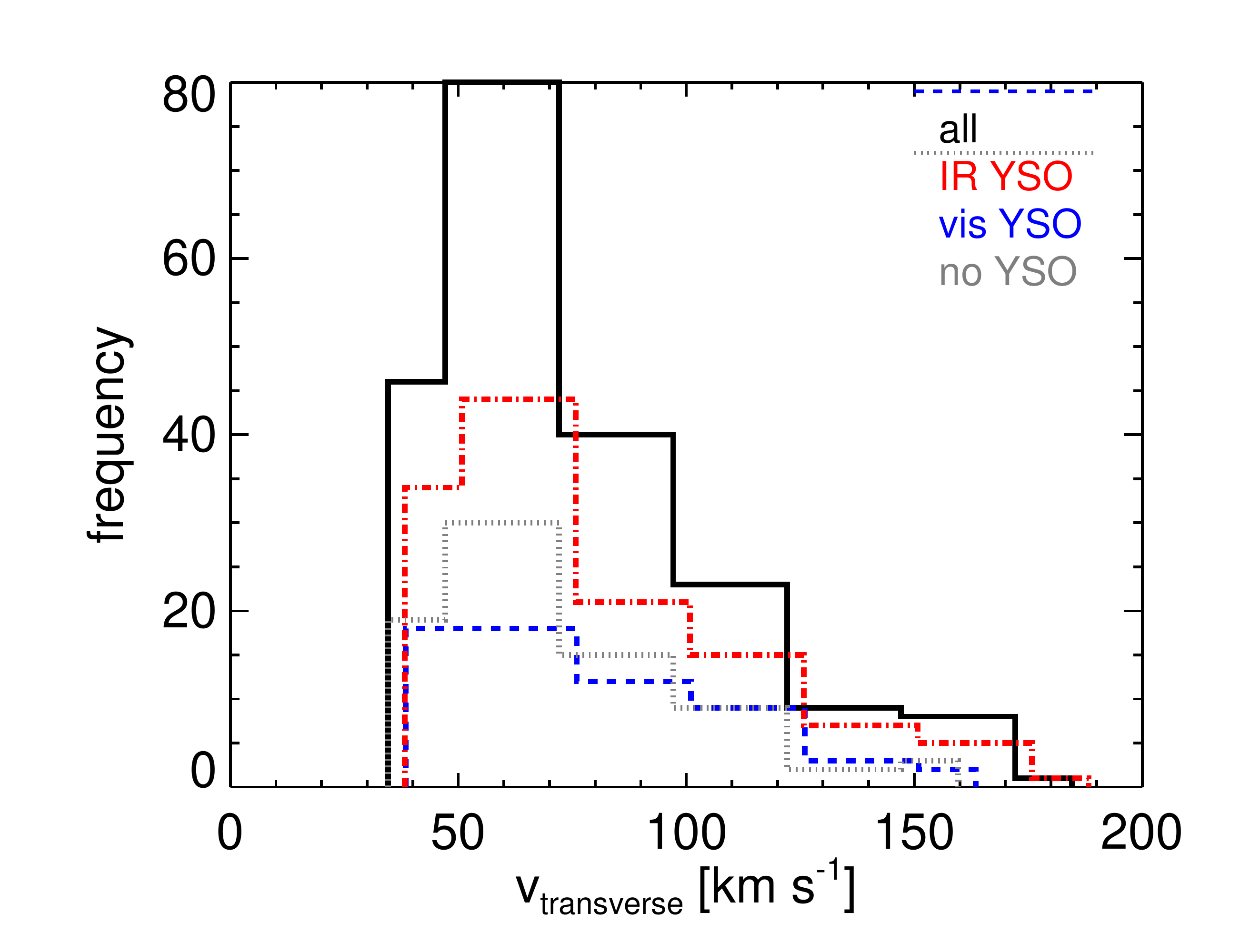} &
\includegraphics[trim=15mm 10mm 15mm 10mm,angle=0,scale=0.35]{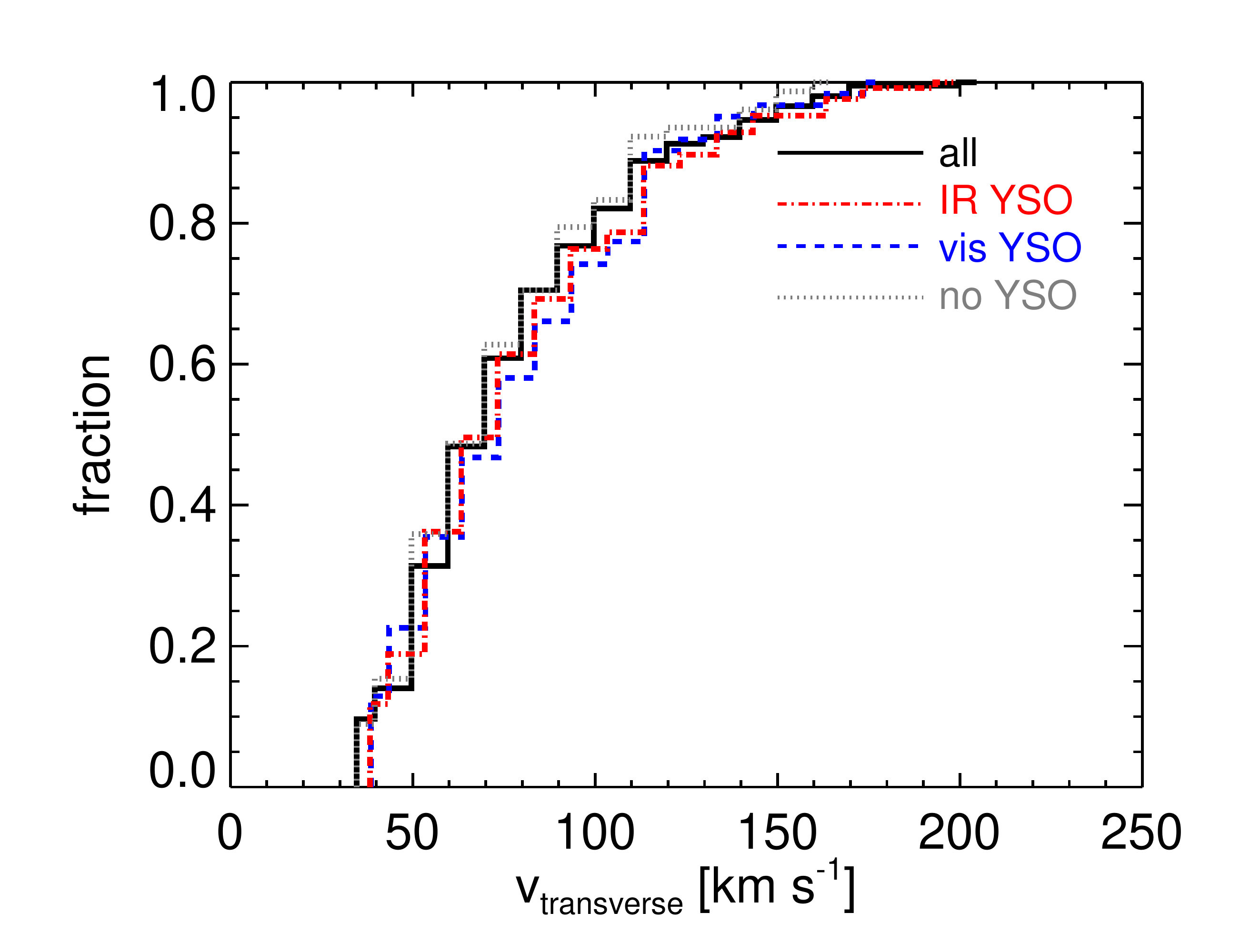} \\
\end{array}$
\caption{
\textit{Left:} 
Histogram of all proper motions measured in the 37 jets with two epochs of \emph{HST}/ACS images (black line). 
The distribution of knot velocities measured in jets driven by embedded protostars (red dash-dot line), unobscured stars (blue dashed line), and sources without a candidate driving source identified (gray dotted line) are overplotted. 
\textit{Right:} 
Cumulative distribution function comparing jet velocities measured from embedded, unobscured, and not-detected driving sources. 
}\label{fig:pm_hist} 
\end{figure*}

\subsection{Jet driving sources}\label{ss:driving_ysos}

Half of the jets presented in this paper can be matched to their driving source (20/37; see Table~\ref{t:jet_props}). 
This includes the 11 jet-driving sources that \citet{rei13} and \citet{rei16} identified from the morphology of near-IR [Fe~{\sc ii}] emission. 
An additional eight jet-driving sources can be identified in H$\alpha$ images where the jet extends from an unobscured star. 
Most of the jet-driving sources (18/20) were included in the PCYC \citep{pov11}. 
For those sources, we list the stellar properties derived from model fits to the IR SED in Table~\ref{t:jet_props}. 
\citet{ohl12} also identified the jet-driving sources for many of these HH jets. 
\citet{rei16} found that for sources present in both catalogs, the derived YSO properties agree to within a few percent.

The HH~903 and HH~1173 jet-driving sources are not included in the PCYC. 
\citet{ohl12} identified an embedded source at the base of HH~903 jet; we include their best-fit stellar parameters for this source in Table~\ref{t:jet_props}. 
The two knots of HH~1173 lie next to an candidate YSO, but move \textit{toward} the IR-excess source \citep[identified as the driving source by][]{ohl12}. 
Kinematics suggest that the jet originated from another nearby star that \citet{smi10} identify as a proplyd. 
Neither \citet{pov11} nor \citet{ohl12} identify this source as a YSO based on the shape of its IR SED.

For the remaining 18 driving sources that are included in the PCYC, we find an average mass of $\sim 3$~M$_{\odot}$.  
Most sources are classified as Stage 0/I (8/18), Stage II (4/18) with the remaining sources (6/18) given an ambiguous evolutionary classification based on the available IR photometry.

\subsection{Estimated dynamical ages}\label{ss:ages} 

We compute dynamical ages assuming that each jet feature has been traveling at the measured transverse velocity since launch. 
Estimated dynamical ages are listed in Table~\ref{t:jet_pm}. 
Jet ages range from a few hundred to $\sim 10^5$~yr, with a median age of $2\times 10^4$~yr. 
The longest jets have the oldest features (e.g., HH~666 and 903, see Figures~\ref{fig:hh666_boxes} and \ref{fig:hh903_boxes}, respectively and Table~\ref{t:jet_pm}), as expected. 
Shocks have likely altered the velocity structure of the jets. 
If the current jet velocity is slower than the launch velocity, then the dynamical age of the jet will be overestimated. 
Alternately, 
if younger jet knots are the easiest to see, the estimated dynamical age may not be a tight constraint on the total duration of outflow activity.


\section{Discussion}\label{s:discussion}

We present proper motions in 37 HH jets in the Carina Nebula measured in two epochs of H$\alpha$ images obtained with \emph{HST}/ACS. 
Most jets have at least one knot with a transverse velocity $\gtrsim 50$~km~s$^{-1}$ that traces fast, jet-like motions. 
A few streams of H$\alpha$ emission identified as jets by \citet{smi10} have transverse velocities $\lesssim 15$~km~s$^{-1}$ (HH~1005, HH~1007, and HH~1011), calling into question whether they are associated with a jet. 
However, we do measure transverse velocities $\gtrsim 15$~km~s$^{-1}$ in all 15 candidate jets and each appears to trace a distinct outflow. 
With the eight newly-confirmed jets reported here, in addition to previous confirmations \citep[see][]{rei13,rei16}, 
this brings the total number of known outflows in Carina to at least 60 \citep[including the 26 jets and outflows discovered by][]{har15}.
Given the extended jet morphology in images, we estimate that these sources lie nearly perpendicular to the line of sight (a tilt angle $< 45^{\circ}$ away from the plane of the sky). 
Transverse velocities in Table~\ref{t:jet_pm} are therefore lower limits; the full 3D space velocities may be faster by as much as a factor of $\sqrt{2}$.

HH~1011, a one-sided microjet that extends off of a small globule in Tr15, is the only source in our sample that does not have any features that clearly move between the two epochs (transverse velocities $<15$~km~s$^{-1}$). 
This upper limit is slower than typical jet velocities and a factor of $\gtrsim 3$ slower than the rest of the sample. 
Thus, the true nature of this object remains unclear.

Figure~\ref{fig:pm_hist} shows a histogram and cumulative distribution function of all transverse velocities. 
Knot velocities from jets driven by embedded protostars (red dash-dot line), unobscured driving sources (blue dashed line), and those where the protostar remains undetected (gray dotted line) are substantially similar and statistically indistinguishable (the maximum deviation between the distributions is less than $\sim 0.1$). 
Jets driven by embedded protostars may interact with more material circumstellar material as they exit their natal globules, so we might expect slower velocities than observed from jets driven by unobscured YSOs. However, the two histograms show no such difference. 
The plotted distributions include all knots from a given source, both the fastest knots in the inner jet and more distant, slower features. 
We discuss the knot structure of these jets and the transfer of momentum in detail in Sections~\ref{ss:knot_strct} and \ref{ss:momentum}, respectively.

\begin{figure*}
\centering
\includegraphics[trim=0mm 5mm 0mm 0mm,angle=0,scale=0.875]{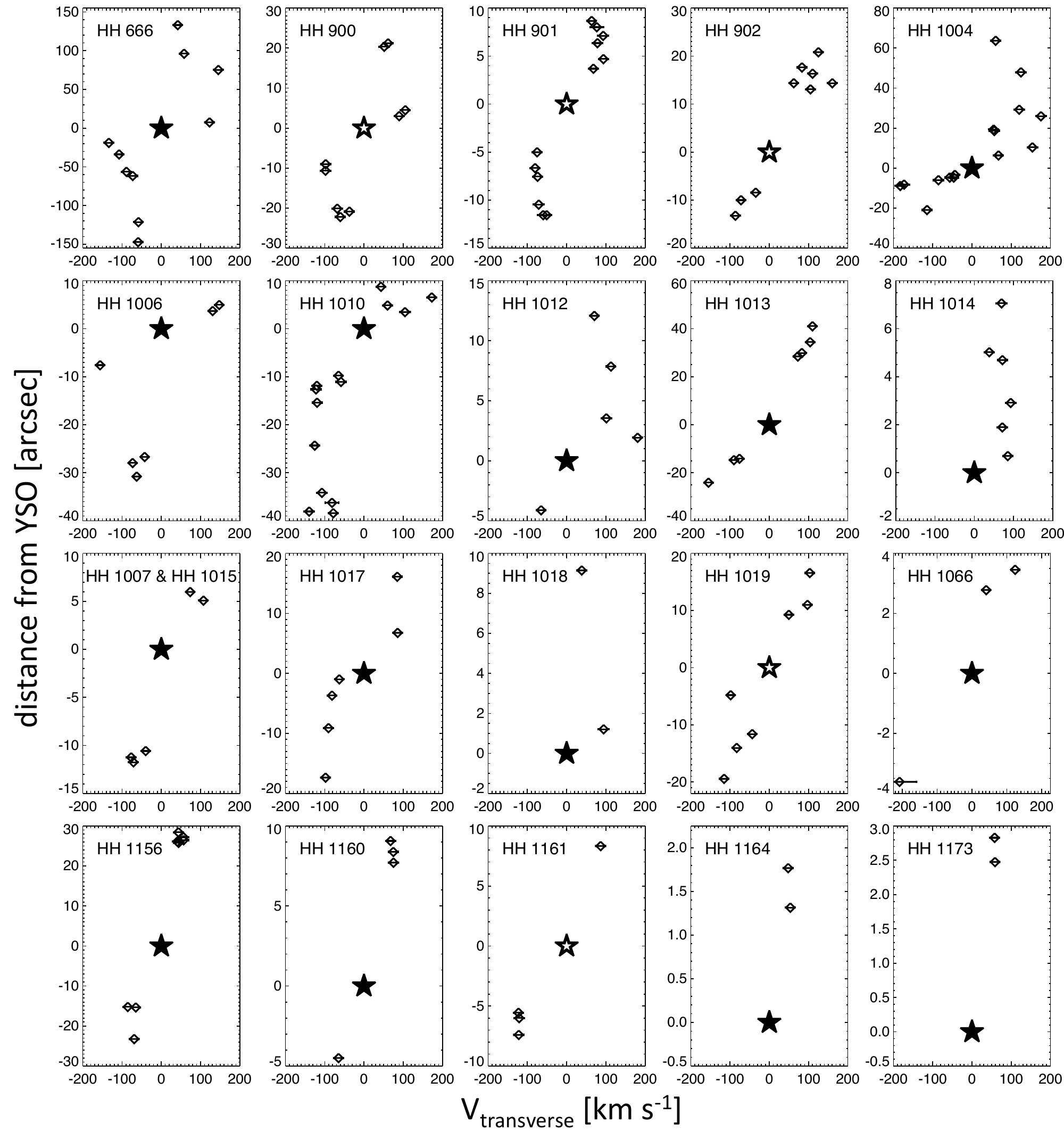}  
\caption{Position-velocity diagrams of jets with a well-localized driving source. Stars denote the YSO position identified in H$\alpha$ and/or IR images (filled stars) or inferred from knot proper motions (outlined stars). 
}\label{fig:pv_diagrams} 
\end{figure*}

Position-velocity diagrams for jets with a well-localized driving source (either detected, or constrained to be within a small area based on jet proper motions) are shown in Figure~\ref{fig:pv_diagrams}. 
More distant knots have lower velocities in many of the jets (e.g., HH~666, 900, 901, 1004, 1006, and 1010), similar to the velocity structure seen in HH~34 \citep{dev97}. 
Less complete sampling of the outflow may obscure a similar velocity structure in those jets where only a few knots can be measured (e.g., HH~1014, 1066, and 1156).
Overall, the ensemble of these diagrams does not show a net trend in knot velocities with distance from the driving source, unlike that seen in some other HH jets (e.g., HH~34, \citealt{dev97}, HH~666 in this sample, see also \citealt{smi04,rei14}). 
Knots that do not, in general, show lower speeds at larger distances from the driving source indicate that the HH objects are much denser than the environment they are passing through. 

Knots that are slightly offset from the main jet axis (as in e.g., HH~666 and HH~903) have slower velocities than those along the main jet body. 
\citet{rei15b,rei15a} reported examples of two-component outflows seen in Carina.  
In both HH~666 and HH~900, [Fe~{\sc ii}] emission traces a fast, collimated jet that is surrounded by a wider-angle H$\alpha$ outflow.
Similar sheaths of H$\alpha$ emission encompass the [Fe~{\sc ii}] jets in HH~903, 1004, and 1164 \citep[see][]{rei16}.
Not all jets driven by embedded protostars exhibit this same two-component structure.
Nevertheless, we see a difference in the velocity of features detected on- and off-axis in sources where we have detected only one outflow component. 
The fastest H$\alpha$ proper motions are those directly tracing the \textit{jet} ($\gtrsim 100$~km~s$^{-1}$, e.g., HH~1006, 1012, 1017). 
Shock-like features (e.g., HH~1008, HH~1009, and the southwest limb of HH~1004) have lower velocities overall ($\sim 50$~km~s$^{-1}$).

\subsection{Knot structure}\label{ss:knot_strct}

Dynamical ages for all sources are less than 1~Myr, and most are less than $10^5$~yr (see Table~\ref{t:jet_pm} and Section~\ref{ss:ages}). 
The median outflow age is $\sim 10^4$ yrs. 
This is only a few percent of the Class 0 and Class I lifetimes estimated for low-mass sources \citep[0.16 Myr and 0.54 Myr, respectively;][]{eva09}. 
However, both the source luminosities and jet mass-loss rates suggest that the jet-driving sources are intermediate-mass protostars (see Table~\ref{t:jet_props}), which will evolve faster. 
For an intermediate-mass ($M \sim 2-8$~M$_{\odot}$) star, the Kelvin-Helmholtz time $t_{KH} = GM^2/LR \sim M^{-2}$ will be 1-2 orders of magnitude shorter than in the low-mass case. 
If the corresponding Class 0 and Class I lifetimes are also 1-2 orders of magnitude shorter, then these jets trace a significant fraction of the accretion history of the driving source.

For jets in this sample with multiple well-separated knots (e.g. HH~666, HH~1006, HH~1010, and HH~1017, see Figures~\ref{fig:hh666_boxes},~\ref{fig:hh1006_boxes},~\ref{fig:hh1010_boxes}, and \ref{fig:hh1017_boxes}), we find knot separations of $\sim 5-30$\arcsec. 
Assuming that all knots have been traveling at a constant velocity since launch, inter-knot separations correspond to $\sim 1000$ yrs between ejection events. 
This is consistent with the duty cycle between accretion/outflow bursts inferred from knot spacing in jets from low-mass stars \citep[e.g.][]{har90,rei92,bal94}.

\citet{rei15b} have also argued for episodic accretion and outflow based on smooth, steady velocities seen in the inner jet of HH~666 and the morphology of the entrained outflow shell. 
Where the morphology of the continuous inner jet is smooth (suggesting no strong shocks), we use the length of the continuous inner jet to estimate the duration of the current/recent accretion outburst that powers the jet (see Section~\ref{ss:ages} and Table~\ref{t:jet_props}). 
Inner jet lengths range from $\sim 1-17$\arcsec, corresponding to bursts that have been ongoing for $\sim 60-1500$ yrs. 
The median outburst duration is $\sim 700$ yr. 
Only the shortest jets are consistent with the observed duration of FU Orionis outbursts \citep{hk96}. 
In order for the observed length of the continuous inner jet to be consistent with the decay times observed in FU~Orionis-like outbursts ($\gtrsim 100$~yr) jet velocities would have to be an order of magnitude higher than we measure.

It is unclear how the typical decay time of an accretion burst from an intermediate-mass star compares to an FU~Orionis outburst, as accretion outbursts have only been detected in a few intermediate-mass sources \citep[e.g.,][]{hin13}. 
Indeed, the length of the continuous inner jet hints at long decay times that would be poorly constrained with modern observations. 
Nevertheless, there is growing evidence that episodic accretion may be a ubiquitous feature of star formation affecting stars from low to high masses \citep[e.g.,][]{fuj15,ste16,hos16,car16,mey17}.

\subsection{Comparison to other jets}\label{ss:comp_others}
\begin{figure}
\centering
\includegraphics[trim=5mm 20mm 5mm 15mm,angle=0,scale=0.35]{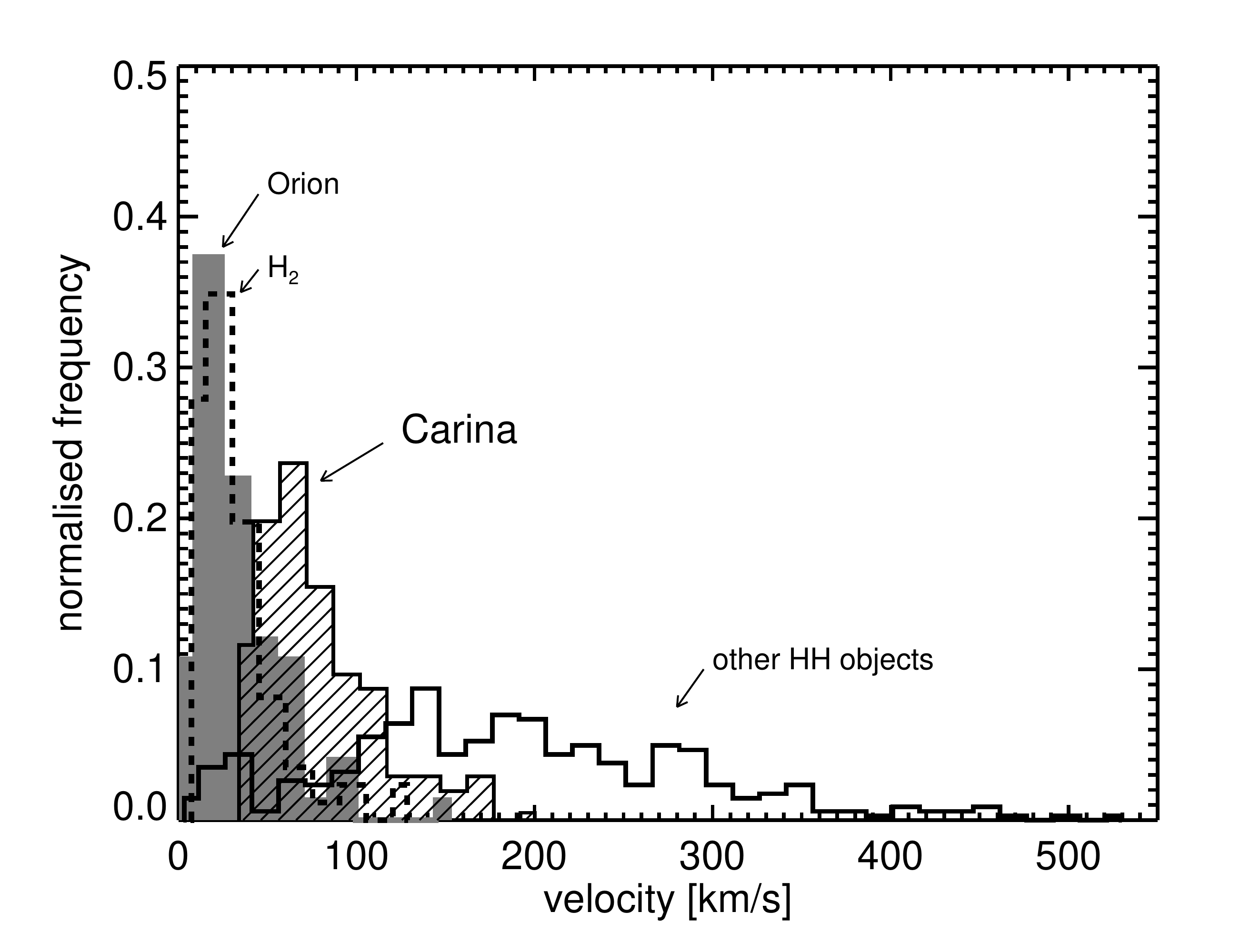} 
\caption{
Normalised histograms comparing jet proper motions measured in Carina (cross-hashed) with those in Orion \citep[shown in gray, from][]{ode03}, H$_2$ jets \citep[dashed line, from][]{zha13} and other HH objects \citep[solid line, from][]{bal02,bal12,dev97,dev09,har01,har05,har07,kad12,mcg07,nor01,rei02,smi05,yus05}. 
}\label{fig:comp_pm_hist} 
\end{figure}

Many of the HH jets in Orion are observed to be one-sided, with no clear counterjet \citep[e.g.,][]{rei98,bal00}. 
A few jets in Carina are also apparently one-sided (e.g., HH~1018, 1163, 1164, 1172, 1173) and all are driven by a relatively unobscured protostar that can clearly be seen in H$\alpha$ images. 
\citet{rei98} suggest that uneven illumination of the disk in the H~{\sc ii} region may explain the asymmetry seen in more exposed microjets. 
In contrast, most of the embedded jets in Carina are bipolar, although in some cases the obscured counterjet can only be seen at longer wavelengths \citep[e.g., HH~1014, see][]{rei16}.  
Velocities in the two limbs of the bipolar jets (see Figure~\ref{fig:pv_diagrams}) are not markedly asymmetric as has been observed in some jets from low-mass stars \citep[e.g.,][]{hir94}.

Figure~\ref{fig:comp_pm_hist} shows a comparison of the transverse velocities measured from jets in Carina with those in other regions. 
The comparison sample of jets represents primarily famous jets driven by low-mass protostars \citep[e.g., HH~1-2, HH~7-11, HH~47, HH~111, from][respectively]{bal02,nor01,har05,har01} with a few sources in Orion \citep[from][]{smi05,bal12}, and an irradiated jet in the Trifid \citep[HH~399,][]{yus05}. 
We present transverse velocities that have \textit{not} been corrected for the tilt angle away from the plane of the sky. 
Transverse velocities in the Carina jets overlap with those measured in other jets reported in the literature (solid line histogram), but do not sample the high-velocity tail of the distribution. 
Instead the distribution in Carina peaks around a velocity $\sim 100$~km~s$^{-1}$. 
We also plot a histogram of proper motions measured in H$_2$ by \citet{zha13} for comparison. 
These molecular jets presumably trace more embedded sources, and indeed, the peak of the H$_2$ transverse velocities is a factor $\sim 2\times$ slower than the jets in Carina.  

For a sample of HH jets from low-mass stars in an environment more similar to Carina, we look to Orion and the H$\alpha$ proper motions measured by \citet{ode03}. 
Knot transverse velocities are a factor of 2 slower than the jets in Carina, with a distribution similar to the H$_2$ jets. 
Recent work placing Orion at 388~pc \citep{kou17} instead of the 460~pc assumed by \citet{ode03} will make the discrepancy larger. 
\citet{smi10} find a similar offset between Carina and Orion when comparing the jet mass-loss rates. 
As argued in that paper, the more vigorous jets in Carina likely reflect the higher mass of the driving sources. 
Slower, lower-mass-loss rate jets are absent in the Carina sample due to the decreased sensitivity to slow/fainter jets at the larger distance.

Higher-mass driving sources may be invoked to explain the faster velocities measured in the jets in Carina compared to Orion. 
However, jets in both Orion and Carina tend to be slower than the transverse velocities measured in a heterogeneous sample of local jets reported in the literature. 
One obvious possibility is selection bias -- that the biggest, brightest, and fastest jets are the best studied. 
The large-scale environment may also affect the jet velocity.
The majority of jets in the comparison sample do not live in an H~{\sc ii} region like Carina or Orion.
\citet{daG96} consider the propagation of jets into stratified media and find that jets moving into increasing pressure will slow down. 
Subsequent work looking at the overall velocity structure of giant HH jets lead \citet{daG01} to suggest that environmental interaction cannot be the primary mechanism to slow jets. 
It is unclear whether the lower velocities measured in jets in H~{\sc ii} regions reflect higher pressure in the environment and possibly more compression in the molecular gas. 
Regardless, the offset is peculiar, given that we do not find a significant difference in Carina between the transverse velocities of jets from stars embedded in a cloud and those exposed in the H~{\sc ii} region.

\subsection{Environmental interaction}\label{ss:momentum}

Jets and outflows are an important source of feedback and may help maintain cloud-scale turbulence in star-forming regions \citep[e.g.,][]{plu13,plu15}. 
Momentum and energy from jets and outflows contributes more locally, helping to reshape and clear the circumstellar envelope \citep[e.g.,][]{arc05,jor07,off11,zha16}. 
Outflow feedback may be particularly important for high-mass star formation, as it clears a cavity along the jet axis, thereby providing a pathway for the escape of thermal radiation. 
Several authors have considered the role that protostellar outflows may play in facilitating non-spherical accretion \citep[e.g.,][]{kru05,cun11,kui15,kui16}.

Combining kinematics presented in this work with mass-loss rates from previous papers \citep{smi10,rei16} allows us to compute the jet kinetic energy and momentum. 
\citet{dio16} compared the energy and momentum of atomic jets in NGC~1333 to the energy and momentum of the molecular outflows associated with the same sources. 
They find that the atomic jets have a small fraction of the momentum of the corresponding outflows, but that the two components inject comparable kinetic energy.

Many of the jets in Carina are composed of multiple distinct knots, reflecting previous ejection events. 
We compute the kinetic energy of the inner jet where we have estimates of both the mass-loss rate and velocity. 
This represents only a fraction of the total energy injected by the jet over its lifetime. 
Estimating the kinetic energy of previous bursts is difficult as shock processing and photoevaporation in the H~{\sc ii} region have likely reduced both the mass and velocity of that ejecta.

To estimate the jet mass, we use the mass-loss rates derived from the H$\alpha$ emission measure by \citet{smi10}. 
This gives the mass of ionized gas in the jet, which \citet{rei13,rei16} show is a lower limit on the true mass-loss rate of the jet. 
In addition, \citet{rei15b} showed that H$\alpha$ and [Fe~{\sc ii}] emission may also trace different kinematics, so we prefer mass-loss rates and velocities estimated from the same line to ensure that they trace the same gas. 
We therefore use the H$\alpha$ mass-loss rate in order to compare the jet kinetic energy across the sample, but note that the energy may be an order of magnitude or more higher if the neutral material in the jet is taken into account.

\begin{figure}
\centering
\includegraphics[trim=5mm 10mm 5mm 15mm,angle=0,scale=0.35]{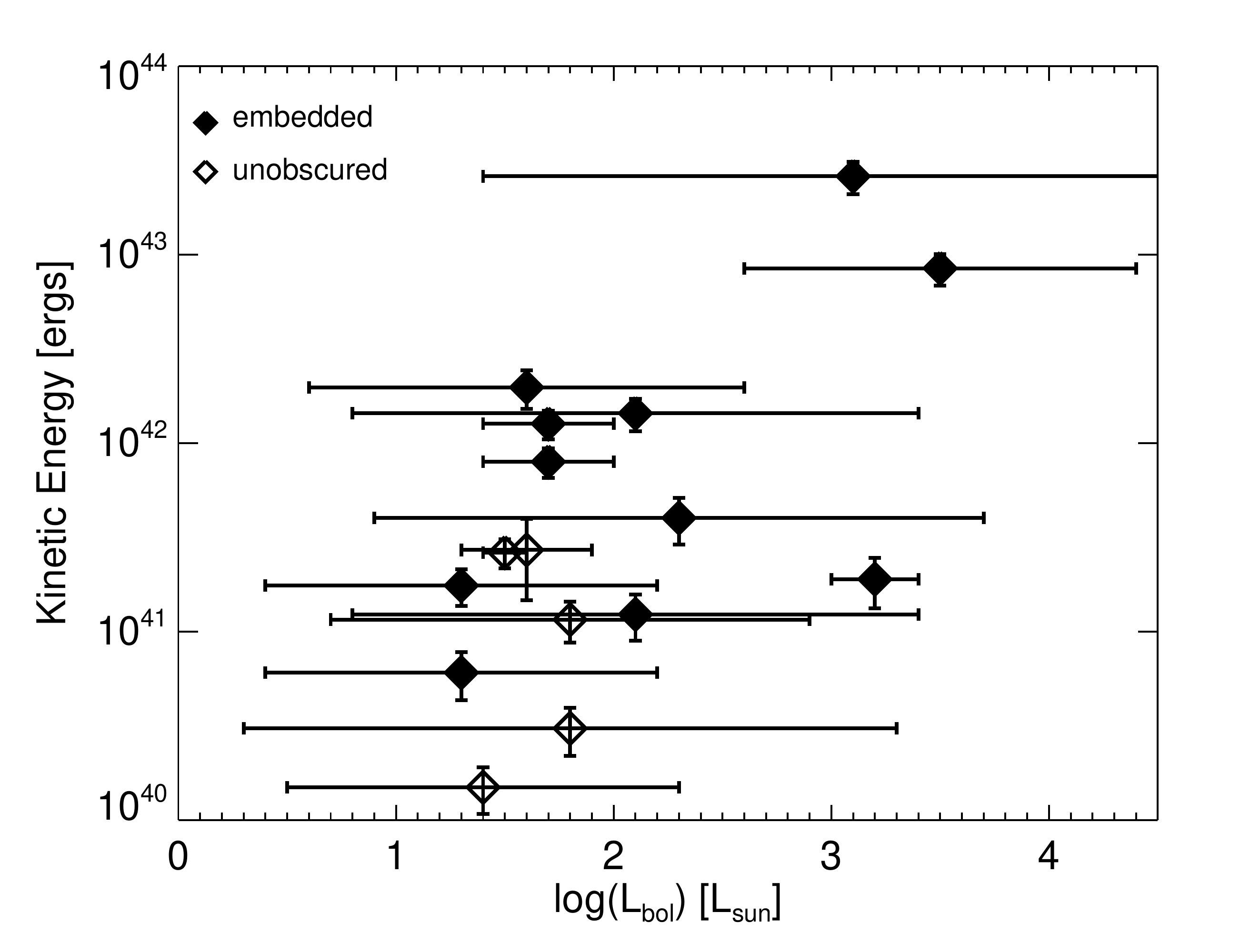} 
\caption{
Jet kinetic energy compared to the bolometric luminosity of the driving source. Embedded sources, shown in filled symbols, tend to have higher kinetic energy than jets driven by unobscured protostars (shown with open symbols). 
}\label{fig:jet_kin} 
\end{figure}

Estimated this way, we find jet kinetic energies of $1.5 \times 10^{40} \lesssim \frac{1}{2} M_{jet} v^2_{jet} \lesssim 2.6 \times 10^{43}$~ergs (median $\sim 8 \times 10^{41}$~ergs; see Figure~\ref{fig:jet_kin}). 
To estimate the local impact of the jet, we compare the jet kinetic energy to the gravitational binding energy of the globule. 
Many of these jets emerge from the tips of dust pillars that protrude into the H~{\sc ii} region. 
For those objects, we assume that the globule is a sphere with radius, $R_{glob}$, equal to half the width of the minor axis of the pillar head. 
Not every jet emerges from a pillar head, however. 
In cases where a jet emerges from the middle of the pillar (e.g., HH~903) or multiple jets emerge from the same globule (e.g., HH~1159-1164 in the Seahorse globule), we take the globule radius, $R_{glob}$, to be half the distance between the two points where the jet emerges from the cloud.

\citet{pre12} presented a column density map of the Carina Nebula based on 70~\micron\ and 160~\micron\ data from \emph{Herschel}. 
Few of the pillars and globules with jets are resolved in 160~\micron\ image ($\sim 10$\arcsec\ resolution). 
This prohibits a robust column density determination for each individual source. 
Guided by the results of \citet{pre12}, we instead adopt three values of the column density to describe jet-driving sources that are (1) visible in H$\alpha$ images, (2) can be identified in IR images, and (3) protostars that remain obscured in an opaque cloud. 
For jets where the driving source can be identified in an H$\alpha$ image, we assume an $A_V \sim 1.0$~mag, corresponding to a column density log$(N_H) \approx 21.2$~cm$^{-2}$. 
For embedded driving sources that can be identified in the IR but are not seen in the optical, we assume $A_V \sim 3.0$~mag for a column density log$(N_H) \approx 21.75$~cm$^{-2}$. 
Lastly, if a source is completely obscured in an opaque cloud, we assume $A_V \sim 10.0$~mag corresponding to a column density log$(N_H) \approx 22.3$~cm$^{-2}$. \citet{rei13} estimated that $A_V \gtrsim 9$~mag obscures the unseen HH~901 driving source. 
This is consistent with an $A_V \gtrsim 10$~mag to the stellar surface estimated for most Stage 0/I YSOs in the PCYC \citep[see][]{pov11}.

Assuming that each jet-driving source is embedded in a sphere of uniform density, we compute the globule mass as 
\begin{equation}
M_{glob} = \mu m_H \frac{N_H}{R_{glob}} \frac{4}{3} \pi R^3_{glob}
\end{equation} 
where 
$\mu = 1.35$ is the mean molecular weight and 
$m_H$ is the mass of Hydrogen. 
The gravitational binding energy, $U_{grav} = \frac{3}{5} \frac{GM^2}{R_{glob}}$, for these globules ranges from $1.1 \times 10^{39} \lesssim U_{grav} \lesssim 1.3 \times 10^{43}$~ergs (median $\sim 1 \times 10^{41}$~ergs). 
Most globules likely have a larger binding energy than these coarse estimates. 
\citet{rei15a} estimate the mass of the HH~900 globule based on the $A_V$ required to hide a $\sim 2$~M$_{\odot}$ protostar and find a globule mass roughly an order of magnitude higher than we estimate here. 
Using model fits to the IR SED (including $\lambda > 100$~\micron\ photometry), \citet{ohl12} estimate envelope masses $\sim 1-3$ orders of magnitude larger than our estimated globule masses.

With these uncertainties in mind, we find that the jet kinetic energy is a factor of a few larger than the gravitational binding energy in most sources (11/15). 
This coarse estimate suggests that the jet has adequate energy to disrupt the core along the outflow axis, creating the disturbed morphology along the ionization front in, e.g., HH~903 and HH~1004, and may even reshape small globules like HH~900.  
In fact, the median kinetic energy of jets driven by embedded protostars is an order of magnitude higher than the median energy of jets driven by unobscured stars ($1.3 \times 10^{42}$~ergs compared to $1.2 \times 10^{41}$~ergs, respectively). 
This hints at a decrease in the jet energy with protostar evolution, as expected for accretion and mass-loss rates that decrease with time \citep[e.g.,][]{bon96,cog12,ant14}. 
However, we note that jet velocities are similar whether or not the driving source is embedded. 
Differences in the jet kinetic energy therefore reflect a difference in the mass-loss rate. 
Jets driven by embedded sources may entrained cloud material which will increase the jet mass in a way that is uncorrelated with the evolutionary stage of the driving source.

Radiative losses reduce the amount of jet energy that is deposited into the star-forming region. 
In contrast, the momentum carried by the jet and outflow are injected locally and conserved within the cloud. 
We consider the momentum flux carried by the jets in Carina in the next section. 

\subsection{Comparison to other types of outflows / quantifying the outflow force}\label{ss:fout_comp}

\begin{figure}
\centering
\includegraphics[trim=10mm 10mm 10mm 10mm,angle=0,scale=0.375]{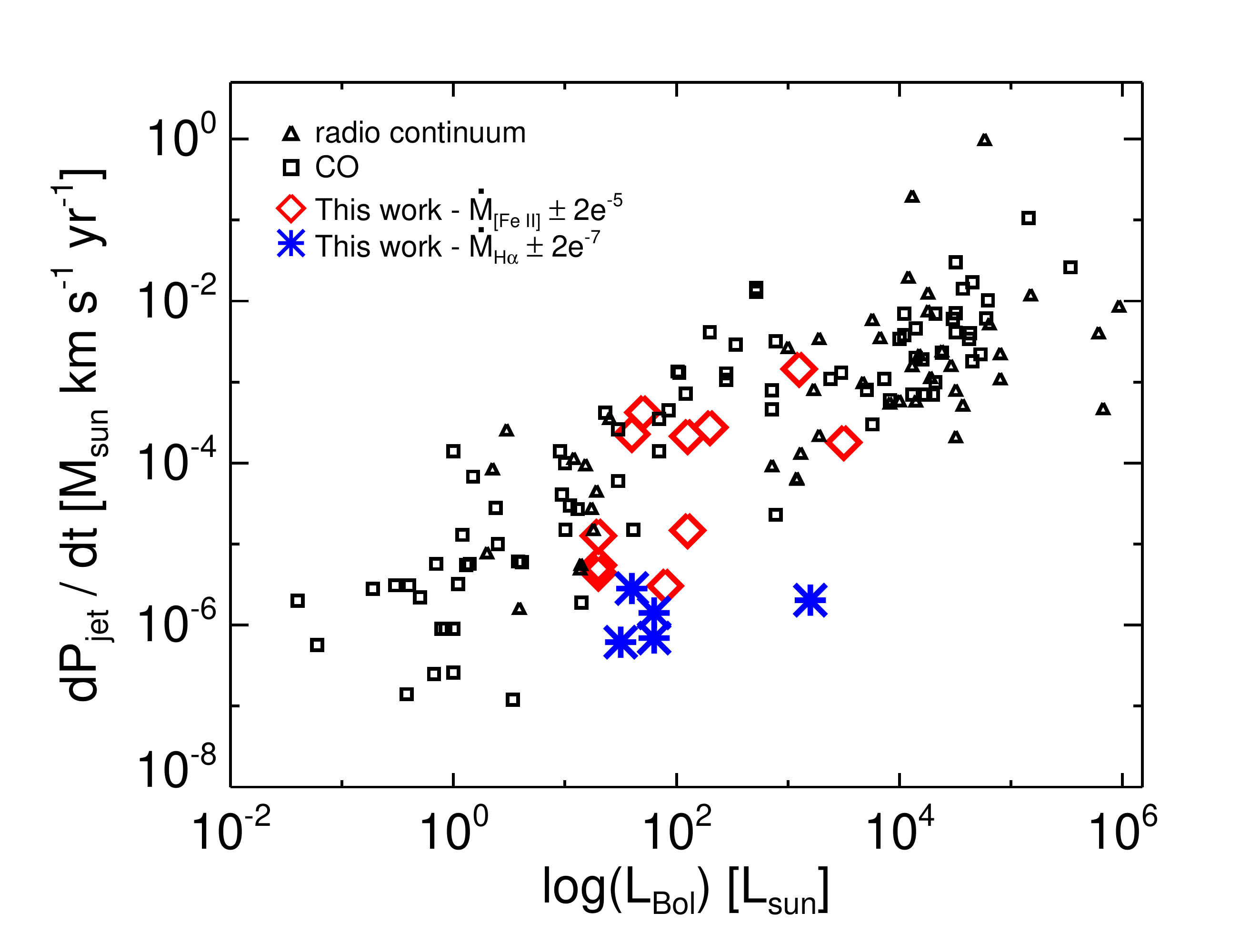} 
\caption{Comparison of the outflow force as a function of the driving source luminosity. 
  Sources with a mass-loss rate estimated from the survival of [Fe~{\sc ii}] \citep{rei16} are shown with red diamonds; for sources without [Fe~{\sc ii}] data, we use the mass-loss rate derived from the H$\alpha$ emission measure \citep{smi10}, shown with blue asterisks.
  We note the typical uncertainty for these two methods of estimating the outflow force alongside the appropriate symbol in the plot legend. 
For comparison, we plot the outflow force estimates derived from radio continuum observations (triangles) by \citet{ain12,mos16,pur16} and 
the outflow force estimated from low-J CO lines (squares) by 
\citet{bon96,van13,dua13,mau15,van16}. }\label{fig:lbol_fdot_jet} 
\end{figure}

The momentum carried by protostellar jets has been investigated as a way to clear the protostellar envelope and drive turbulence in the larger star forming cloud. 
Outflows may dominate feedback in low-mass star-forming regions, but in a high-mass region like Carina, the contribution of winds and radiation from dozens of O-type stars will exceed that of jets and outflows by orders of magnitude \citep[see, e.g.,][]{balIAUS11,bre12}. 
Nevertheless, estimating the momentum injected by jets from higher-mass protostars is useful for understanding how that feedback may affect clusters composed primarily of low-mass stars. 
In addition, comparing the momentum flux or outflow force, dP$_{jet}$/dt, between different sources has been used to investigate the underlying outflow launching mechanism. 
Smooth scaling of the relationship between the outflow force and the bolometric luminosity of the driving source has been used to argue that the same physical mechanism governs outflow from low- and high-mass stars \citep[e.g.][]{kon99,mau15}.

\citet{rei13,rei14,rei16} argue that the irradiated HH jets seen in Carina provide a unique view of outflow physics common to the formation of more massive protostars. 
In embedded star-forming regions, the highly collimated jet cores will remain unseen behind the large column of gas and dust typical of more massive star-forming regions. 
\citet{rei14} estimate that four jets in Carina have enough momentum to power a molecular outflow comparable to those observed from intermediate-mass protostars \citep[e.g.,][]{bel08}. 
With both mass-loss rates and velocities, we can test whether the outflow force of the jets in Carina is comparable to that measured with other tracers in other environments.

We compute the outflow force in the inner jet using the transverse velocity measured from H$\alpha$ proper motions with the mass-loss rate estimated in that portion of that jet. 
Where available, we used the jet mass-loss rates that \citet{rei16} estimated from the minimum density to shield Fe$^{+}$ in the jet (red diamonds in Figure~\ref{fig:lbol_fdot_jet}). 
For jets not included in the \citet{rei16} sample, we used the mass-loss rates estimated from the H$\alpha$ emission measure by \citet{smi10} (blue asterisks). 
Mass-loss rates estimated from the survival of Fe$^+$ are roughly an order of magnitude higher than estimated from the H$\alpha$ emission measure; thus the outflow force is also an order of magnitude higher. 
Velocity differences between H$\alpha$ and [Fe~{\sc ii}] may increase the estimated outflow force by a factor of a few. 
We compare the outflow force to the bolometric luminosity of the source estimated from model fits to the IR SED by \citep[][see Table~\ref{t:jet_props}]{pov11} in Figure~\ref{fig:lbol_fdot_jet}.

As with the jet kinetic energy, we find that embedded sources drive more forceful outflows. 
The median outflow force of embedded sources is a factor of two higher than the median outflow force of jets driven by unobscured protostars.
While this is a coarse distinction between evolutionary stages, this trend is in general agreement with the results of \citet{bon96} for outflows from low-mass stars.

Several other authors have measured the mechanical force of CO outflows \citep[e.g.][]{bon96,van13,mau15,van16} and radio continuum jets \citep[e.g.][]{ain12,mos16,pur16}. 
We include these sources for comparison in Figure~\ref{fig:lbol_fdot_jet}. 
The jets in Carina presented here lie on the same relation between outflow force and source luminosity as the outflows from low- and high-mass stars if the most complete (and therefore higher) mass-loss rate is used. 
Smooth behavior of the outflow force as a function of $L_{bol}$ from low- to high-masses provides further evidence for a common outflow mechanism. 
We note that the jets in Carina are more analogous to the low-mass sources in that we are measuring the contribution of one jet from one protostar. 
This is in contrast to the high-mass end where the outflow force reflects the contribution of everything within that high-mass clump. 
Finally, placing the jets in Carina on this relation and finding good agreement with the outflow force supports the argument of \citet{rei13,rei14,rei16} that the jets in Carina trace a scaled-up version of low-mass star formation, and a different view of the same underlying outflow physics that create outflows from more embedded regions.

\section{Conclusions}\label{s:conclusions}

We present proper motions of well-defined knots in 37 jets and HH objects in the Carina Nebula. 
With two epochs of H$\alpha$ \emph{HST}/ACS images obtained $\sim 10$~yrs apart, we are sensitive to nebular motions moving faster than $\sim 15$~km~s$^{-1}$. 
For 20/37 jets, proper motion confirm the intermediate-mass protostar (median mass $\sim 3$~M$_{\odot}$) identified along the jet axis as the driving source.

Transverse velocities in these jets from intermediate-mass stars are similar to those measured in jets driven by low-mass stars (the median knot velocity is $\sim 75$~km~s$^{-1}$). 
We do not observe a significant difference between the knot velocities in jets driven by embedded protostars and unobscured stars. 
Despite their similar velocities, we find that jets from embedded sources tend to be more energetic and have higher momentum than their unencumbered counterparts that lay naked in the H~{\sc ii} region.

Dynamical ages of all jets (median $\sim 10^4$~yr) are shorter than the Class 0/I lifetimes estimated for low-mass stars \citep{eva09}. 
Many sources have extended inner jets with dynamical ages of a few hundred years that suggest a recent/ongoing accretion burst. 
This provides additional evidence that episodic accretion may be important for the formation of stars of all masses and hints that the burst duration may be longer for higher mass stars.

We combine the kinematics measured in this paper with mass-loss rates estimated by \citet{smi10} and \citet{rei16} to compute the outflow force. 
Previous measurements suggest a smooth scaling of the outflow force with the bolometric luminosity of the driving source. 
The jets presented in this work affirm this relationship. 
Sources lie closest to the locus of high- and low-mass outflows when the outflow force is computed using the higher mass-loss rates obtained by \citet{rei16}, who include the neutral jet core in the estimated mass-loss rate. 
Unlike other measurements of higher-luminosity sources, we measure the outflow force of a single jet, rather than the cumulative force of all outflows from a cluster of forming stars.

Smooth scaling of the outflow force with source luminosity has also been used to argue in favor of a common physical mechanism underlying the production of outflows regardless of driving source mass. 
That the jets in Carina fall on the relation between low- and high-mass stars in an outflow force diagram offers strong additional evidence that high-mass stars can form via a scaled-up version of low-mass star formation.


\section*{Acknowledgments}
We wish to thank Jay Anderson for helping us quantify the additional uncertainty introduced by the $180^{\circ}$ rotation between epochs. 
Thanks also to Bo Reipurth for helpful conversations. 
MR would like to thank John Bieging, Tom Haworth, and Chris Miller. 
HH numbers are assigned by Bo Reipurth in order to correspond with the catalogue of HH objects maintained at http://ifa.hawaii.edu/reipurth/; see also Reipurth B., \& Reiter M., 2017, \textit{A General Catalog of Herbig-Haro Objects}, 3$^{rd}$ Edition, in prep. 
Support for this work was provided by NASA through grants AR-12155, GO-13390, and GO-13391 from the Space Telescope Science Institute. 
This work is based on observations made with the NASA/ESA Hubble Space Telescope, obtained from the Data Archive at the Space Telescope Science Institute, which is operated by the Association of Universities for Research in Astronomy, Inc., under NASA contract NAS 5-26555. These \textit{HST} observations are associated with programs GO~10241, 10475, 13390, and 13391.


\bibliographystyle{mnras}
\bibliography{bibliography_hh_objs}


\clearpage 

\begin{landscape}
\begin{table}
\caption{Properties of jets and their driving sources}
\footnotesize
\begin{tabular}{lrrrrrrrrrrrrrr}
\hline\hline
Name    & L$_{inner}$  & L$_{inner}$  & Burst      & Duty    & $\dot{M}_{H\alpha}$  & $\dot{M}_{[Fe~II]}$   & log(K.E.) & log(dP/dt) & log(dP/dt)    & Two   & PCYC & log(L$_{bol}$) & Mass & Stage  \\
        & [\arcsec]  & [pc]        & Duration   & Cycle   & $M_{\odot}$ yr$^{-1}$ & $M_{\odot}$ yr$^{-1}$ & [ergs]    & H$\alpha$  & [Fe~{\sc ii}] & comp? &  & [L$_{\odot}$] & $M_{\odot}$ &  \\
\hline 
HH~666  &  16.7   & 0.19   & 1400 (89)  & 11000 &  1.90e-7 & 1.2e-5 & 43.4 & -4.5 & -2.8 & y &  345 & 3.1 (1.7) & 6.3 (1.3) &  II  \\
HH~901  & 4.3     & 0.05   & 1185 (259) &   300 &   4.0e-8 & 4.7e-6 & 42.1 & -5.1 & -3.5 & ? &  ... & ...       & ...       & ...  \\
HH~902  & 12.9    & 0.14   & 1341 (108) &   400 &  1.76e-7 & 1.0e-5 & 42.9 & -4.7 & -3.0 & ? &  ... & ...       & ...       & ...  \\
HH~1066 &  2.1    & 0.02   &  185 (88)  &  ...  &  1.44e-7 & 1.7e-6 & 42.2 & -4.7 & -3.7 & y &  429 & 2.1 (1.3) & 2.8 (1.6) &  A   \\
\hline
HH~900  &  6.5    & 0.07   & 1002 (113) &  1400 &  5.68e-7 & 1.7e-5 & 43.2 & -4.3 & -2.7 & y &  ... & ...       & ...       & ...  \\
HH~903  & 12.8    & 0.14   & 1501 (119) &  ...  & 1.72e-7 & 1.0e-6 & 43.0  & -4.8 & -4.0 & y &  $*$ & 2.4 (0)   &  4.3 (0)  & ...  \\
HH~1004 &  9.1    & 0.10   &  675 (75)  &  ...  &  1.20e-7 & 1.2e-6 & 42.9 & -4.8 & -3.7 & y & 1198 & 3.5 (0.9) & 7.5 (3.0) &  A   \\
HH~1005 &  ...    & ...    & ...        &  ...  &  1.24e-7 & 1.2e-6 & ...  & ...  & -4.0 & ? &  ... & ...       & ...       & ...  \\
HH~1006 & 5.4     & 0.06   &  423 (82)  &  5000 &   2.8e-8 & 3.0e-6 & 42.0 & -5.5 & -3.4 & n & 1173 & 1.7 (0.3) & 1.8 (0.9) & 0/I  \\
HH~1007$^{\dagger}$ &  ... & ...    & ...  & ...  &   3.5e-8 & ...    & ...  & ...  & ...  & n &  ... & ...       & ...       & ...  \\
HH~1008 &  ...    & ...    & ...        &  ...  & ...    & ...      & ...  & ...  & ...  & n &  ... & ...       & ...       & ...  \\
HH~1009 &  ...    & ...    & ...        &  ...  & ...    & ...      & ...  & ...  & ...  & n &  ... & ...       & ...       & ...  \\
HH~1010 &  8.1    & 0.09   &  928 (115) &  1400 &   6.8e-8 & 2.4e-6 & 42.3 & -5.3 & -3.6 & n &   55 & 1.6 (1.0) & 1.8 (1.4) &  A   \\
HH~1011 & 1.35    & 0.02   & ...       &  ...   &   5.2e-8 & ...    & 36.9 & -7.6 & ...  & n &  ... & ...       & ...       & ...  \\
HH~1012 &  0.5    & 0.006  & ...        &   400 &   1.3e-8 & ...    & 40.2 & ...  & ...  & n &  610 & 1.4 (0.9) & 2.1 (0.6) &  II  \\
HH~1013 & 0.96    & 0.01   &  68 (70)   &   200 &   2.1e-8 & ...    & 41.4 & -6.2 & ...  & ? &  490 & 1.5 (0.1) & 2.4 (0.2) &  II  \\
HH~1014 &  4.3    & 0.05   &  668 (156) &  ...  &   4.4e-8 & 3.9e-6 & 41.6 & -5.5 & -3.6 & n &  984 & 2.3 (1.4) & 2.5 (1.8) & 0/I  \\
HH~1015$^{\dagger}$ & 7.2 & 0.08 & 868 (126) & ... & 8.4e-9 & 1.4e-7 & 41.2  & -6.1 & -4.9 & n &  538 & 1.3 (0.9) & 1.5 (1.2) & 0/I  \\
HH~1016 &  ...    & ...  & ...       &  ...     &   3.6e-8 & ...  & ...    & -5.4 & ...  & n &  ... & ...       & ...       & ...  \\
HH~1017 &  1.3    & 0.01   & 225 (174)  &   600 &   1.1e-8 & ...    & 40.5 & -6.2 & ...  & n &  634 & 1.8 (1.5) & 2.5 (1.2) &  A   \\
HH~1018 &  1.6    & 0.02   & 185 (117)  &  2500 &   1.5e-8 & ...    & 41.1 & -5.8 & ...  & n &  705 & 1.8 (1.1) & 2.2 (1.2) &  A   \\
HH~1019 & 11.4    & 0.13   & ...        &   500 & ...    & ...      & ...  & ...  & ...  & n &  ... & ...       & ...       & ...  \\
HH~1156 &  ...    & ...    & ...        &  ...  &      ... & 1.4e-6 & ...  & ...  & ...  & n &  986 & 1.4 (0.9) & 1.4 (1.1) & 0/I  \\
HH~1159 &  ...    & ...    & ...        &  ...  & ...    & ...      & ...  & ...  & ...  & n &  ... & ...       & ...       & ...  \\
HH~1160 &  9.2    & 0.10   & 1384 (152) &  ...  &   1.2e-8 & 2.1e-7 & 41.1 & -6.1 & -4.8 & ? &  787 & 2.1 (1.3) & 3.8 (1.3) & 0/I  \\
HH~1161 & 10.5    & 0.12   & 1296 (335) &  ...  &   4.0e-8 & 2.1e-7 & 42.2 & -5.3 & -4.6 & ? &  ... & ...       & ...       & ...  \\
HH~1162 &  ...    & ...    & ...        &  ...  & ...    & ...      & ...  & ...  & ...  & n &  ... & ...       & ...       & ...  \\
HH~1163 &  1.1    & 0.01   & 195 (177)  &  ...  &      ... & 7.2e-8 & ...  & ...  & -5.4 & y &  803 & 1.3 (0.2) & 1.8 (0.9) &  A   \\
HH~1164 &  1.7    & 0.2    & 365 (215)  &  ...  &      ... & 6.0e-8 & ...  & ...  & -5.5 & y &  790 & 1.9 (0.6) & 3.2 (1.2) & 0/I  \\
HH~1166  &  5.4    & 0.06   & 958 (180)  &  ...  &   3.3e-8 & ...    & 41.3 & -5.7 & ...  & n &  666 & 3.2 (0.2) & 7.3 (1.4) & 0/I  \\
HH~1167  &  2.8    & 0.03   & 467 (167)  &  ...  &   1.3e-8 & 8.4e-8 & 40.8 & -6.1 & -5.3 & n &  760 & 1.3 (0.9) & 1.4 (1.3) & 0/I  \\
HH~1168  &  ...    & ...    & ...        &  ...  & ...      & ...    & ...  & ...  & ...  & n &  ... & ...       & ...       & ...  \\
HH~1169 &  ...    & ...    & ...        &  ...  & ...      & ...    & ...  & ...  & ...  & n &  ... & ...       & ...       & ...  \\
HH~1170 &  ...    & ...    & ...        &  ...  & ...      & ...    & ...  & ...  & ...  & n &  ... & ...       & ...       & ...  \\
HH~1171 &  ...    & ...    & ...        &  ...  & ...      & ...    & ...  & ...  & ...  & n &  ... & ...       & ...       & ...  \\
HH~1172 & 0.75    & 0.01   &   61 (85)  &  ...  &   1.2e-8 & ...    & 41.4 & -5.5 & ...  & n &  401 & 1.6 (0.3) & 2.4 (0.5) &  II  \\
HH~1173 &  ...    & ...    & ...        &  ...  &   1.3e-8 & ...    & ...  & -6.1 & ...  & n & $\star$ & ...       & ...       & ...  \\
\hline
\multicolumn{15}{l}{$*$ YSO model fit for J104556.4-600608 from \citet{ohl12} } \\ 
\multicolumn{15}{l}{$^{\dagger}$ two sides of the same jet } \\ 
\multicolumn{15}{l}{$\star$ driving source identified as candidate proplyd 104354.91-593245 in \citet{smi10} } \\ 
\end{tabular}
\label{t:jet_props}
\end{table}
\end{landscape}


\clearpage 

\begin{onecolumn}
\begin{center}
\begin{longtable}{lrrrr}
\caption[Proper motions and transverse velocities]{Proper motions and transverse velocities} \label{t:jet_pm} \\
\hline\hline
\vspace{5pt}
Knot & $\delta x$ & $\delta y$ & v$_T$      & age \\ 
name & mas       & mas       & [km s$^{-1}$] & [yr] \\ 
\endfirsthead
\hline \hline 
\multicolumn{5}{c}{ \textbf{HH~666} } \\ 
\hline 
HH~666~D & -35 (10) & -11 (10) & 42 (12) & 34548 (8539) \\
HH~666~A & -46 (10) & -21 (10) & 58 (12) & 18048 (2835) \\
HH~666~E & -110 (10) & -62 (10) & 145 (12) & 5642 (162) \\
HH~666~M & -100 (10) & -39 (10) & 123 (12) & 661 (292) \\
HH~666~M1$^{\dagger}$ &  -125 (1) &  -34 (1) &    159 (4)  & 1108 (34) \\
HH~666~M14$^{\dagger}$ &  -122 (3) &  -30 (2) &    154 (5) & 1097 (42) \\
HH~666~M13$^{\dagger}$ &  -122 (1) &  -33 (3) &    155 (4) & 850 (27) \\ 
HH~666~M2$^{\dagger}$ &  -124 (1) &  -40 (1) &    159 (4)  & 789 (25) \\
HH~666~M3$^{\dagger}$ &  -113 (0.5) & -32 (0.2) &  143 (3) & 779 (24) \\
HH~666~M4$^{\dagger}$ &  -131 (1) &  -25 (2) &    163 (4)  & 593 (19) \\
HH~666~M9$^{\dagger}$ &  -126 (2) &  -20 (4) &    156 (4)  & 568 (19) \\ 
HH~666~M8$^{\dagger}$ &  -123 (1) &  -41 (1) &    159 (4)  & 537 (17) \\
HH~666~M5a$^{\dagger}$ &  -117 (1) &  -31 (2) &    149 (3) & 576 (18) \\ 
HH~666~M10$^{\dagger}$ &  -114 (2) &  -37 (1) &    147 (4) & 472 (16) \\
HH~666~M7$^{\dagger}$ &  -124 (1) &  -37 (1) &    158 (4)  & 405 (13) \\ 
HH~666~M11$^{\dagger}$ &  -95 (3) &  -19 (3) &    118 (5)  & 462 (21) \\ 
HH~666~M12$^{\dagger}$ &  -52 (0.1) &  -7 (0.3) &   64 (1) & 1193 (37) \\
HH~666~O1$^{\dagger}$ &   107 (2) &   66 (1) &    154 (4)  & 741 (24) \\
HH~666~O10$^{\dagger}$ &   123 (3) &   47 (3) &    162 (5) & 895 (35) \\
HH~666~O2$^{\dagger}$ &   88 (3) &   63 (6) &    132 (6)   & 1249 (66) \\
HH~666~O9$^{\dagger}$ &   117 (3) &   94 (5) &    184 (7)  & 989 (41) \\
HH~666~O8$^{\dagger}$ &   134 (2) &   67 (2) &    184 (5)  & 1017 (34) \\
HH~666~O3$^{\dagger}$ &   108 (1) &   54 (1) &    148 (4)  & 1363 (44) \\
HH~666~O4$^{\dagger}$ &   91 (1) &   59 (2) &    133 (3)   & 1564 (52) \\
HH~666~O5$^{\dagger}$ &   108 (2) &   47 (2) &    144 (4)  & 1434 (52) \\
HH~666~O6$^{\dagger}$ &   50 (67) &   52 (52) &  88 (73)   & 2156 (1792) \\ 
HH~666~O7$^{\dagger}$ &   9 (3) &   35 (2) &    45 (3)     & 3210 (194) \\
HH~666~O & 110 (10) & 38 (10) & 134 (12) & 1546 (190) \\
HH~666~N & 82 (10) & 45 (10) & 108 (12) & 3436 (31) \\
HH~666~Ia & 73 (10) & 24 (10) & 89 (12) & 6919 (420) \\
HH~666~Ib & 57 (10) & 27 (10) & 73 (12) & 9247 (873) \\
HH~666~C & 47 (10) & 19 (10) & 58 (12) & 22750 (3766) \\
HH~666~U & 51 (10) & 8 (10) & 59 (12) & 27181 (4575) \\
HH~666~S & 15 (10) & 26 (10) & 34 (12) & 54191 (17053) \\
\hline 
\multicolumn{5}{c}{ \textbf{HH~901} } \\ 
\hline 
HH~901~Y  & 44 (10)  & 5 (11)    & 51 (12) & 2482 (284) \\
HH~901~Ya & 51 (12)  & -8 (11)   & 60 (14) & 2110 (243) \\
HH~901~L  & 62 (11)  & 7 (11)    & 71 (13) & 1605 (314) \\
HH~901~S  & 64 (10)  & -2 (10)   & 74 (12) & 1113 (414) \\
HH~901~U  & 70 (11)  & 3 (10)    & 80 (12) & 904 (405) \\
HH~901~O  & 65 (10)  & -7 (10)   & 75 (12) & 726 (465) \\
HH~901~I  & -59 (10) & -0.6 (10) & 68 (12) & 591 (539) \\
HH~901~R  & -81 (10) & -7 (10)   & 94 (12) & 547 (3972) \\
HH~901~A  & -68 (11) & -9 (10)   & 79 (13) & 883 (409) \\
HH~901~F  & -79 (12) & -19 (11)  & 93 (13) & 833 (349) \\
HH~901~E  & -67 (15) & -3 (11)   & 77 (17) & 1138 (315) \\
HH~901~N  & -55 (10) & 0.3 (10)  & 64 (12) & 1486 (416) \\
\hline 
\multicolumn{5}{c}{ \textbf{HH~902} } \\ 
\hline 
HH~902~S & 75 (10) & -3 (10) & 86 (12) & 1672 (280) \\
HH~902~U & 62 (10) & -11 (10) & 72 (12) & 1513 (362) \\
HH~902~E & 27 (10) & -13 (10) & 35 (12) & 2657 (377) \\
HH~902~V & -88 (10) & 23 (10) & 105 (12) & 1361 (264) \\
HH~902~Ia & -133 (11) & 45 (10) & 161 (13) & 976 (196) \\
HH~902~Ib & -53 (10) & 13 (10) & 62 (12) & 2515 (232) \\
HH~902~L & -94 (10) & 21 (15) & 110 (12) & 1618 (219) \\
HH~902~B & -71 (10) & 14 (10) & 83 (12) & 2314 (201) \\
HH~902~O & -106 (10) & 26 (10) & 125 (12) & 1815 (178) \\
\hline 
\multicolumn{5}{c}{ \textbf{HH~1066} } \\ 
\hline 
HH~1066~W & -102 (10) & -36 (10) & 124 (12) & 304 (323) \\
HH~1066~Wa & -28 (10) & 22 (11) & 41 (12) & 742 (851) \\
HH~1066~E & 87 (40) & 159 (43) & 208 (49) & 190 (166) \\ 
\hline \hline
\multicolumn{5}{c}{ } \\ 
\hline \hline
\multicolumn{5}{c}{ \textbf{HH~900$^*$} } \\ 
\hline 
HH~900~A$^*$ &   43 (1) &  -24 (0.4) &    60 (1) & 3523 (115) \\
HH~900~B$^*$ &   31 (1) &  -2 (2)    &    37 (2) & 5364 (307) \\
HH~900~C$^*$ &   55 (0.1) &  -11 (1) &    68 (1) & 2803 (87) \\ 
HH~900~D$^*$ &   74 (1) &  -31 (4)   &    97 (3) & 859 (34) \\
HH~900~E$^*$ &   76 (2) &  -25 (1)   &    97 (3) & 682 (25) \\
HH~900~F$^*$ &  -60 (1) &   42 (2)   &    89 (2) & 691 (24) \\
HH~900~G$^*$ &  -77 (1) &   39 (1)   &    104 (2) & 746 (24) \\
HH~900~H$^*$ &  -40 (3) &   14 (1)   &    51 (3) & 5379 (365) \\
HH~900~I$^*$ &  -42 (1) &   28 (1)   &    61 (2) & 4664 (174) \\
\hline 
\multicolumn{5}{c}{ \textbf{HH~903} } \\ 
\hline 
HH~903~B & 51 (10) & 24 (10) & 69 (12) & 6190 (474) \\
HH~903~C & 49 (10) & 22 (10) & 66 (12) & 6013 (462) \\
HH~903~E & 83 (11) & 23 (11) & 106 (13) & 3090 (27) \\
HH~903~F & 112 (10) & 24 (10) & 140 (13) & 1231 (200) \\
HH~903~G & 64 (10) & 24 (10) & 84 (12) & 2347 (173) \\
HH~903~H & 28 (10) & 77 (10) & 101 (12) & 2236 (157) \\
HH~903~I & -75 (10) & -14 (10) & 93 (12) & 7455 (529) \\
HH~903~J & -73 (10) & -25 (10) & 95 (12) & 7460 (517) \\
HH~903~K & -49 (10) & 11 (10) & 61 (12) & 15164 (2338) \\
HH~903~L & -64 (10) & 9 (10) & 79 (12) & 12897 (1469) \\
HH~903~M & -94 (10) & 27 (10) & 120 (13) & 8815 (557) \\
HH~903~N & -59 (10) & 26 (10) & 79 (12) & 15043 (1806) \\
HH~903~O & -62 (10) & 12 (10) & 78 (12) & 15549 (1915) \\
HH~903~P & -71 (10) & 24 (10) & 92 (13) & 13363 (1343) \\
HH~903~Q & -30 (10) & -14 (10) & 40 (12) & 30824 (8385) \\
HH~903~R & -49 (10) & 10 (10) & 61 (12) & 21030 (3556) \\
HH~903~S & -47 (10) & 48 (10) & 83 (12) & 15100 (1738) \\
HH~903~T & -52 (10) & 14 (10) & 65 (12) & 19725 (3054) \\
HH~903~U & -47 (10) & 26 (10) & 66 (12) & 20281 (3121) \\
HH~903~V & -26 (10) & -24 (10) & 43 (12) & 31611 (8063) \\
HH~903~W & -48 (10) & 28 (10) & 68 (12) & 20112 (3031) \\
HH~903~X & -40 (10) & 48 (10) & 76 (12) & 17594 (2292) \\
HH~903~Z & -38 (10) & 59 (10) & 86 (12) & 14500 (1591) \\
HH~903~AA & -63 (10) & 38 (10) & 90 (12) & 13584 (1394) \\
HH~903~BB & -48 (10) & 15 (10) & 61 (12) & 19376 (3185) \\
HH~903~II & 52 (10) & 27 (10) & 72 (12) & 6549 (521) \\
HH~903~JJ & -64 (10) & 7 (10) & 79 (12) & 7995 (699) \\
HH~903~KK & -86 (10) & -10 (10) & 106 (12) & 6345 (335) \\
HH~903~LL & -50 (10) & 14 (10) & 64 (12) & 15465 (2304) \\
\hline 
\multicolumn{5}{c}{ \textbf{HH~1004} } \\ 
\hline 
HH~1004~A & 94 (10) & -13 (10) & 114 (13) & 1992 (163) \\
HH~1004~B & 138 (10) & -61 (10) & 183 (13) & 529 (202) \\
HH~1004~C & 128 (11) & -63 (10) & 173 (13) & 517 (212) \\
HH~1004~D & 56 (11) & -42 (11) & 85 (13) & 765 (393) \\
HH~1004~F & 38 (10) & -2 (10) & 46 (12) & 1101 (651) \\
HH~1004~G & 47 (10) & -2 (10) & 57 (12) & 896 (577) \\
HH~1004~H & 24 (10) & -27 (10) & 44 (12) & 833 (768) \\
HH~1004~I & -50 (10) & 25 (10) & 68 (12) & 1030 (459) \\
HH~1004~J & -119 (10) & 45 (10) & 154 (13) & 737 (222) \\
HH~1004~L & -24 (10) & 12 (10) & 33 (12) & 5682 (773) \\
HH~1004~M & -47 (10) & -1 (10) & 57 (12) & 3568 (1) \\
HH~1004~N & -45 (10) & 2 (10) & 55 (12) & 3819 (54) \\
HH~1004~O & -48 (10) & -4 (10) & 58 (12) & 3545 (6) \\
HH~1004~P & -128 (10) & 69 (10) & 176 (13) & 1612 (131) \\
HH~1004~R & -99 (10) & 1 (10) & 120 (13) & 2651 (85) \\
HH~1004~U & -101 (10) & 22 (10) & 125 (13) & 4187 (71) \\
HH~1004~X & -50 (10) & -2 (10) & 60 (12) & 11518 (1614) \\
\hline 
\multicolumn{5}{c}{ \textbf{HH~1005} } \\ 
\hline 
HH~1005~Q & -50 (10) & -26 (10) & 69 (12) & ...  \\
HH~1005~S & -126 (10) & 13 (10) & 153 (13) & ...  \\
HH~1005~T & -50 (10) & 45 (10) & 81 (12) & ...  \\
HH~1005~V & -101 (10) & -0.007 (10) & 123 (12) & ...  \\
HH~1005~W & -125 (10) & 49 (10) & 162 (13) & ...  \\
HH~1005~Y & 1 (10) & 1 (10) & 1 (12) & ...  \\
\hline 
\multicolumn{5}{c}{ \textbf{HH~1006} } \\ 
\hline 
HH~1006~A & -23 (11) & -150 (11) & 148 (11) & 366 (267) \\
HH~1006~B & -37 (10) & -129 (10) & 131 (10) & 308 (309) \\
HH~1006~C & 31 (11) & 157 (11) & 156 (11) & 531 (241) \\
HH~1006~D & 0.14 (11) & 44 (10) & 42 (10) & 6863 (627) \\
HH~1006~E & -3 (16) & 75 (11) & 73 (11) & 4172 (4) \\
HH~1006~F & -9 (10) & 64 (10) & 63 (10) & 5355 (148) \\
\hline 
\multicolumn{5}{c}{ \textbf{HH~1007} } \\ 
\hline 
HH~1007~A & 24 (10) & 57 (10) & 71 (12) & 1806 (312) \\
HH~1007~B & 32 (10) & 59 (10) & 77 (12) & 1600 (318) \\
HH~1007~C & -35 (10) & 0.09 (11) & 40 (12) & 2901 (231) \\
\hline 
\multicolumn{5}{c}{ \textbf{HH~1008} } \\ 
\hline 
HH~1008~B & -27 (10) & 48 (10) & 67 (12) & ...  \\
HH~1008~C & -21 (10) & 38 (10) & 53 (12) & ...  \\
HH~1008~F & -25 (10) & 27 (10) & 45 (12) & ...  \\
\hline 
\multicolumn{5}{c}{ \textbf{HH~1009} } \\ 
\hline 
HH~1009~A & 38 (10) & 5 (10) & 46 (12) & ...  \\
HH~1009~B & 87 (10) & 39 (11) & 115 (13) & ...  \\
HH~1009~C & 40 (10) & 8 (10) & 49 (13) & ...  \\
HH~1009~D & 74 (12) & 25 (11) & 95 (15) & ...  \\
HH~1009~E & 43 (12) & 24 (10) & 60 (14) & ...  \\
HH~1009~F & 50 (10) & 12 (10) & 62 (12) & ...  \\
HH~1009~G & 57 (10) & 34 (10) & 80 (12) & ...  \\
HH~1009~H & 73 (11) & 70 (12) & 122 (14) & ...  \\
HH~1009~J & 55 (13) & -28 (12) & 75 (15) & ...  \\
\hline 
\multicolumn{5}{c}{ \textbf{HH~1010} } \\ 
\hline 
HH~1010~P & -1 (10) & 38 (10) & 43 (12) & 2199 (421) \\
HH~1010~B & 100 (10) & -112 (10) & 173 (12) & 412 (224) \\
HH~1010~C & 37 (10) & -36 (11) & 60 (13) & 882 (543) \\
HH~1010~D & 66 (11) & -62 (12) & 105 (13) & 363 (372) \\
HH~1010~E & -21 (10) & -22 (10) & 35 (12) & 2382 (468) \\
HH~1010~F & -48 (10) & 30 (10) & 65 (12) & 1645 (379) \\
HH~1010~G & -14 (13) & 49 (11) & 59 (13) & 2061 (295) \\
HH~1010~H & -51 (11) & 91 (10) & 120 (12) & 1080 (255) \\
HH~1010~I & -75 (11) & 76 (11) & 122 (13) & 1128 (238) \\
HH~1010~J & -55 (11) & 88 (10) & 119 (12) & 1406 (224) \\
HH~1010~K & -50 (11) & 97 (11) & 126 (13) & 2114 (136) \\
HH~1010~L & -56 (11) & 75 (11) & 107 (13) & 3478 (8) \\
HH~1010~M & -3 (11) & 71 (14) & 81 (17) & 4853 (455) \\
HH~1010~N & -71 (11) & 99 (11) & 140 (13) & 2976 (42) \\
HH~1010~O & 19 (10) & 65 (10) & 78 (12) & 5357 (247) \\
HH~1010~A & 18 (12) & 25 (14) & 35 (16) & 13267 (4686) \\
\hline 
\multicolumn{5}{c}{ \textbf{HH~1011} } \\ 
\hline 
HH~1011~A & -0.3(10) & 0.05 (10) & 0.5 (14) & ...  \\
HH~1011~B & 6 (10) & 3 (10) & 9 (14) & ...  \\
\hline 
\multicolumn{5}{c}{ \textbf{HH~1012} } \\ 
\hline 
HH~1012~A & 50 (10) & 33 (10) & 65 (12) & 690 (548) \\
HH~1012~B & -18 (11) & -164 (10) & 181 (12) & 116 (234) \\
HH~1012~C & 5 (10) & -92 (11) & 101 (13) & 382 (383) \\
HH~1012~D & 45 (10) & -93 (10) & 113 (11) & 759 (310) \\
HH~1012~E & -4 (10) & -64 (10) & 71 (11) & 1864 (322) \\
\hline 
\multicolumn{5}{c}{ \textbf{HH~1013} } \\ 
\hline 
HH~1013~A & 61 (10) & -80 (10) & 110 (11) & 4070 (23) \\
HH~1013~B & 49 (10) & -81 (10) & 104 (11) & 3611 (29) \\
HH~1013~D & 24 (10) & -72 (10) & 83 (11) & 3935 (17) \\
HH~1013~F & 40 (10) & -53 (10) & 73 (11) & 4275 (61) \\
HH~1013~H & -65 (10) & 126 (10) & 155 (12) & 1697 (153) \\
HH~1013~I & -37 (11) & 58 (10) & 76 (12) & 2028 (265) \\
HH~1013~J & -48 (10) & 66 (10) & 90 (11) & 1780 (262) \\
\hline 
\multicolumn{5}{c}{ \textbf{HH~1014} } \\ 
\hline 
HH~1014~A & -55 (11) & -16 (11) & 69 (14) & 1109 (410) \\
HH~1014~B & -32 (14) & -22 (11) & 47 (16) & 1291 (503) \\
HH~1014~C & -22 (10) & -23 (10) & 38 (12) & 1432 (679) \\
HH~1014~D & -55 (10) & -24 (10) & 72 (12) & 715 (486) \\
HH~1014~E & -24 (10) & 7 (10) & 30 (12) & 1255 (952) \\
HH~1014~F & -72 (10) & -26 (10) & 93 (12) & 342 (425) \\
HH~1014~G & -56 (11) & -19 (10) & 71 (13) & 289 (557) \\
HH~1014~H & -66 (10) & -26 (10) & 85 (12) & 89 (498) \\
\hline 
\multicolumn{5}{c}{ \textbf{HH~1015} } \\ 
\hline 
HH~1015~D & -103 (11) & -39 (11) & 107 (11) & 517 (352) \\
HH~1015~E & -47 (10) & -59 (10) & 74 (10) & 887 (471) \\
\hline 
\multicolumn{5}{c}{ \textbf{HH~1016} } \\ 
\hline 
HH~1016~A & -21 (10) & -42 (12) & 58 (14) & ...  \\
HH~1016~C & -21 (12) & -41 (10) & 56 (13) & ...  \\
HH~1016~D & -37 (10) & -49 (10) & 75 (12) & ...  \\
HH~1016~E & -73 (14) & -96 (14) & 146 (18) & ...  \\
HH~1016~F & -67 (15) & -39 (11) & 94 (17) & ...  \\ 
\hline 
\multicolumn{5}{c}{ \textbf{HH~1017} } \\ 
\hline 
HH~1017~NE2-A & 49 (10) & -56 (10) & 85 (12) & 2071 (229) \\
HH~1017~NE2-B & -3 (10) & -1 (10) & 4 (12) & 37951 (98166) \\
HH~1017~NE1 & 51 (10) & -55 (10) & 86 (12) & 860 (392) \\
HH~1017~jet & -25 (10) & 49 (10) & 63 (12) & 171 (663) \\
HH~1017~SW1 & -47 (10) & 53 (10) & 81 (12) & 495 (466) \\
HH~1017~SW2 & -26 (10) & 74 (10) & 90 (12) & 1098 (341) \\
HH~1017~SW3 & -49 (10) & 70 (10) & 98 (12) & 1934 (214) \\
\hline 
\multicolumn{5}{c}{ \textbf{HH~1018} } \\ 
\hline 
HH~1018~A & -48 (11) & -62 (11) & 94 (14) & 140 (443) \\
HH~1018~B & -19 (10) & -26 (10) & 38 (12) & 2592 (311) \\
\hline 
\multicolumn{5}{c}{ \textbf{HH~1019} } \\ 
\hline 
HH~1019~A$^{\ddagger}$ & 38 (10) & -86 (10) & 103 (11) & 1754 (232) \\
HH~1019~B$^{\ddagger}$ & -0.07 (10) & -45 (10) & 50 (11) & 2030 (427) \\
HH~1019~C$^{\ddagger}$ & 3 (10) & 90 (10) & 98 (11) & 532 (383) \\
HH~1019~D$^{\ddagger}$ & -1 (10) & 40 (10) & 43 (11) & 2912 (266) \\
HH~1019~E$^{\ddagger}$ & -15 (10) & 74 (10) & 83 (11) & 1842 (279) \\
HH~1019~F$^{\ddagger}$ & -26 (10) & 102 (10) & 115 (11) & 1841 (199) \\
HH~1019~G$^{\ddagger}$ & -13 (10) & -88 (10) & 97 (11) & 1234 (307) \\
\hline 
\multicolumn{5}{c}{ \textbf{HH~1156} } \\ 
\hline 
HH~1156~K & -36 (11) & 12 (10) & 44 (12) & 7107 (1015) \\
HH~1156~L & -42 (12) & 26 (11) & 56 (13) & 5263 (486) \\
HH~1156~M & -47 (12) & 16 (10) & 57 (13) & 5089 (423) \\
HH~1156~N & -34 (10) & 12 (10) & 41 (12) & 6966 (931) \\
HH~1156~O & -29 (10) & 26 (11) & 44 (12) & 6391 (773) \\
HH~1156~I & 36 (10) & -65 (10) & 85 (12) & 1943 (246) \\
HH~1156~J & 25 (10) & -51 (10) & 65 (12) & 2571 (206) \\
HH~1156~P & 10 (10) & -59 (10) & 69 (12) & 3671 (2) \\
\hline 
\multicolumn{5}{c}{ \textbf{HH~1159} } \\ 
\hline 
HH~1159~A & 19 (10) & 51 (10) & 65 (12) & ...  \\
HH~1159~B & -1 (10) & 24 (10) & 29 (12) & ...  \\
HH~1159~C & -1 (10) & 25 (10) & 31 (12) & ...  \\
HH~1159~D & -1 (10) & 46 (10) & 56 (12) & ...  \\
HH~1159~E & -2 (10) & 50 (10) & 61 (12) & ...  \\
\hline 
\multicolumn{5}{c}{ \textbf{HH~1160} } \\ 
\hline 
HH~1160~A & -51 (10) & -23 (10) & 68 (12) & 1473 (382) \\
HH~1160~B & -24 (10) & -57 (10) & 75 (12) & 1235 (384) \\
HH~1160~C & -37 (10) & -49 (10) & 75 (12) & 1130 (400) \\
HH~1160~D & 37 (10) & 38 (10) & 65 (12) & 765 (534) \\
\hline 
\multicolumn{5}{c}{ \textbf{HH~1161} } \\ 
\hline 
HH~1161~A & -4 (10) & 100 (10) & 121 (12) & 547 (308) \\
HH~1161~B & -1 (10) & 101 (10) & 123 (12) & 498 (308) \\
HH~1161~C & -14 (10) & 100 (10) & 122 (12) & 668 (292) \\
HH~1161~D & 60 (10) & -39 (10) & 87 (12) & 1059 (357) \\
HH~1161~E & 1 (10) & -27 (10) & 32 (12) & 3551 (13) \\
\hline 
\multicolumn{5}{c}{ \textbf{HH~1162} } \\ 
\hline 
HH~1162~A & -1 (10) & 23 (10) & 28 (12) & ... \\
HH~1162~B & -8 (10) & 25 (10) & 32 (12) & ... \\
HH~1162~C & 13 (10) & 43 (10) & 55 (12) & ... \\
HH~1162~D & 11 (10) & 12 (10) & 20 (12) & ... \\
HH~1162~E & -25 (10) & 26 (10) & 45 (12) & ... \\
HH~1162~F & 4 (10) & 48 (10) & 59 (12) & ... \\
HH~1162~G & 23 (10) & 48 (10) & 65 (12) & ... \\
\hline 
\multicolumn{5}{c}{ \textbf{HH~1163} } \\ 
\hline 
HH~1163~A & 45 (10) & -24 (10) & 62 (12) & 209 (674) \\
\hline 
\multicolumn{5}{c}{ \textbf{HH~1164} } \\ 
\hline 
HH~1164~A & 36 (10) & 25 (10) & 54 (12) & 270 (759) \\
HH~1164~B & 32 (10) & 24 (10) & 48 (12) & 405 (811) \\
\hline 
\multicolumn{5}{c}{ \textbf{HH~1166} } \\ 
\hline 
HH~1166~A & -12 (10) & -49 (10) & 61 (12) & 20641 (3393) \\
\hline 
\multicolumn{5}{c}{ \textbf{HH~1167} } \\ 
\hline 
HH~1167~A & -49 (10) & 22 (10) & 65 (12) & 428 (591) \\
\hline 
\multicolumn{5}{c}{ \textbf{HH~1168} } \\ 
\hline 
HH~1168~A & 36 (10) & -10 (10) & 41 (11) & ... \\
\hline 
\multicolumn{5}{c}{ \textbf{HH~1169} } \\ 
\hline 
HH~1169~A & -42 (10) & 37 (10) & 69 (13) & 3302 (33) \\
\hline 
\multicolumn{5}{c}{ \textbf{HH~1170} } \\ 
\hline 
HH~1170~A & -49 (10) & 49 (10) & 84 (12) & ... \\
\hline 
\multicolumn{5}{c}{ \textbf{HH~1171} } \\ 
\hline 
HH~1171~A & -23 (10) & 46 (10) & 62 (12) & ... \\
HH~1171~B & -47 (13) & 49 (11) & 82 (14) & ... \\
HH~1171~C & -30 (11) & 141 (12) & 174 (15) & ... \\
HH~1171~D & 4 (12) & 27 (11) & 34 (13) & ... \\
HH~1171~E & 31 (12) & 74 (16) & 98 (19) & ... \\
\hline 
\multicolumn{5}{c}{ \textbf{HH~1172} } \\ 
\hline 
HH~1172~A & -137 (87) & 134 (41) & 235 (84) & 31 (174) \\
\hline 
\multicolumn{5}{c}{ \textbf{HH~1173} } \\ 
\hline 
HH~1173~A & -36 (10) & -35 (10) & 58 (12) & 534 (648) \\
HH~1173~B & -50 (10) & -10 (10) & 59 (12) & 461 (653) \\
\hline
\multicolumn{5}{l}{$^{\dagger}$ proper motions measured by \citet{rei15b}} \\ 
\multicolumn{5}{l}{ \textbf{$^*$} proper motions measured by \citet{rei15a}} \\ 
\multicolumn{5}{l}{$^{\ddagger}$ proper motions measured by \citet{rei17}} \\ 
\end{longtable}
\end{center} 
\end{onecolumn}

\label{lastpage}

\end{document}